\documentclass[amssymb,
               amsmath,
               floatfix,
               pre,
               onecolumn,
               notitlepage,
               longbibliography,
               superscriptaddress]{revtex4-1}

\usepackage{libertine}
\usepackage[libertine,
            slantedGreek,
            bigdelims,
            vvarbb
            ]{newtxmath}
\usepackage{graphicx}
\usepackage{dcolumn}
\usepackage{bm}
\usepackage{color}
\usepackage{xcolor}
\usepackage{soul}

\usepackage{bbm}

\usepackage{hyperref}
\hypersetup{
    colorlinks=true, linktocpage=true, pdfstartpage=1, pdfstartview=FitV,
    breaklinks=true, pdfpagemode=UseNone, pageanchor=true, pdfpagemode=UseOutlines,%
    plainpages=false, bookmarksnumbered, bookmarksopen=true, bookmarksopenlevel=1,%
    hypertexnames=true, pdfhighlight=/O,
    urlcolor=webbrown, linkcolor=RoyalBlue, citecolor=webgreen,
    pdftitle={Dissipation in Hydrodynamics},%
    pdfauthor={Danilo Forastiere, Francesco Avanzini, Massimiliano Esposito}%
    pdfsubject={},%
    pdfkeywords={},%
    pdfcreator={pdfLaTeX},%
    pdfproducer={LaTeX REVTeX}%
  }

\makeatletter
\DeclareFontFamily{OMX}{MnSymbolE}{}
\DeclareSymbolFont{MnLargeSymbols}{OMX}{MnSymbolE}{m}{n}
\SetSymbolFont{MnLargeSymbols}{bold}{OMX}{MnSymbolE}{b}{n}
\DeclareFontShape{OMX}{MnSymbolE}{m}{n}{
    <-6>  MnSymbolE5
   <6-7>  MnSymbolE6
   <7-8>  MnSymbolE7
   <8-9>  MnSymbolE8
   <9-10> MnSymbolE9
  <10-12> MnSymbolE10
  <12->   MnSymbolE12
}{}
\DeclareFontShape{OMX}{MnSymbolE}{b}{n}{
    <-6>  MnSymbolE-Bold5
   <6-7>  MnSymbolE-Bold6
   <7-8>  MnSymbolE-Bold7
   <8-9>  MnSymbolE-Bold8
   <9-10> MnSymbolE-Bold9
  <10-12> MnSymbolE-Bold10
  <12->   MnSymbolE-Bold12
}{}

\let\llangle\@undefined
\let\rrangle\@undefined
\DeclareMathDelimiter{\llangle}{\mathopen}%
                     {MnLargeSymbols}{'164}{MnLargeSymbols}{'164}
\DeclareMathDelimiter{\rrangle}{\mathclose}%
                     {MnLargeSymbols}{'171}{MnLargeSymbols}{'171}
\makeatother
\providecommand*{\davg}[1]{\left\llangle#1\right\rrangle}

\DeclareMathOperator{\e}{\mathrm{e}} 
\renewcommand*{\i}{\mathrm{i}}

\newcommand*{\abs}[1]{\left\lvert{#1}\right\rvert}

\renewcommand*{\vec}[1]{\bm{#1}} 

\newcommand{\tr}{^{\mathsf{T}}}

\newcommand{\pos}{\vec{x}}

\newcommand{\force}{\vec F}

\newcommand{\vel}{{\vec{v}}}
\newcommand{\vele}{{{v}}}

\newcommand{\bath}{{b}}

\newcommand{\bk}[1]{\left\langle#1\right\rangle}

\newcommand{\mucost}{{\overline{\mu}}}
\newcommand{\tpotmacro}{Y}
\newcommand{\tpotmeso}{\bk{y}}

\newcommand{\aden}{{\bk{\rho}}}
\newcommand{\avel}{{\bk{\vec v}}}
\newcommand{\avele}[1]{{\bk{v_{#1}}}}

\newcommand{\aten}{{\bk{e}}}
\newcommand{\aien}{{\bk{u}}}
\newcommand{\aent}{{\bk{s_\text{B}}}}
\newcommand{\ament}{{{S_\text{B}}}}

\newcommand{\apres}{{\bk{P}}}
\newcommand{\aprese}[1]{{\bk{P_{#1}}}}
\newcommand{\apreseoff}[1]{{\bk{\Pi_{#1}}}}
\newcommand{\aheat}{{\bk{\vec J_q}}}

\newcommand{\aepr}{{\bk{\dot\sigma_\text{B}}}}
\newcommand{\aef}{{\bk{\vec J_s}}}

\newcommand{\ef}{{\dot{S}_\varepsilon}}

\makeatletter
\renewcommand*{\d}%
{\@ifnextchar^{\DIfF}{\DIfF^{}}}
\def\DIfF^#1{%
  \mathop{\mathrm{\mathstrut d}}%
  \nolimits^{#1}\gobblespace
}
\def\gobblespace{%
  \futurelet\diffarg\opspace}
\def\opspace{%
  \let\DiffSpace\!%
  \ifx\diffarg(%
  \let\DiffSpace\relax
  \else
  \ifx\diffarg\[%
    \let\DiffSpace\relax
    \else
    \ifx\diffarg\{%
    \let\DiffSpace\relax
    \fi\fi\fi\DiffSpace
  }

\DeclareMathOperator{\Tr}{Tr}

\providecommand*{\avg}[1]{\left\langle#1\right\rangle}

\definecolor{webgreen}{rgb}{0,.5,0}
\definecolor{webbrown}{rgb}{.6,0,0}
\definecolor{grigio}{rgb}{.85,.85,.85} 
\definecolor{RoyalBlue}{rgb}{0.0, 0.14, 0.4}
\definecolor{skyblue1}{rgb}{0.45,0.62,0.81}
\definecolor{skyblue2}{rgb}{0.2,0.39,0.64}
\definecolor{skyblue3}{rgb}{0.13,0.29,0.53}
\definecolor{scarlet1}{rgb}{0.93,0.16,0.16}
\definecolor{scarlet2}{rgb}{0.8,0,0}
\definecolor{scarlet3}{rgb}{0.64,0,0}

\begin{document}
\title{ Dissipation in Hydrodynamics from Micro- to Macroscale:\\
Wisdom from Boltzmann and Stochastic Thermodynamics}

\newcommand\unilu{\affiliation{Complex Systems and Statistical Mechanics, Department of Physics and Materials Science, University of Luxembourg, L-1511 Luxembourg City, Luxembourg}}
\newcommand\unipdfis{\affiliation{Department of Physics and Astronomy ``Galileo Galilei'', University of Padova, Via F. Marzolo, 8, I-35131 Padova, Italy}}
\newcommand\infn{\affiliation{INFN, Sezione di Padova, via Marzolo 8, I-35131 Padova, Italy}}
\newcommand\unipdchim{\affiliation{Department of Chemical Sciences, University of Padova, Via F. Marzolo, 1, I-35131 Padova, Italy}}

\author{Danilo Forastiere}
\email{danilo.forastiere@unipd.it}
\unilu 
\unipdfis
\infn

\author{Francesco Avanzini}
\email{francesco.avanzini@unipd.it}
\unilu
\unipdchim

\author{Massimiliano Esposito}
\email{massimiliano.esposito@uni.lu}
\unilu

\date{\today}

\begin{abstract}
We show that macroscopic irreversible thermodynamics for viscous fluids can be derived from exact information-theoretic thermodynamic identities valid at the microscale.
Entropy production, in particular, is a measure of the loss of many-particle correlations in the same way in which it measures the loss of system-reservoirs correlations in stochastic thermodynamics (ST).
More specifically, we first show that boundary conditions at the macroscopic level define a natural decomposition of the entropy production rate (EPR) in terms of thermodynamic forces multiplying their conjugate currents, as well as a change in suitable nonequilibrium potential that acts as a Lyapunov function in the absence of forces.
Moving to the microscale, we identify the exact identities at the origin of these dissipative contributions for isolated Hamiltonian systems.
We then show that the molecular chaos hypothesis, which gives rise to the Boltzmann equation at the mesoscale, leads to a positive rate of loss of many-particle correlations, which we identify with the Boltzmann EPR.
By generalizing the Boltzmann equation to account for boundaries with nonuniform temperature and nonzero velocity, and resorting to the Chapman--Enskog expansion, we recover the macroscopic theory we started from.
Finally, using a linearized Boltzmann equation we derive ST for dilute particles in a weakly out-of-equilibrium fluid and its corresponding macroscopic thermodynamics.
Our work unambiguously demonstrates the information-theoretical origin of thermodynamic notions of entropy and dissipation in macroscale irreversible thermodynamics.  
\end{abstract}
\maketitle

\section{Introduction}
\label{sec:intro}

Hydrodynamics is a theory of crucial importance across science.
It provides tools to deal with systems spanning many orders of magnitude in terms of spatial scale, and thus it is fundamental for our understanding of phenomena ranging from chemical physics and biology to planetary science and astrophysics~\cite{dgm1962nonequilibrium,  de2006hydrodynamic,phillips2012physical, peixoto1992physics, chandrasekhar2013hydrodynamic}. 
One of its simplest formulations describes an isotropic fluid in terms of five dynamical fields, total energy (kinetic, potential and internal), the three components of momentum, and mass.
These are conserved at the microscopic level and thus give rise to balance equations at the macroscopic level~\cite{chaikin1995principles, gaspard2022statistical}.
These equations are however not closed from a dynamical standpoint.
In irreversible thermodynamics, a complete thermodynamic description of a fluid assumes the existence of a local entropy which is solely a function of internal energy, volume and particle number, as in equilibrium. 
This so-called local equilibrium assumption defines temperature, pressure and chemical potential fields as partial derivatives of the entropy field and links them to the total energy, momentum and mass density fields. 
A second assumption, which defines linear irreversible thermodynamics, is that the fluxes featured in the balance equations (heat flux and pressure tensor) can be linearized in terms of the gradients of temperature and velocity. 
This linear regime assumption is essential to make the balance equations a closed dynamical system~\cite{dgm1962nonequilibrium, de2006hydrodynamic, gallavotti2002foundations}.

The microscopic underpinnings of irreversible thermodynamics have been the object of intense scrutiny for more than a century and a half~\cite{chapman1990mathematical, dgm1962nonequilibrium, lanford1975time,  cercignani1988boltzmann, spohn2012large, gaspard1998chaos, gaspard2022statistical}.
The main problems, which remain still open today, are related to understanding the emergence of a macroscopic irreversible behavior, from  microscopic reversible many-body dynamics.
The first problem concerns how to go from the microscopic dynamics of a gas (formulated in terms of the many-body probability distribution of all the fluid particles) which is Hamiltonian and thus reversible, to a mesoscopic, irreversible closed evolution equation for the single-particle probability distribution (for example, via the Boltzmann equation~\cite{chapman1990mathematical,  cercignani1988boltzmann, gaspard2022statistical}).
The second problem is the derivation of irreversible thermodynamics at the macroscopic level from the mesoscopic level (for example, using the Chapman--Enskog expansion~\cite{chapman1990mathematical, prigogine1949domaine, van1987chapman}).
Irreversibility (\emph{i.e.}, a nonnegative EPR) in this context is understood to arise from the molecular chaos hypothesis which neglects correlations amongst particles building after molecular collisions~\cite{spohn2012large, gaspard1998chaos}.

More recently, significant progress has been achieved in ST~\cite{jarzynski1997nonequilibrium, seifert2012stochastic, van2015ensemble, peliti2021stochastic, strasberg2022quantum}, which offers a different approach to deal with the mesoscopic level of description.
Here the typical setup consists of a system in contact with one or multiple reservoirs.
At the microscopic level of description, the full probability distribution contains both the system and reservoirs degrees of freedom~\cite{van1992stochastic, breuer2002theory}. 
Exact identities corresponding to the first and second law of thermodynamics have been derived when considering a class of initial conditions where the system is prepared in an arbitrary state but is uncorrelated from the reservoirs that are at equilibrium.
The entropy production is non-negative and takes the form of a relative entropy between the exact total probability distribution at a given time and the exact system probability distribution times the equilibrium reservoir distribution \cite{jarzynski1999microscopic, esposito2010entropy}.
It thus measures the information lost when both the system-reservoirs correlations and the displacement from equilibrium of the reservoirs are neglected \cite{esposito2010entropy, ptaszynski2019entropy}. 
Various assumptions need to be made to derive a closed dynamics for the system probability distribution only (obtained by tracing out the reservoirs) \cite{breuer2002theory}. 
The most common approach relies on the Born--Markov approximation (assuming fast reservoirs and weak system-reservoir coupling \cite{breuer2002theory, strasberg2022quantum}) which plays a role similar to the molecular chaos hypothesis in the context of fluids~\cite{dgm1962nonequilibrium, cercignani1988boltzmann}. 
It also makes it possible to express all the thermodynamic quantities entering the first and second law in terms of the system probability distribution only and to prove that the EPR is non-negative~\cite{soret2022thermodynamic, strasberg2022quantum}.
The dynamics of the system only perceives the linear response properties of the reservoirs via transition rates --- or drift and diffusion coefficients in a continuous description --- which as a result satisfy the so-called local detailed balance property \cite{esposito2009nonequilibrium, esposito2012stochastic, falasco2021local, maes2021local}.
This latter condition is known to be essential to build a consistent nonequilibrium thermodynamic description for a system described by a stochastic dynamics \cite{kurchan1998fluctuation, crooks1999entropy, seifert2005entropy, rao2018conservation, maes2021local}.
These approaches hold for a large class of mesoscopic models such as Markov jump processes, overdamped and underdamped Fokker--Planck equations~\cite{seifert2005entropy, esposito2010threefaces,  van2010threefaces, esposito2010three}.

Furthermore, the formulation of ST for Markov jump processes enables a decomposition of the EPR which provides a constructive identification of the nonequilibrium potential and of the global thermodynamic forces driving the system out of equilibrium~\cite{rao2018conservation}.
Close to equilibrium, it reproduces the classical results from linear irreversible thermodynamics, such as Onsager reciprocal relations and minimum entropy production principle~\cite{lebowitz1999gallavotti, andrieux2004fluctuation, forastiere2022linear}.
This decomposition of the EPR remains valid for macroscopic systems when assuming local equilibrium but without resorting to the linear regime assumption.  
This fact has been shown for chemical reaction networks with diffusion (but neglecting inertial or viscous effects), where deterministic field theories ensue~\cite{falasco2018information, avanzini2019thermodynamics}.
This construction reveals substantial differences between global and local thermodynamics because all global conservation laws can be broken at the local level~\cite{avanzini2019thermodynamics}.
These macroscopic thermodynamic theories with local equilibrium can be explicitly constructed from an underlying ST by scaling up particle numbers (see Refs.~\cite{rao2018conservationII, lazarescu2019large,  freitas2021stochastic}, for chemical reaction networks, Potts models and electronic circuits, respectively, and Refs. \cite{freitas2022emergent, falasco2024RMP} for the general arguments).
These approaches correspond to overdamped dynamics where inertial effects are neglected, but generalization to underdamped dynamics seems possible and could be used to derive hydrodynamics.

In this paper, 
we revisit the derivation of hydrodynamics from the microscopic to the mesoscopic and macroscopic levels and establish the corresponding thermodynamics at each of these levels, in light of recent progress in ST.
In particular, we give an information-theoretic interpretation to the entropy production and to the thermodynamic potentials of macroscopic hydrodynamics in the same fashion as it is done in ST \cite{esposito2012stochastic, rao2018conservation}. 
Furthermore, we clarify the role of the two distinct assumptions of local equilibrium and local detailed balance, which are of fundamental importance for hydrodynamics and ST, respectively, by constructing an example of mesoscopic evolution that allows us to discuss both. 
More specifically, in the first part of the paper we focus on macroscopic hydrodynamics, where we decompose the EPR in terms of the constraints imposed at the boundaries (e.g., a nonuniform temperature profile).
This allows us to identify i) the global thermodynamic forces driving the system out of equilibrium and ii) the proper thermodynamic potential.
When the thermodynamic forces are switched off and no time-dependent protocol is externally applied, the thermodynamic potential is maximized in the relaxation process to equilibrium, for which it acts as a Lyapunov function of the nonlinear dynamics.
This shows how thermodynamic consistency (\emph{i.e.}, the requirement that the heat flux satisfies the second law) translates into a dynamical result for a fully nonlinear theory.
In the second part of the paper, we focus on microscopic and mesoscopic descriptions behind macroscopic hydrodynamics to reveal the information-theoretic origin of the thermodynamic potential and EPR derived in the first part. 
To do so, we revisit the standard derivation of hydrodynamics which starts from the microscale (Hamiltonian dynamics for particles interacting via pairwise potentials) and goes to the macroscale via the mesoscale (Boltzmann equation) employing the Chapman--Enskog expansion \cite{dgm1962nonequilibrium, cercignani1988boltzmann, spohn2012large, gaspard2022statistical}.
For an isolated system, when considering an initially  factorized distribution in terms of (identical) one-particle distributions, we show that the entropy production (identified with the change in Shannon entropy of the one-particle distribution) is always non-negative. 
Furthermore, in conceptual analogy with the system-reservoir setup, it takes the form of a relative entropy between the exact many-body probability distribution and the product of the one-particle distributions.
Entropy production is thus a direct measure of the information lost when neglecting inter-particle correlations.
When resorting to the molecular chaos hypothesis -- used to derive a mesoscopic description in terms of the Boltzmann equation which ensures the relaxation of the one-particle distribution to equilibrium -- the EPR becomes non-negative. 
Further resorting to the Chapman--Enskog expansion~\cite{dgm1962nonequilibrium, cercignani1988boltzmann, kardar2007statistical} and the relaxation-time approximation~\cite{lundstrom2002fundamentals} used to derive macroscopic hydrodynamics, we identify the parts of the EPR which survive and give rise to the entropy production of irreversible thermodynamics.
We then move to non-isolated systems which we model at the level of the Boltzmann equation by describing the collisions at nonisothermal and possibly moving boundary in a thermodynamically consistent way inspired by Refs.~\cite{bergmann1955new, lebowitz1957irreversible, cercignani1988boltzmann, gaspard2022statistical}.
We show that a first and second law ensue and that in the absence of thermodynamic forces, the nonequilibrium Massieu potential is maximized by the relaxation dynamics towards equilibrium.
Finally, we derive a linear Boltzmann equation, similar to a master equation in the momentum space.
This mesoscale theory provides a formulation of ST underlying macroscopic hydrodynamics in the regime where the Chapman--Enskog approach is justified.
In fact, the system is shown to satisfy  hydrodynamic equations at the macroscale  while also obeying the local detailed balance at the mesoscopic level, providing, in addition, a systematic treatment of its ST in the presence of an extended and weakly out-of-equilibrium reservoir \cite{van2015stochastic, horowitz2016work, falasco2021local}. 
The main outcome of our work is to unambiguously show how microscopic and mesoscopic formulations of nonequilibrium thermodynamics in terms of information-theoretic quantities are consistent with  the macroscopic thermodynamics of non-isolated systems.

\subsection*{Outline} 
The paper is organized as follows. 
In \S~\ref{sec:macro_balances}, we consider the dynamics of continuous systems and their thermodynamics, we write the balance equations for energy and entropy and then identify the proper thermodynamic potentials and forces for isolated and non-isolated systems.
In \S~\ref{sec:microscopic_hamiltonian}, we derive thermodynamic inequalities for the temporal change in information-theoretic entropy and nonequilibrium Massieu potential at the microscopic scale following Hamiltonian dynamics by using a coarse-grained approach based on the one-particle distribution function.
In \S~\ref{sec:mesoscopic_boltzmann}, we move to the mesoscopic scale and revisit the thermodynamic aspects of the microscopic derivation of the hydrodynamics of isolated systems using the Boltzmann equation.
We use the Chapman--Enskog expansion to obtain the local equilibrium condition as the leading approximation in terms of the slowly varying variables describing macroscopic systems.
We furthermore estimate the EPR during the initial stages of the relaxation to the local equilibrium.
In \S~\ref{sec:mesoscopic_boltzmann_open}, we generalize the Boltzmann equation to describe non-isolated systems in a thermodynamically consistent fashion.
By writing explicitly the energy and entropy balance equations,
we show how the nonequilibrium Massieu potential corresponds to the macroscopic thermodynamic potential identified in \S~\ref{sec:macro_balances}.
Finally, in \S~\ref{sec:stochastic_thermo}, we formulate a linear version of the generalized Boltzmann equation that is consistent with both ST and macroscopic hydrodynamics.
Future developments are discussed in \S~\ref{sec:conclusions}.

\section{Macroscopic Scale: Navier-Stokes Equations}
\label{sec:macro_balances}

We revisit here the standard hydrodynamics of viscous fluids~\cite{dgm1962nonequilibrium, kardar2007statistical}.
We start by obtaining the local formulation of thermodynamics (\emph{i.e.}, the one valid in the interior of a domain $\Omega$) and then systematically show how to determine the thermodynamic potential as well as the global thermodynamic forces keeping non-isolated fluids out of equilibrium (which require knowledge of the conditions imposed at the boundary $\partial \Omega)$.

\subsection{Dynamics\label{sub:dynamics}}
The dynamics of fluids is given by the Navier-Stokes equations~\cite{dgm1962nonequilibrium}, 
written in Lagrangian form using the material derivative $D_t \equiv \partial_t + \vec{v}\cdot \nabla_{\pos}$ as
\begin{subequations}
\begin{align}
   D_t \rho &
   = -\rho \nabla_{\pos} \cdot \vec{v} \label{eq:mass_conservation_lagrangian}  \,,\\
    \rho D_t \vec v &= - \nabla_{\pos} \cdot P + \rho \force
     \label{eq:momentum_conservation}\,,
\end{align}
\label{eq:NS}%
\end{subequations}
for the spatio-temporal fields density of mass $\rho(\vec x; t)$, velocity  $\vec{v}(\vec x; t)$, external force (per unit mass) $\force(\vec x; t)$, and pressure tensor $P(\vec x; t)$
(with $(\nabla_{\pos} \cdot P)_{j} = \sum_i \partial_{x_i} P_{ij}$,
and $i$ and $j$ labeling the three spatial components).
The pressure tensor $P$ describes the stress exerted by the fluid element on its surrounding and, in the case of isotropic, compressible fluids, can be written in terms of the scalar pressure $p(\vec x; t)$ and the (symmetric) viscous tensor $\Pi(\vec x; t)$.
In turn, for fluids sufficiently close to thermodynamic equilibrium, $\Pi(\vec x; t)$ is a function of the derivatives of the velocity~\cite{landau1987fluid}:
\begin{align}
    P_{ij}(\vec x; t) 
    = p(\vec x; t) \delta_{ij} - \Pi_{ij}(\vec x; t) 
    =  p(\vec x; t)  \delta_{ij} 
    -\left(\delta_{ij}\lambda \nabla_{\pos}\cdot\vec{v}(\vec x; t) + \nu\rho(\vec x; t) \left(\partial_{x_j} v_i(\vec x; t) + \partial_{x_i}  v_j(\vec x; t) \right) \right)  \,. \label{eq:stress_tensor}
\end{align}
This expression of $P$ is obtained under a series of assumptions, valid in the linear regime.
First, $P$ is assumed to be invariant under global Galilean transformations, \emph{i.e.}, $\vec{v}({\pos};t)\mapsto \vec{v}({\pos};t)+\vec{V}$, implying that it depends on the velocity only through its derivatives.
Second, $P$ is assumed to depend only on the first derivatives $\partial_{x_i}v_j$, which are supposed to be small enough to contribute  only linearly to $P$.
Third, the pressure tensor $P$ is assumed to be symmetric, which is justified when the molecules composing the fluid are spherical or when the fluid density is very low~\cite{dgm1962nonequilibrium}.
Fourth, assuming isotropy, only two different scalar coefficients, ${\nu}$ and $\lambda$, are needed to characterize viscous effects in the fluid (out of six possible ones). 
The coefficient $ {\nu}$, accounting for the shear, is called kinematic viscosity (while  $\rho{\nu}$  is called dynamic viscosity)
and the coefficient $\lambda$, accounting for dissipative compression or expansion, is called second viscosity.
When it is not specified otherwise, we will not rely on the specific expression of $\Pi_{ij}$ in~\eqref{eq:stress_tensor} in the rest of the article.

The mass conservation~\eqref{eq:mass_conservation_lagrangian} can equivalently be written in Eulerian form as 
\begin{align}
\partial_t \rho &= - \nabla_{\pos} \cdot (\rho \vec{v})\,.
\label{eq:mass_conservation_eulerian}
\end{align}
Since we will often switch between the Lagrangian and Eulerian description, we rely on this useful identity valid for any scalar quantity~$o$
\begin{align}
    \rho D_t o = \partial_t (\rho o) + \nabla_{\pos} \cdot (\rho o \vec{v})\,, \label{eq:identity}
\end{align}
which is derived using the definition of $D_t$ and Eq.~\eqref{eq:mass_conservation_eulerian}. 

Note that throughout this manuscript, we use a compact notation in which any spatial and temporal dependence (of, for instance, $\rho$ and $\vec{v}$) is left implicit if no ambiguity can arise.

\subsection{Local Energy Balance}
\label{sec:macro_energy_balances}

The specific total energy, per unit mass, 
\begin{align}
    e = k + \phi + u \,, \label{eq:total_energy_def}
\end{align} 
is given by three different contributions. 
The first, $k(\vec x; t)=\frac{1}{2}(v(\vec x; t))^2$, is  the specific kinetic energy associated to the motion of the fluid element.
The second, $\phi(\vec x; t)$,  is the specific potential energy of the fluid element
which can always be written as the sum 
of two terms:
$\phi(\vec{x};t)=\phi_\mathrm{int}(\vec{x}; t)+\phi_{\mathrm{w}}(\vec{x};t)$.
The term $\phi_\mathrm{int}(\vec{x}; t)$ accounts for the potential energy  in the bulk of the fluid.
An example of this contribution is the gravitational potential energy satisfying $- \nabla_{\pos} \phi_\mathrm{g}(\vec x) = -g \vec{u} = \vec{F}_\mathrm{g}$ 
with $\vec{F}_\mathrm{g}$ the gravitational force (per unit mass), 
$g$ the modulus of the gravitational acceleration,
and $\vec{u}$ the unit vector in the vertical direction.
The term $\phi_{\mathrm{w}}(\vec{x};t)$ accounts instead for the confinement generated by the (moving) boundaries.
We assume that all possible forces acting on a fluid element derive from a suitable  potential energy, and therefore $\vec{F} = -\nabla_{\vec{x}}\phi$
{($\vec{F}_{\mathrm{int}} = -\nabla_{\vec{x}}\phi_\mathrm{int}$ 
and $\vec{F}_{\mathrm{w}} = -\nabla_{\vec{x}}\phi_{\mathrm{w}}$)}
in the following.
The last contribution in Eq.~\eqref{eq:total_energy_def}, $u(\vec x;t)$, is the specific internal energy associated to all the other degrees of freedom. 
The corresponding global quantities are 
\begin{subequations}
\begin{align}
E &\equiv\int \d\vec{x} \rho(\vec{x})e(\vec{x})\,,\\
K&\equiv\int \d\vec{x} \rho(\vec{x})k(\vec{x})\,,\label{eq:global_kinetic_energy}\\
E_\mathrm{pot} &\equiv\int \d\vec{x} \rho(\vec{x})\phi(\vec{x})\,, \label{eq:global_potential_energy}\\
U&\equiv\int \d\vec{x} \rho(\vec{x})u(\vec{x})\,.
\end{align}
        \label{global_energies}%
\end{subequations}

The balance equation for the total energy is
$\rho D_t e  + \nabla_{\pos} \cdot \vec{J}_e = 0 $, as long as the potential energy is time-independent, while it becomes
\begin{align}
    \rho D_t e  + \nabla_{\pos} \cdot \vec{J}_e = \rho \partial_t \phi \,,\label{eq:total_energy_balance}
\end{align}
when the potential energy is time-dependent.
Here, $\vec{J}_e(\vec x; t)$ is the flux of total energy which can be specified by considering how each single contribution to the total energy in Eq.~\eqref{eq:total_energy_def} changes in time using the Navier-Stokes equations~\eqref{eq:NS}.

First, we derive the balance equation for $k(\vec x; t)$ from the momentum conservation equation~\eqref{eq:momentum_conservation} and Eq.~\eqref{eq:stress_tensor}:
\begin{align}
    \rho D_t k + \sum_i \partial_{x_i} \left( \phi \rho v_i + p v_i - \sum_{j} \Pi_{ij}v_j\right)=  \phi \sum_{i} \partial_{x_i} (\rho v_i) + \sum_i p \partial_{x_i} v_i - \sum_{ij}\Pi_{ij}\partial_{x_i} v_j  \label{eq:kinetic_energy_balance}\,,
\end{align}
where a term in the form of a divergence of fluxes has been isolated on the left-hand side.
On the right-hand side, source terms appear, even though at this stage it is still not possible to tell apart dissipative contributions from those that modify the potential energy of the system. 
Moving to the Eulerian frame using the identity~\eqref{eq:identity}, Eq.~\eqref{eq:kinetic_energy_balance} becomes
\begin{align}
     \partial_t (\rho k) + \sum_i \partial_{x_i} \left( \rho (k +\phi) v_i +  p v_i - \sum_{j} {\Pi_{ij}v_j}\right)=  \phi \sum_{i} \partial_{x_i} (\rho v_i) + \sum_i p \partial_{x_i} v_i - \sum_{ij}\Pi_{ij}\partial_{x_i} v_j \,. \label{eq:eulerian_kinetic_energy_balance}
\end{align}
Second, we consider the balance for the potential energy. 
By using Eq.~\eqref{eq:mass_conservation_eulerian}, we find \begin{align}
    \partial_t (\rho \phi)  = \rho \partial_t \phi  - \phi \nabla_{\pos} \cdot (\rho \vec{v}) = \rho \partial_t \phi  - \phi \sum_{i} \partial_{x_i} (\rho v_i)  \,.\label{eq:potential_energy_balance}
\end{align} 
By adding Eqs.~\eqref{eq:eulerian_kinetic_energy_balance} and~\eqref{eq:potential_energy_balance}, we obtain the  mechanical energy balance in the Eulerian frame:
\begin{align}
    \partial_t (\rho (k + \phi)) + \sum_i \partial_{x_i} \left( \rho (k +\phi) v_i +  p v_i - \sum_{j} \Pi_{ij}v_j\right) =  \rho \partial_t \phi +  \sum_i p \partial_{x_i} v_i - \sum_{ij}\Pi_{ij}\partial_{x_i} v_j  \,. \label{eq:mechanical_energy_balance}
\end{align}
Notice that the first term on the right-hand side of Eq.~\eqref{eq:eulerian_kinetic_energy_balance}, representing the nondissipative interconversion rate of kinetic energy into potential due to compression and advection with the velocity field $\vec{v}$, is entirely compensated by the corresponding term in the potential energy balance equation~\eqref{eq:potential_energy_balance} and, therefore, does not enter the balance equation for the mechanical energy~\eqref{eq:mechanical_energy_balance}. 
 In Lagrangian form (obtained using~\eqref{eq:identity}), Eq.~\eqref{eq:mechanical_energy_balance} reads
\begin{align}
    \rho D_t (k + \phi)  + \sum_i \partial_{x_i} \left( p v_i - \sum_{j} \Pi_{ij}v_j\right) =  \rho \partial_t \phi+   \sum_i p \partial_{x_i} v_i - \sum_{ij}\Pi_{ij}\partial_{x_i} v_j  \,. \label{eq:mechanical_energy_balance_lagrangian}
\end{align}
Having derived the balance equation for the total mechanical energy, we use Eq.~\eqref{eq:total_energy_balance}  to obtain the balance equation for the internal energy $u(\vec x; t)$:
\begin{align}
    \rho D_t u = \rho \partial_t \phi- \rho  D_t (k + \phi) - \nabla_{\pos} \cdot \vec{J}_e\,. \label{eq:total_energy_balance_II}
\end{align} 
Notice that the first term on the right-hand side is due to a time-dependent change in potential energy. It does not directly contribute to the balance equation of the internal energy because it is compensated by the corresponding term in the mechanical energy balance~\eqref{eq:mechanical_energy_balance_lagrangian}.
Nevertheless, it contributes  to the internal energy indirectly since it enters the Navier-Stokes equations (via Eq.~\eqref{eq:momentum_conservation}) and thus affects the state of the system.

The expression of total energy flux $\vec{J}_e$ can now be given as the sum of the fluxes on the left-hand side of Eq.~\eqref{eq:mechanical_energy_balance_lagrangian} and the heat flux $\vec{J}_{q}$ (for now unspecified)
\begin{align}
    \vec{J}_{e} =  p \vec{v} -  {\Pi\vec{v}}  + \vec{J}_{q} \label{eq:total_energy_flux} \,.
\end{align}
Indeed, this leads to a local balance equation for the internal energy (\emph{i.e.}, the first law), obtained
using Eq.~\eqref{eq:mechanical_energy_balance_lagrangian} and~\eqref{eq:total_energy_flux} in Eq.~\eqref{eq:total_energy_balance_II},
\begin{align}
    \rho D_t u  + \nabla_{\pos} \cdot \vec{J}_q =  - p \nabla_{\pos} \cdot \vec{v} + \Tr\Pi\nabla_{\pos} \vec{v}   \label{eq:internal_energy_balance_lagrangian} \,,
\end{align}
where the divergence only acts on the heat flux, while a source term appears on the right-hand side.
Here, we define $\Tr \Pi \nabla_{\pos} \vec{v} \equiv \sum_{ji}\Pi_{ij}\partial_{x_j} v_i$.
Notice that the last term in Eq.~\eqref{eq:internal_energy_balance_lagrangian} can be equivalently written in terms of the strain-rate tensor $\epsilon = \frac{1}{2}(\partial_{x_i}v_j + \partial_{x_j}v_i)$ since $\Pi$ is symmetric:
$\Tr \Pi \nabla_{\vec{x}} \vec{v} = \frac{1}{2} \sum_{ij}\Pi_{ij}(\partial_{x_i}v_j + \partial_{x_j}v_i) = \sum_{ij}\Pi_{ij}\epsilon_{ji}$.
In Eulerian form, Eq.~\eqref{eq:internal_energy_balance_lagrangian} becomes
\begin{align}
    \partial_t (\rho u) + \nabla_{\pos} \cdot (\rho u \vec{v} + \vec{J}_q)=  - p \nabla_{\pos} \cdot \vec{v} + \Tr \Pi\epsilon   \,. \label{eq:internal_energy_balance}
\end{align}
Equation~\eqref{eq:internal_energy_balance_lagrangian} and~\eqref{eq:internal_energy_balance} constitute two equivalent (local) formulations of the first law of thermodynamics for fluids. 
Indeed, once they are integrated over a volume element to obtain their global counterparts, one can recognize that the internal energy changes because of the exchanged heat at the boundary, or the mechanical expansion of the system, (accounted on the left-hand side of Eq.~\eqref{eq:internal_energy_balance}) or the viscous and other dissipative effects that take place inside the volume (accounted on the right-hand side of Eq.~\eqref{eq:internal_energy_balance}).

Finally, by substituting  the expression of the total energy flux~\eqref{eq:total_energy_flux} in the balance equation~\eqref{eq:total_energy_balance} for the total energy, we obtain
\begin{align}
     \rho D_t e  =  \rho \partial_t \phi - \nabla_{\pos} \cdot \vec{J}_q   - \nabla_{\pos} \cdot \left( p\vec{v} - \Pi \vec{v}\right)\,, \label{eq:first_law_alternative}
\end{align}
which will be used to prove the existence of a global Lyapunov function in systems with a well-defined equilibrium state (\S~\ref{sec:systems_with_reservoirs}).

\subsection{Local Equilibrium and Local Entropy Balance}
\label{sec:local_eq}

The second balance equation that the system satisfies concerns the entropy.
To derive it in the macroscopic theory, we resort to the following assumption:
each fluid element $\omega$ (macroscopic, but small compared to the domain $\Omega$) becomes locally equilibrated with the surroundings after a very short time scale corresponding to microscopic collisions.
This local equilibrium has well-defined values of local temperature~$T$, pressure~$p$ and chemical potential~$\mu$, and the fluid element is thus characterized by the extensive quantities entropy~$S_\omega$, internal energy~$U_\omega$, volume~$V_\omega$ and number of particles~$N_\omega$, which can all vary in time.
Furthermore, the particle density inside the volume is uniform. 
Thus, the fundamental relation of equilibrium thermodynamics holds for the fluid element and can be written in the form 
\begin{align}
     T  \d_t S_\omega &= \d_t U_\omega +  p \d_t V_\omega - M\mu \d_t N_\omega\,, \label{eq:local_eq_extensive}
\end{align}
also called local equilibrium condition.  Note that here $M$ is the molecular mass and we used the chemical potential per unit mass $\mu = -T(\partial S_\omega/\partial N_\omega)/M$.

To formulate Eq.~\eqref{eq:local_eq_extensive} in the same way as the balance equation~\eqref{eq:internal_energy_balance_lagrangian} for the internal energy, we need to introduce the specific entropy $s$, the specific internal energy $u$ and the specific volume $\rho^{-1}$ (that coincides with the inverse of the local mass density).
This, together with Reynolds' theorem (see appendix~\ref{app:reynolds}) and the identity~\eqref{eq:identity}, leads to
\begin{subequations}
\begin{align}
    &\d_t S_\omega = \d_t\int_{\omega(t)} \d{\pos}\, \rho s = \int_{\omega(t)} \d{\pos}\, \left(\partial_t(\rho s) +  \nabla_{\pos}\cdot \left(\rho s \vec{v}\right)\right)=\int_{\omega(t)} \d{\pos}\, \rho D_t s\,,\\
    &\d_t U_{\omega} =\d_t \int_{\omega(t)} \d{\pos}\, \rho u = \int_{\omega(t)} \d{\pos}\, \left(\partial_t(\rho u) +  \nabla_{\pos}\cdot \left(\rho u \vec{v}\right)\right) =\int_{\omega(t)} \d{\pos}\, \rho D_t u \,,\\
    &\d_t V_\omega=\d_t \int_{\omega(t)} \d{\pos}\, \rho \rho^{-1} = \int_{\omega(t)} \d{\pos}\, (\partial_t(\rho \rho^{-1}) + \nabla_{\pos}\cdot (\rho \rho^{-1}\vec{v})) = \int_{\omega(t)} \d{\pos} \, \rho D_t \rho^{-1} \,,\\
    &\d_t N_\omega =\d_t \int_{\omega(t)} \d{\pos}\, \rho M^{-1} = \int_{\omega(t)} \d{\pos}\, (\partial_t(\rho M^{-1}) + \nabla_{\pos}\cdot (\rho M^{-1}\vec{v})) =\int_{\omega(t)} \d{\pos}\,  \rho D_t M^{-1}=0 \,,\label{eq:local_eq_particle_number}
    \end{align}
\end{subequations}
where Eq.~\eqref{eq:local_eq_particle_number} means that particles do not leave or enter the fluid element, which moves according to the velocity field~$\vec{v}$.
The local equilibrium condition~\eqref{eq:local_eq_extensive} in terms of specific quantities becomes
\begin{align}
\int_{\omega(t)} \d {\pos} 
     \rho \left(T  D_t s - D_t u - p D_t \rho^{-1}\right)  &=  0\,. \label{eq:local_eq_intermediate}
\end{align}
 Since all variables are assumed to be uniform inside the fluid element $\omega$, we can also write the specific variables as $s=S_\omega /(MN_\omega)$, $u=U_\omega/(MN_\omega)$, $\rho^{-1} = V_\omega/(MN_\omega)$, with $MN_\omega$ being the total mass  of the fluid element.
Equation~\eqref{eq:local_eq_intermediate} thus becomes 
\begin{align}
    T  D_t s = D_t u +  p D_t \rho^{-1} \,. \label{eq:local_eq}
\end{align}

The extensivity of the entropy, \emph{i.e.}, $S_\omega(\lambda U_\omega ,  \lambda V_\omega, {\lambda N_\omega}) = \lambda S_\omega(U_\omega, V_\omega, {N_\omega})$ for all positive values of $ \lambda$, can be combined with Eq.~\eqref{eq:local_eq} to provide a constraint on the intensive fields $T$, $p$, $\mu$, which are therefore not independent from each other. 
Indeed, by taking the derivative of $S_\omega(\lambda U_\omega ,  \lambda V_\omega, {\lambda N_\omega})$ in $\lambda$ evaluated at $\lambda=1$ and dividing everything by the total mass $MN_\omega$, we get the Euler relation
\begin{align}
      s = \frac{u}{T} + \frac{p}{T}\rho^{-1} -\frac{\mu}{T}\,. \label{eq:local_extensivity} 
\end{align}
Then, by subtracting Eq.~\eqref{eq:local_eq} from the material derivative of $s$ in Eq.~\eqref{eq:local_extensivity}, we obtain the Gibbs-Duhem equation, that constrains the evolution of intensive quantities
\begin{align}
    u D_t \left( \frac{1}{T}\right) + \rho^{-1} D_t \left(\frac{p}{T}\right)  - D_t \left(\frac{\mu}{T}\right) =0\,.\label{eq:gibbs-duhem-closed}
\end{align}

Finally, the (local) entropy balance is obtained by using the first law~\eqref{eq:internal_energy_balance_lagrangian} and the continuity equation~\eqref{eq:mass_conservation_lagrangian} in the local equilibrium~\eqref{eq:local_eq}:
\begin{align}
    T \rho D_t s 
    =   - \nabla_{\pos} \cdot \vec{J}_q + \Tr\Pi\nabla_{\pos} \vec{v}  \,,
\end{align}
or, equivalently,
\begin{align}
     \rho D_t s + \nabla_{\pos} \cdot \left(\frac{\vec{J}_q}{T}\right)
       =  \dot{\sigma}_{\mathrm{macro}} \,. \label{eq:entropy_balance}
\end{align}
In Eq.~\eqref{eq:entropy_balance} we identify the entropy flux $\vec{J}_s\equiv \vec{J}_q/T$ and the (macroscopic) EPR per unit volume
\begin{subequations}
\begin{align}
    \dot{\sigma}_{\mathrm{macro}}
    &=  \vec{J}_q \cdot \nabla_{\pos} \left(\frac{1}{T}\right)   + \frac{1}{T}\Tr\Pi\nabla_{\pos} \vec{v}  \label{eq:epr_local} \\
   &=\underbrace{\dot{\sigma}_{\mathrm{hc}}}_{\geq0} +\underbrace{\dot{\sigma}_{\mathrm{vf}}}_{\geq0} 
   {  \geq 0}\label{eq:epr_local_defs}\,. 
\end{align}
\end{subequations}

The EPR is given by the sum of two different contributions corresponding to different local dissipative mechanisms: heat conduction ($\dot{\sigma}_{\mathrm{hc}}$) and viscous friction ($\dot{\sigma}_{\mathrm{vf}}$).
In the macroscopic theory, the assumption that the two dissipative contributions in Eq.~\eqref{eq:epr_local_defs} are decoupled and separately positive constitutes a formulation of the second law of thermodynamics, and it is related to the so-called Curie principle's \cite{dgm1962nonequilibrium}.
This assumption is the rationale behind the inequalities in Eq.~\eqref{eq:epr_local_defs}.
In the microscopic and mesoscopic theory, one is instead interested in showing the positivity of the EPR combining hypotheses concerning the dynamics and the statistical behavior of a large number of particles (see \S\S~\ref{sec:microscopic_hamiltonian},~\ref{sec:mesoscopic_boltzmann} and~\ref{sec:mesoscopic_boltzmann_open}).

Equations~\eqref{eq:mass_conservation_lagrangian},~\eqref{eq:momentum_conservation} and \eqref{eq:internal_energy_balance_lagrangian} form a closed dynamical system for the spatio-temporal fields $(\rho,u,\vec{v})$ when (i) the intensive fields $p$, $T$, $\mu$ are also expressed as functions of $(\rho,u,\vec{v})$ from the knowledge of the entropy appearing in Eq.~\eqref{eq:local_extensivity} and (ii) when an expression for the nonequilibrium fluxes $\vec{J}_q$ and $\Pi$ is available.
The latter condition is achieved when confining the analysis to the linear regime, where $\vec{J}_q$ and $\Pi$ are linear functions of $\nabla_{\vec{x}}T$ and $\nabla_{\vec{x}}\vec{v}$, see \emph{e.g.} the second equality in Eq.~\eqref{eq:stress_tensor}.
In this case, the assumption that the distinct contributions to the entropy production are positive translates into the positivity of some coefficients characterizing the material (see \S~\ref{sec:linear_response}).

In the following, we will exploit the global conservation laws (\emph{i.e.}, those obtained upon integration over the whole volume) resulting from the boundary conditions to obtain stronger dynamical results valid beyond the linear regime.
We will demonstrate how different  boundary conditions (see Fig.~\ref{fig:chamber}) give rise to different global thermodynamic forces and thermodynamic potentials.
The latter act as a Lyapunov function for systems in contact with an equilibrium environment.

\subsection{Entropy Balance and Thermodynamic Potential for Isolated Systems \label{sec:entropy_balance_isolated}} 

We formulate the global entropy balance for an isolated fluid contained in a domain~$\Omega$ and determine its thermodynamic potential.
We label $\partial\Omega$ and $\vec{n}$ the boundary and the outward pointing unit vector, respectively.
On this boundary, the heat flow and the velocity field must satisfy \begin{subequations}
\begin{align}
     \left.\vec{J}_q ({\pos};t)\cdot \vec{n}\right|_{\partial \Omega}  &= 0\label{eq:no_flux}\,,\\ 
       \left.\vec{v}({\pos};t)\cdot \vec{n} \right|_{\partial \Omega} &= 0\,,\label{eq:definition_boundary}
\end{align}%
\label{eq:boundary_isolated}%
\end{subequations}
namely, heat flux and velocity are orthogonal to the boundary 
which physically means that no energy or momentum can be exchanged with the boundaries.
Note that Eq.~\eqref{eq:definition_boundary} is trivially satisfied if one imposes a no-slip condition at the boundary, \emph{i.e.}, $\left.\vec{v}({\pos};t)\right|_{\partial \Omega}=\vec{0}$.
 We now take the time derivative of the global entropy $S \equiv \int_{\Omega} \rho s \d {\pos}$ using the local entropy balance~\eqref{eq:entropy_balance}, Eq.~\eqref{eq:identity} and the boundary conditions in Eq.~\eqref{eq:boundary_isolated}: 
\begin{align}
    \d_t S = \int_\Omega \d {\pos}\,   \partial_t(\rho s)  &= -\int_{\partial \Omega} \d \vec{n} \cdot \left(  \frac{\vec{J}_q}{T} + \rho s \vec{v} \right) + \int_\Omega \d {\pos} \,\dot{\sigma}_{\mathrm{macro}} =\int_\Omega \d {\pos}\,\dot{\sigma}_{\mathrm{macro}}\equiv \dot{\Sigma}_{\mathrm{macro}}    \geq 0 \,,\label{eq:total_entropy_isolated}
\end{align}
where we introduced the global EPR $\dot{\Sigma}_{\mathrm{macro}}$.
The inequality in Eq.~\eqref{eq:total_entropy_isolated} (resulting from Eq.~\eqref{eq:epr_local_defs})
implies that entropy is the thermodynamic potential that is maximized when an isolated system reaches equilibrium~\cite{callen1948application}.
Prigogine introduced the notion of ``evolution criterion'' for an inequality of the type in Eq.~\eqref{eq:total_entropy_isolated}, where the time derivative of a thermodynamic potential has a definite sign~\cite{glansdorff1974thermodynamic, glansdorff1971thermodynamic}.
If such a potential has an upper bound, then it plays the role of the (negative of the) Lyapunov function for the system.
In \S~\ref{sec:systems_with_reservoirs}, we obtain analogous relations for non-isolated systems.

\subsection{Entropy Balance and Thermodynamic Potential for Non-Isolated Systems}
\label{sec:systems_with_reservoirs}

\begin{figure}
    \centering
    \includegraphics[scale=0.15]{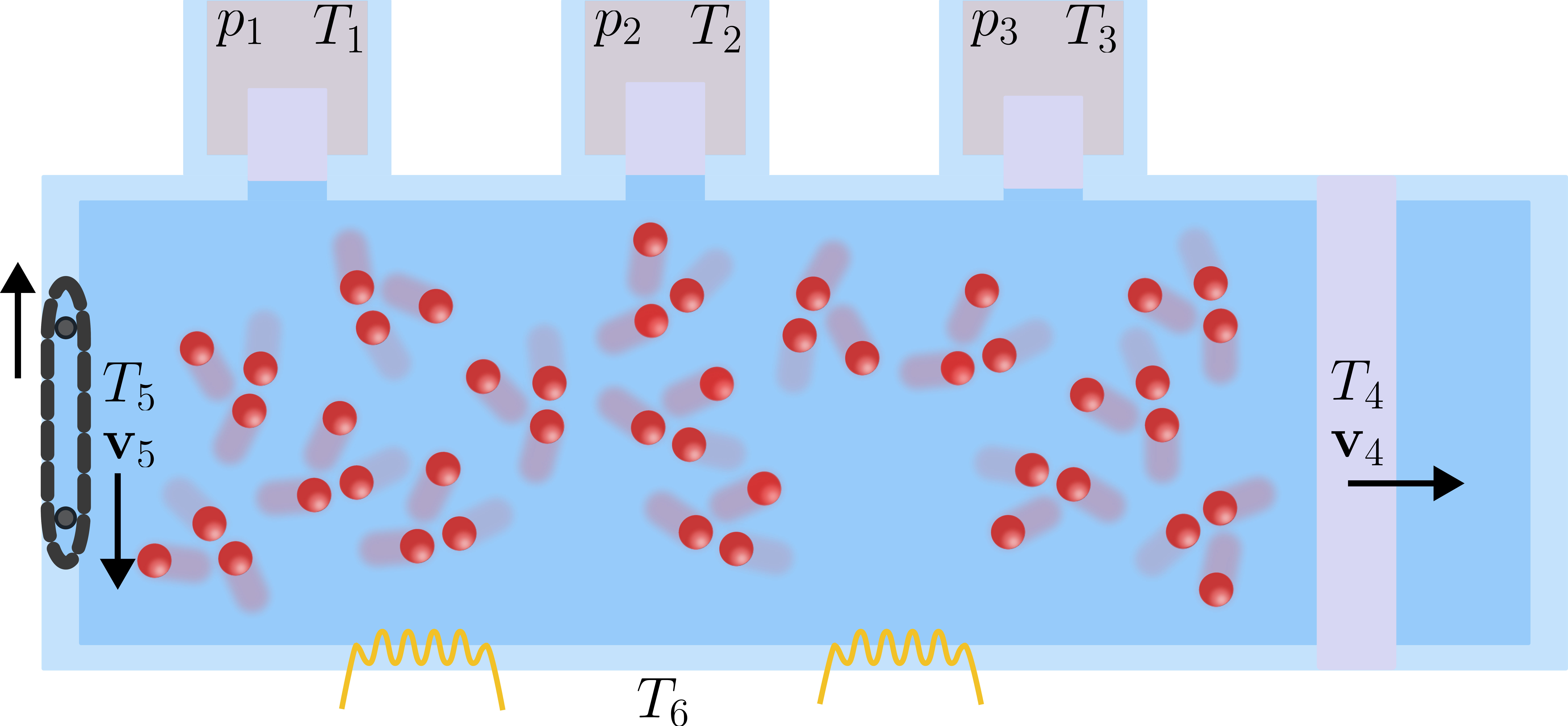}
    \caption{
    { Illustration of a fluid in contact with    6 different regions of the environment corresponding to different types of boundary conditions.
    Regions 1-3 correspond to movable pistons coupling the system to reservoirs at a fixed temperature and pressure.
    The corresponding EPR is given in Eq.~\eqref{eq:epr_balance_open_full_td_2}.
    Regions 4-5 are characterized by an externally fixed velocity on the boundary, which results in   volume changes (4) or shear (5).
    The corresponding EPR is given in Eq.~\eqref{eq:epr_closed_velocity_I}.
    On region 6, the wall is not moving and a heat flux can be maintained using external thermal devices, \emph{e.g.}, resistors coupled to an electric circuit.
    Notice that the external temperature should be defined everywhere on the boundary to allow for equilibration, which occurs when all the regions of the boundary have the same temperature.}
 }
    \label{fig:chamber}
\end{figure}

In non-isolated systems, the existence of a Lyapunov function satisfying the equivalent of Eq.~\eqref{eq:total_entropy_isolated} depends on the boundary conditions.
We now show how the thermodynamic potential which is maximized at equilibrium is modified when different experimentally relevant boundary conditions are imposed.

\subsubsection{Closed, rigid systems.}

Consider a closed but non-isolated system  in a rigid domain $\Omega$ with constant volume $V$, delimited by a boundary $\partial \Omega$ satisfying 
\begin{subequations}
\begin{align}
    \left.T({\pos};t)\right|_{\partial \Omega} &= T_B \,,\label{eq:bcs-closed_temp}\\
    \left.\vec{v}({\pos};t)\right|_{\partial \Omega} &= \vec{0}\,, \label{eq:bcs-closed_velocity}
\end{align}\label{eq:boundary_cloded_rigid}%
\end{subequations}
namely, with fixed temperature and a no-slip condition.
The system can exchange heat with the environment via a nonvanishing heat flux $\vec{J}_q$ at the boundary. 
If the potential $\phi$ as well as the temperature $T_B$ are constant in time, the system will relax to the equililbrium compatible with the external temperature. 
However, the global entropy $S = \int_{\Omega} \rho s \d {\pos}$ is not the thermodynamic potential anymore. 
Indeed, by using the local entropy balance equation~\eqref{eq:entropy_balance}, Eq.~\eqref{eq:identity} and the boundary conditions in Eq.~\eqref{eq:boundary_cloded_rigid}, we find 
\begin{align}
     \d_t S  =- \int_{\partial \Omega} \d \vec{n} \cdot \left(\frac{\vec{J}_q}{T} + \rho s \vec{v}\right) + \int_\Omega \d{\pos}\, \dot{\sigma}_{\mathrm{macro}} 
    =\frac{{\dot{Q}}}{T_B} + \dot{\Sigma}_{\mathrm{macro}} \,,\label{eq:entropy_balance_closed}
\end{align}
where ${\dot{Q}}\equiv- \int_{\partial \Omega} \d \vec{n} \cdot \vec{J}_q $ is the exchanged heat at the boundary (positive when it flows from the environment to the system) which is related to the global total energy $E \equiv \int_{\Omega} \rho e \d {\pos}$ via 
\begin{align}
     \d_t E = -\int_{\partial \Omega} \d \vec{n} \cdot \left(\vec{J}_q + p \vec{v} - \Pi\vec{v} + \rho e \vec{v} \right) = {\dot{Q}}\,,
    \label{eq:integrated_total_energy_conservation_rigid}
\end{align}
using Eqs.~\eqref{eq:first_law_alternative},~\eqref{eq:identity} and~\eqref{eq:boundary_cloded_rigid}.
Hence, 
the thermodynamic potential that is maximized at equilibrium can be written as
\begin{align}
    \d_t \left( S - \frac{E }{T_B} \right) =\dot{\Sigma}_{\mathrm{macro}}
     \geq 0 \,,\label{eq:epr_closed_rigid_system}%
\end{align}%
where the inequality follows from Eq.~\eqref{eq:epr_local_defs}.
This means that for a closed, rigid system in contact with an external thermostat, the thermodynamic potential is the difference between the global entropy and the global total energy divided by the temperature of the boundary.

\subsubsection{Closed system under pressure control.}
\label{sec:macro_close_pressure}

Closed systems where the volume $V(t)$ can change in time due to control on the pressure are exemplified by the idealized case of a fluid enclosed by a flexible membrane with negligible surface tension or put in a vessel with a movable, weightless piston which is in contact with a gas in equilibrium at a given temperature and pressure (see Fig.~\ref{fig:chamber}).
Mathematically, the temperature on the boundary is externally fixed according to
\begin{align}
    \left.T({\pos};t)\right|_{\partial \Omega(t)} &= T_B \,\label{bcs-nonrigid_temp}\,.
\end{align}
On the other hand, the pressure  $P({\pos};t)$ of the fluid and that of the environment $p_B({\pos})$ do not have to coincide on the boundary.
In fact, the imbalance between the two pressures generates the external force in the Navier-Stokes equation~\eqref{eq:momentum_conservation},  \emph{i.e.}, the force $\boldsymbol{F}_\mathrm{w}$ is different from zero only on the boundary $\partial\Omega(t)$ and satisfies $\int_\Omega \d\vec{x}\rho \boldsymbol{F}_{\mathrm{w}} = \int_{\partial\Omega}  \d\vec{n}\cdot(P- p_B\mathbb{1})$ 
(where $\mathbb{1}$ is the identity matrix).
To analyze the dynamics of the boundary, we  assume the no-slip condition on the boundary, such that it moves with a velocity $\vec{v}_B(\vec{x})$ that 
coincides with the velocity field $\vec v(\vec{x})$ of the fluid evaluated on the boundary
\begin{align}
    \left.(\vec{v}(\vec{x})-\vec{v}_B(\vec{x})) \right|_{\partial\Omega(t)}= \vec{0}\,.
    \label{eq:bcs-velocity}
\end{align}
Hence, the volume changes in time according to
\begin{align}
    \d_t V =   \int_{\partial \Omega(t)} \d \vec{n} \cdot  \vec{v}_B (\vec{x})\, \label{bcs-nonrigid_velocity}\,,
\end{align} 
as prescribed by Reynolds' theorem {(see Eq.~\eqref{eq:reynolds} in Appendix \ref{app:reynolds})}.
To compute the mechanical power note that the only contribution to the time derivative of the potential energy per unit mass is the one due to the moving wall, which is localized on the boundary, \emph{i.e.}, $\partial_t\phi(\vec{x})=\partial_t\phi_{\mathrm{w}}(\vec{x}-\vec{x}_B(t))$, the latter being proportional to $\delta(\vec{x}-\vec{x}_B(t))$.
Using the chain rule, since the wall is moving with velocity $\vec{v}_B$, we have $\partial_t\phi_{\mathrm{w}}(\vec{x}-\vec{x}_B(t)) = -\nabla_{\vec{x}}\phi_{\mathrm{w}}(\vec{x})\cdot \vec{v}_B(\vec{x})=\vec{F}_\mathrm{w}\cdot\vec{v}_B$.
Therefore, 
\begin{equation}
    \int_{\Omega(t)} \d\vec{x}\, \rho(\vec{x}) \partial_t \phi(\vec{x}) = 
    \int_{\partial\Omega(t)} \d\vec{n}\cdot(P(\vec{x}) - p_B (\vec{x})\mathbb{1})\vec{v}_B(\vec{x}) \,,
    \label{eq:varpot}
\end{equation}
where we used that the force $\vec{F}_{w}$ is different from zero only on the boundary and results from the pressure unbalance. 
 
Using Eqs.~\eqref{eq:first_law_alternative},~\eqref{eq:entropy_balance} and Eq.~\eqref{eq:varpot}, we can write the balance equations for the total energy $E = \int_{\Omega} \rho e \d {\pos}$ and entropy $S = \int_{\Omega} \rho s \d {\pos}$ which read
\begin{subequations}
\begin{align}
       \d_t E = \int_{\Omega(t)} \d {\pos}\, \rho D_t e   
       =\dot{Q} - \int_{\partial\Omega(t)} \d\vec{n}\cdot p_B \vec{v}_B\,,
\label{eq:integrated_total_energy_conservation_impermeable}
\end{align}
\begin{align}
 \d_t S  =\int_{\Omega(t)} \d {\pos}\, \rho D_t s =  
    \frac{\dot{Q}}{T_B} 
    + \dot{\Sigma}_{\mathrm{macro}}\,, \label{eq:entropy_balance_impermeable}
\end{align}
\end{subequations}
respectively. 
By combining now Eq.~\eqref{eq:integrated_total_energy_conservation_impermeable} with Eq.~\eqref{eq:entropy_balance_impermeable},
we obtain the following expression for the global EPR
\begin{align}
    \dot{\Sigma}_{\mathrm{macro}}  
     &= \d_t  \left( S - \frac{E}{T_B}\right) - \frac{1}{T_B} \int_{\partial\Omega(t)} \d\vec{n}\cdot p_B \vec{v}_B \geq 0
    \,, \label{eq:epr_closed_flexible_Ia}
\end{align}
where the inequality follows from Eq.~\eqref{eq:epr_local_defs}.

If the pressure on the boundary is constant, namely, $ \int_{\partial\Omega(t)} \d\vec{n}\cdot p_B \vec{v}_B = p_B \d_t V$ (from Eq.~\eqref{bcs-nonrigid_velocity}), 
Eq.~\eqref{eq:epr_closed_flexible_Ia} becomes
\begin{align}
    \dot{\Sigma}_{\mathrm{macro}}  
     &= \d_t  \left( S - \frac{E + p_BV}{T_B}\right)  \geq 0
    \,, \label{eq:epr_closed_flexible_I}
\end{align}
which allows us to identify the proper thermodynamic potential
of a closed systems
in contact with an external thermostat 
and exposed to a constant pressure
as the difference between
the global entropy and
the sum of the global energy plus $p_B V$ divided by the temperature of the boundary.

For a reversible transformation, the EPR vanishes,  $\dot{\Sigma}_{\mathrm{macro}}=0$,
and Eq.~\eqref{eq:epr_closed_flexible_Ia} identifies the maximum mechanical work which can be extracted  as~\cite{landau2013statistical}
\begin{align}
W^{\mathrm{max}}_{\mathrm{mech}}\equiv p_B \Delta V_\mathrm{eq}=  T_B \Delta S_\mathrm{eq} - \Delta E_\mathrm{eq}\,.
\end{align}
This maximum available work or available free energy is also known as ``exergy'' in engineering~\cite{bejan2016advanced, kotas2012exergy}.

\subsubsection{Closed system under velocity control.}
\label{sec:closed_velocity_control}

The volume of closed systems can also change when the velocity profile on the boundary $\vec{v}_B$ is externally controlled according to Eq.~\eqref{bcs-nonrigid_velocity}.
If we consider again a uniform temperature on the boundary as in Eq.~\eqref{bcs-nonrigid_temp}, 
we can combine the balance equations for
the global total energy $E$ and entropy $S $ 
in Eqs.~\eqref{eq:first_law_alternative}, and~\eqref{eq:entropy_balance}, respectively, to express the global entropy production as
\begin{align}
    \dot{\Sigma}_{\mathrm{macro}} &= \d_t \left(S - \frac{E}{T_B}\right) + \frac{1}{T_B} \int_{\Omega(t)} \d\vec{x} \, (\rho \partial_t\phi -\nabla_{\vec{x}}\cdot P\vec{v})\,,\label{eq:epr_closed_velocity_I}\\  
   & = \d_t \left(S - \frac{E}{T_B} \right) + \frac{1}{T_B} \underbrace{\left(\int_{\partial\Omega(t)} \d \vec{x} \, \rho \vec{F}_{\mathrm{w}}\cdot  \vec{v}_B  - \int_{\partial\Omega(t)} \d \vec{n}\cdot P\vec{v}_B \right)}_{{ \equiv}\dot{W}} \geq 0\,, \label{eq:epr_closed_velocity_II}
\end{align}
where the inequality follows from Eq.~\eqref{eq:epr_local_defs} 
and we used that the force $\boldsymbol{F}_\mathrm{w}$ is different from zero only on the boundary $\partial\Omega(t)$.
Here, as well as in \S~\ref{sec:macro_close_pressure},
the potential energy density is time-dependent because of the moving boundary.
The last equality is obtained after noticing that the contribution of the time derivative $\partial_t \phi=\partial_t \phi_{\mathrm{int}}+\partial_t \phi_{\mathrm{w}}$ is localized on the boundary, as  $\partial_t \phi_{\mathrm{int}}=0$, $\partial_t \phi_w = -\vec{v}_B\cdot \nabla_{\vec{x}} \phi_{\mathrm{w}}(\vec{x}-\vec{x}_B(t))  = \vec{v}_B\cdot\vec{F}_{\mathrm{w}}$.

The thermodynamics behind Eq.~\eqref{eq:epr_closed_velocity_II} is transparent.
If the velocity profile at the boundary is nonuniform,
the system remains out of equilibrium and undergoes either expansion, or compression, or shearing.
From a thermodynamic standpoint, this results from the environment continuously performing mechanical work.
When shearing at constant volume far from the turbulent regime (\emph{i.e.}, when the nonlinear terms in \eqref{eq:momentum_conservation} are negligible), 
the system approaches a nonequilibrium steady state.
Finally,  
if the boundary has uniform  and constant speed $\vec{v}_B$
and the center of mass (CM) of the system moves at the same velocity as the boundary,
i.e., $\vec{v}_{\text{CM}}=\vec{v}_{B}$,
the mechanical power in Eq.~\eqref{eq:epr_closed_velocity_II} reduces to the rate of change of the energy of the center of mass, namely $\dot{W}=\d_t E_{\text{CM}}$, (see Eq.~\eqref{eq:macro_E_cm_rate_velocity_control}), 
provided that the internal potential energy $\phi_{\text{int}}$ is linear in $\vec{x}$,
namely $\phi_{\text{int}}(\vec{x})=\vec{g}\cdot\vec{x}$ 
(as it is for the gravitational potential).
Therefore the system relaxes towards equilibrium and the EPR reduces to the total derivative $\dot{\Sigma}_{\mathrm{macro}} = \d_t \left(S - { (E-E_{\text{CM}})}/{T_B}\right)$, which coincides with Eq.~\eqref{eq:epr_closed_rigid_system} as the two situations are equivalent under a Galilean transformation.

\subsubsection{Closed system with multiple heat baths and time-dependent driving.\label{par:general_EPR}}

We now consider a closed non-isolated system
contained in a domain $\Omega$ of  volume $V$ whose boundary $\partial \Omega$
is composed of $b=1,\dots,B$ distinct regions $\partial \Omega_{\bath}$ called reservoirs, or baths, \emph{i.e.}, $\partial \Omega = \cup_b \partial \Omega_{\bath}$.
The temperature of each region $T_{\bath}(t)$ and the (uniform) pressure each region is exposed to $p_b(t)$ are externally controlled and may depend on time following arbitrary protocols (see Fig.~\ref{fig:chamber}).
On each region $b$, the fluid temperature is assumed to be equilibrated with the environment, namely,
\begin{align}
    \left.T({\pos};t)\right|_{\partial \Omega_{\bath}} &= T_{\bath}(t)\,,\label{bcs-open_temp}
\end{align}
while its pressure differs in general from the external pressure, \emph{i.e.}, $P(x;t)|_{\partial \Omega_{\bath}}\neq p_b(t)\mathbb 1$.
As we already discussed in \S~\ref{sec:macro_close_pressure},
the mismatch of pressure at each region implies that 
i) the boundaries move according to Eq.~\eqref{eq:bcs-velocity} (which now holds at the level of each single region $b$),
ii) the volume changes according to Eq.~\eqref{bcs-nonrigid_velocity} (which can be split into the contribution of each single region)
and iii) the potential energy density changes in time according to Eq.~\eqref{eq:varpot} (which now holds at the level of each single region $b$).

From a thermodynamic standpoint, by using Eqs.~\eqref{eq:first_law_alternative},~\eqref{eq:entropy_balance} and Eq.~\eqref{eq:varpot} together with $\partial \Omega = \cup_b \partial \Omega_{\bath}$,
the balance equations for the total energy $E = \int_{\Omega} \rho e \d {\pos}$ and entropy $S = \int_{\Omega} \rho s \d {\pos}$ read
\begin{subequations}
\begin{align}
       \d_t E =
       \sum_{b=1}^B\dot{Q}_b 
       - \sum_{b=1}^B p_b\d_tV_b \,,
    \label{eq:energy_balance_open}
\end{align}
\begin{align}
 \d_t S  =  
    \sum_{b=1}^B\frac{\dot{Q}_b}{T_b} 
    + \dot{\Sigma}_{\mathrm{macro}}\,, \label{eq:entropy_balance_open}
\end{align}
\end{subequations}
respectively, 
with $\dot{Q}_{\bath}\equiv- \int_{\partial \Omega_{{\bath}}} \d \vec{n} \cdot \vec{J}_q $
and $\d_t V_b \equiv \int_{\partial\Omega_b(t)} \d\vec{n}\cdot \vec{v}_b $.
Without loss of generality, we choose the reservoir $b=B$ as the reference for the equilibrium temperature and pressure.
Thus, by combining now Eq.~\eqref{eq:energy_balance_open} with Eq.~\eqref{eq:entropy_balance_open},
the global EPR can be written as
\begin{equation}
    \dot{\Sigma}_{\mathrm{macro}} =
    \d_t S  - \frac{\d_t E}{T_B} 
    + \sum_{b=1}^B \dot{Q}_b \bigg(\frac{1}{T_B} - \frac{1}{T_b}\bigg)
    - \frac{1}{T_B}\sum_{b=1}^B p_b\d_tV_b \geq 0    \,,
\end{equation}
or, by summing and subtracting $(p_B\d_t V)/T_B$ (with $\d_t V = \sum_b \d_t V_b$)
together with collecting all the time derivatives in a single contribution, as
\begin{equation}
    \dot{\Sigma}_{\mathrm{macro}} =
    \d_t \bigg( S  - \frac{ E +p_B V }{T_B} \bigg)
    + \sum_{b=1}^B \dot{Q}_b \bigg(\frac{1}{T_B} - \frac{1}{T_b}\bigg)
    + \frac{1}{T_B}\underbrace{\sum_{b=1}^B (p_B - p_b )\d_t V_b}_{\dot{W}_{\mathrm{mech}}}
    + (E +p_B V)\d_t\frac{1}{T_B} + \frac{V}{T_B}\d_tp_B \geq 0
    \,,
\label{eq:epr_balance_open_full_td}
\end{equation}
where the inequality follows from Eq.~\eqref{eq:epr_local_defs}.
Equation~\eqref{eq:epr_balance_open_full_td} is the general balance equation for the global EPR in terms of (from the right to the left)
i) time-dependent driving of the reference pressure $p_B$ and temperature $T_B$,
ii) global thermodynamic forces (\emph{i.e.}, pressure and temperature differences between the boundaries), 
and iii) the time derivative of a thermodynamic potential,
\begin{equation}
    \tpotmacro(t) \equiv S - \frac{E + p_{B}(t)V(t)}{T_{B}(t)}\,.
    \label{eq:macro_termo_pot}
\end{equation}
If the system is coupled to a single reservoir with no time-dependent driving of $T_B$ and $p_B$, then the thermodynamic potential $\tpotmacro(t)$ would be maximized at the (equilibrium) steady state.

Note that, by using the thermodynamic potential~\eqref{eq:macro_termo_pot}, 
Equation~\eqref{eq:epr_balance_open_full_td} reads
\begin{equation}
    \dot{\Sigma}_{\mathrm{macro}} =
    \d_t \tpotmacro
    + \sum_{b=1}^B \dot{Q}_b \bigg(\frac{1}{T_B} - \frac{1}{T_b}\bigg)
    + \frac{\dot{W}_\mathrm{mech}}{T_B}
    -\frac{\partial\tpotmacro}{\partial(1/T_B)} \d_t\frac{1}{T_B} 
    - \frac{\partial\tpotmacro}{\partial p_B}\d_tp_B \geq 0
    \,.
\label{eq:epr_balance_open_full_td_2}
\end{equation}

The systematic derivation of the decomposition of the EPR in terms of the global thermodynamic forces, the time-dependent driving and the derivative of a thermodynamic potential is a significant result of the present work.
Indeed, as it is well known, the classic formulation of nonequilibrium thermodynamics~\cite{prigogine1949domaine, dgm1962nonequilibrium} is unable to differentiate systems that relax toward equilibrium from those who are actively maintained away from it by just looking at the expression of the EPR, in which only local thermodynamic forces appear~\cite{gallavotti2002foundations}.
The systematic procedure to obtain the decomposition of the EPR that we outlined here for the case of a single component fluid, but which can be easily extended to more complex situations, is strictly related to the one used in engineering thermodynamics~\cite{bejan2016advanced, kotas2012exergy} in the context of the so-called ``exergy method'' of analysis.
However, the latter is introduced in a phenomenological way that may disregard some contributions, like those accounting for time-dependent driving on the boundaries.
Equation~\eqref{eq:epr_balance_open_full_td_2} may thus provide a useful point of view to interpret new numerical and laboratory experiments in fluid mechanics~\cite{yang2020periodically, urban2022thermal} in which time-dependent driving at the boundaries translates into nontrivial effects on macroscopic heat fluxes affecting the efficiency of heat transport.

Notice that an  alternative formulation of Eq.~\eqref{eq:epr_balance_open_full_td_2} can be obtained by representing each reservoir $b=1,\dots,B$ as an ideal gas at equilibrium at temperature $T_b$.
Indeed, this allows us to quantify the energy exchanges between the system and the $b$ reservoir according to the equilibrium first law:
\begin{align}
    \d_t E_b = \dot{Q}_b - p_b \d_t V_b \,,
    \label{eq:1law_ideal_b_gas}
\end{align}
where $-\d_t E_b$ is the energy variation of the $b$ reservoir.
Thus, by using Eq.~\eqref{eq:1law_ideal_b_gas} in Eq.~\eqref{eq:epr_balance_open_full_td_2} and treating $1/T_B$ and $p_B/T_B$ as independent intensive variables of the thermodynamic potential $Y(t)$ in Eq.~\eqref{eq:macro_termo_pot}, Eq.~\eqref{eq:epr_balance_open_full_td_2} becomes
\begin{small}
\begin{equation}
     \dot{\Sigma}_{\mathrm{macro}} =
     \d_t Y + 
     \sum_{b=1}^{B} \left[\d_t{E}_{{\bath}}\left( \frac{1}{T_{B}}  -  \frac{1}{T_{\bath}} \right) +  \left(\frac{p_B}{T_B}-\frac{p_\bath}{T_\bath}\right) \d_t V_\bath  \right]
     -\frac{\partial\tpotmacro}{\partial(1/T_B)} \d_t\frac{1}{T_B} 
     - \frac{\partial Y}{\partial (p_B/T_B)}\d_t \left(\frac{p_{B}}{T_{B}}\right)\geq 0
     \,.
\label{eq:epr_balance_open_full_td_3}
\end{equation}
\end{small}
Equation~\eqref{eq:epr_balance_open_full_td_3} gives the rationale for the expressions of the intensive fields used in ST, see \emph{e.g.} \cite{rao2018conservation}.

\subsection{Linear regime and minimum entropy production principle}
\label{sec:linear_response}

We conclude the discussion on the macroscopic aspects of the thermodynamics of non-isolated systems by showing that systems close to equilibrium satisfy a minimum entropy production principle in a form similar to the one obtained in the framework of ST~\cite{forastiere2022linear}.
In turn, the latter is akin to the adiabatic-nonadiabatic decomposition \cite{esposito2010three} specialized in the linear regime.
For related work on decompositions of the EPR in macroscopic systems see also Ref.~\cite{maes2015revisiting}.
To obtain the aforementioned minimum entropy production principle,
we write the macroscopic EPR $\dot{\Sigma}_{\mathrm{macro}}$ obtained by
integrating Eq.~\eqref{eq:epr_local} in the linear regime, namely, when $\vec{J}_{\mathrm{q}} = \kappa \nabla_{\pos}(1/T)$ and $\Pi$ are evaluated at the equilibrium density $\rho_\mathrm{eq}$ and temperature $T_\mathrm{eq}$:
 \begin{align}
      \dot{\Sigma}_{\mathrm{macro}}  \simeq \int \d {\pos} \,
      { 
      \begin{pmatrix}
      \partial_{x_i}\left(1/T\right)\\
     \epsilon_{ij}
     \end{pmatrix}\tr}
     \underbrace{\begin{pmatrix}
      \kappa\, \delta_{ij}   & 0\\
      0 & L_{(ij),(i'j')}
     \end{pmatrix}}_{\equiv \mathbb{O}}
     \begin{pmatrix}
      \partial_{x_j}\left(1/T\right)\\
      \epsilon_{i'j'}
     \end{pmatrix}\label{eq:epr_linear_regime}\,.
 \end{align}
Here, we use a notation such that the vectors' entries are the three components of $\partial_{x_i}\left(1/T\right)$ and the nine components of the strain tensor $\epsilon_{ij}\equiv(\partial_{x_i} v_j+\partial_{x_j} v_i) / 2$.
Furthermore, we introduce the $9\times9$ positive-definite matrix whose entries are 
$L_{(ij),(i'j')} \equiv (\lambda \delta_{ij} \delta_{i'j'} + 2 \rho_{\mathrm{eq}} \nu  \delta_{ii'} \delta_{jj'})/{T_{\mathrm{eq}}}$
which together with the $3\times3$ positive-definite matrix with entries $\kappa\delta_{ij}$ defines the positive-definite linear response matrix $\mathbb{O}$.
{By restricting ourselves to the case of an incompressible fluid
(and thus neglecting phenomena like pressure waves or thermal expansion),}
in Appendix \ref{app:mep} we prove that Eq.~\eqref{eq:epr_linear_regime} can be alternatively written in terms of the steady state EPR $\dot{\Sigma}^\mathrm{ss}_{\mathrm{macro}}$ according to the following decomposition
\begin{align}
    \dot{\Sigma}_{\mathrm{macro}}
     &\simeq\dot{\Sigma}^\mathrm{ss}_{\mathrm{macro}}  + \int \d {\pos}\, 
     { 
      \begin{pmatrix}
      \partial_{x_i}\left(1/T-1/T^{\mathrm{ss}}\right)\\
      \epsilon_{ij}-\epsilon_{ij}^{\mathrm{ss}}
     \end{pmatrix}
     \tr}
     \begin{pmatrix}
      \kappa\, \delta_{ij} & 0\\
      0 & L_{(ij),(i'j')}
     \end{pmatrix}
     \begin{pmatrix}
         \partial_{x_j}\left(1/T-1/T^{\mathrm{ss}}\right)\\
      \epsilon_{i'j'}-\epsilon_{i'j'}^{\mathrm{ss}}
     \end{pmatrix}\label{eq:epr_ss_decomposition}\,.
\end{align} 
Thus, since the response matrix $\mathbb{O}$ is positive-definite, the state that minimizes the total EPR coincides with the steady-state solution.
In Appendix~\ref{app:mep} we also prove that the EPR is a Lyapunov function of the dynamics close to equilibrium (without resorting to further assumptions). 
We remark that  $\dot{\Sigma}^\mathrm{ss}_{\mathrm{macro}}$ can be computed in terms of quantities only defined at the boundaries using for example Eq.~\eqref{eq:epr_balance_open_full_td_2}, which specializes, for systems without time-dependent driving, to
\begin{align}
\dot{\Sigma}_{\mathrm{macro}}^{\mathrm{ss}} =
     \sum_{b=1}^{B-1} {\dot{Q}}_{\bath}^{ \mathrm{ss}} \left(\frac{1}{T_{B}^{\mathrm{ss}}}-\frac{1}{T_{\bath}^{{\mathrm{ss}}}}\right)
      \label{eq:ss_epr}\,.
 \end{align}

We further stress that the formulation of the minimum entropy production principle given in Eq.~\eqref{eq:epr_ss_decomposition} is different from the version proved in Ref.~\cite{glansdorff1971thermodynamic} (where no dynamics is directly involved).
Equation~\eqref{eq:epr_ss_decomposition} has the same form of the decomposition of the EPR proposed by Oono and Paniconi \cite{oono1998steady} on phenomenological grounds.
In Appendix \ref{app:mep} we provide the conditions under which it can be expected to hold.
Equation~\eqref{eq:epr_ss_decomposition} is a remarkable property of close to equilibrium steady states, which has a direct analogue in ST and can be interpreted as a geometric formulation of Prigogine's minimum entropy production principle~\cite{dgm1962nonequilibrium, glansdorff1974thermodynamic}.
However, notice that the positivity of the quadratic form in the local thermodynamic forces appearing on the right-hand side of Eq.~\eqref{eq:epr_ss_decomposition} is only ensured by the equilibrium properties of the response coefficients, and thus cannot be generalized to situations in which the steady state is not close to equilibrium.

\section{Microscopic Scale: Hamiltonian Dynamics}
\label{sec:microscopic_hamiltonian}

In this Section, we derive thermodynamic identities (and the related inequalities) for isolated systems made of particles obeying Hamiltonian dynamics.
We use the symbol $\vec{x}_{n}$ (resp. $\vec{p}_n$) for the spatial coordinates (resp. momentum) of the $n$-th microscopic particle, while the symbol ${\pos}$ (resp. $\vec{p}$) is reserved to indicate a generic value of position (resp. momentum).

\subsection{Dynamics}
We start from a system of $N$ identical particles whose dynamics follows Hamilton's equations
\begin{align}
    \d_t \vec{x}_n = \nabla_{\vec{p}_n} H \,, \qquad \d_t \vec{p}_n=-\nabla_{\vec{x}_n}H \,,
    \label{eq:hamilton}
\end{align}
for $n=1,\dots, N$, with Hamiltonian $H=H(\Gamma_N;t)$ and $\Gamma_N = \{\vec{x}_1,\dots,\vec{x}_N;\vec{p}_1,\dots,\vec{p}_N\}$.
The evolution of the corresponding phase-space density $P_N(\Gamma_N;t)$ follows the Liouville equation~\cite{gaspard2022statistical} 
\begin{align}
    \partial_t P_N
    =  \sum_{n=1}^N \left( \nabla_{\vec{x}_n} H \cdot \nabla_{\vec{p}_n} P_N - \nabla_{\vec{p}_n} H \cdot \nabla_{\vec{x}_n} P_N \right)= \{H, P_N \}\,, \label{eq:liouville}
\end{align}
where we introduced the Poisson brackets $\{A,B\}=\sum_n \nabla_{\vec{x}_n}A\cdot\nabla_{\vec{p}_n}B-\nabla_{\vec{p}_n}A\cdot\nabla_{\vec{x}_n}B$.
By extending the definition of material derivative to the phase space, 
\emph{i.e.}, $D_t = \partial_t + \sum_n(\nabla_{\vec{p}_n} H\cdot \nabla_{\pos_n} - \nabla_{\pos_n} H\cdot \nabla_{\vec{p}_n})$, 
and using the Hamilton equations~\eqref{eq:hamilton}, the Liouville equation becomes
\begin{align}
    D_t P_N = \partial_t P_N + \sum_n \nabla_{\vec{p}_n} H \cdot \nabla_{\vec{x}_n}P_N  - \sum_n \nabla_{\pos_n} H\cdot \nabla_{\vec{p}_{n}} P_N = 0\,.
\end{align}

A coarse-grained description of the system in terms of a one-particle probability density $P_1(\vec{x},\vec{p};t)$ is obtained by marginalization over the $N-1$ positions and momenta of the other particles.
Since the $N$ particles are identical and thus there are $N$ identical ways to choose one particle, it is convenient to define the one-particle distribution function
\begin{align}
    f_1(\gamma_1;t)  = N \int \d \Gamma_{N-1} P_N(\Gamma_N;t )\,, \label{eq:one_particle_f_def}
\end{align}
which is clearly normalized to $N$.
Here and in the following, the volume element is $\d \Gamma_l =\prod_{n=N-l+1}^N 
{ \d \gamma_n}$, where $\gamma_n=(\vec{x}_n,\vec{p}_n)$  and $\d \gamma_n = (2\pi \hbar)^{-3} \d \vec{x}_n\d \vec{p}_n$ are  the vector of canonical coordinates of the $n$-th particle and its associated volume element, respectively.

The evolution equation of the one-particle distribution function is obtained from marginalizing the Liouville equation~\eqref{eq:liouville}. 
When $H$ can be written as $H = \sum_n \frac{1}{2 M } \vec{p}_n^2 +\sum_{n}(  M\phi(\vec{x}_n) + \sum_{m>n} \psi(\abs{\vec{x}_n -\vec{x}_m}))$, with $\phi$ the external single-particle potential per unit mass and $\psi$ the interaction potential between pairs of particles,
the time evolution of $f_1$ is given by the first equation of the Bogoliubov--Born--Green--Kirkwood--Yvon (BBGKY) hierarchy~\cite{kardar2007statistical}:
\begin{align}
    \partial_t f_1(\vec{x},\vec{p})
    &= \{H_1, f_1\}+\int  \d \gamma_2\, \nabla_{\vec{x}}\psi(\abs{\vec{x}-\vec{x}_2})\cdot \nabla_{\vec{p}} f_2(\vec{x},\vec{p};\vec{x}_2,\vec{p}_2)\,.\label{eq:one-particle_ev}
\end{align}
In obtaining Eq.~\eqref{eq:one-particle_ev}, we used that $P_N$ vanishes at the boundary of the integration domain.
Equation~\eqref{eq:one-particle_ev} shows that the evolution of $f_1$ splits into 
i) the free evolution generated by a single-particle Hamiltonian $H_1(\gamma) = \frac{1}{2M}\vec{p}^2+M\phi(\vec{x})$ and 
ii) a collision integral depending on the two-particle distribution function
\begin{align}
    f_2 (\vec{x},\vec{p};\vec{x}_2,\vec{p}_2{ ;t}) =  N(N-1) \int \d\Gamma_{N-2} P_N(\Gamma_N{ ;t})\,.
\end{align}
This procedure can be iterated to obtain the BBGKY hierarchy of equations specifying the evolution of the $l$-particles distribution function
$f_l(\gamma_1,\dots,\gamma_l{ ;t}) =  \frac{N!}{(N-l)!}\int \d \Gamma_{N-l}\, P_N(\Gamma_N{ ;t} )$
in terms of the $l$-particles Hamiltonian $H_l(\Gamma_l)=\sum_{n=1}^l( H_1(\gamma_n) + \sum_{n<m\leq l}\psi(\abs{\vec{x}_n-\vec{x}_m}))$ and the corresponding collision term~\cite{kardar2007statistical}.
If the hierarchy is not truncated, all the information of the microscopic dynamics is retained
and the system evolves according to the original Hamiltonian dynamics. 
Nevertheless, the one-particle distribution function, the corresponding Boltzmann entropy and the corresponding nonequilibrium free energy satisfy exact thermodynamic identities, as we  proceed to show in \S~\ref{p:entropy_micro}.

\subsection{Thermodynamic identities} 
\label{p:entropy_micro}

We  derive exact thermodynamic identities for the Hamiltonian dynamics, 
and use them to discuss the emergence of macroscopic thermodynamics by means of coarse-graining and dynamical chaoticity.

We start by introducing two definitions of entropy.
The Shannon entropy
\begin{align}
    S_\mathrm{tot}(t) \equiv - k_\mathrm{B} \int \d \Gamma_N\, P_N(\Gamma_N;t) \ln P_N(\Gamma_N;t)   \, \label{eq:shannon_entropy}
\end{align}
is defined using the probability density $P_N(\Gamma_N;t)$ over the whole phase-space,
while the Boltzmann entropy~\cite{dgm1962nonequilibrium}
\begin{align}
     S_\mathrm{B}(t) \equiv - k_\mathrm{B} \int \d \gamma_1 \,f_1(\gamma_1;t) (\ln f_1(\gamma_1;t)-1) 
     \label{eq:entropy_kinetic_def}\,
\end{align} 
is defined using the one-particle distribution function~\eqref{eq:one_particle_f_def}.
On the one hand, the Shannon entropy~\eqref{eq:shannon_entropy} is not a thermodynamic entropy: it has not the same properties of the macroscopic entropy introduced by the local equilibrium~\eqref{eq:local_eq_extensive}.
For instance, $S_\mathrm{tot}$ is a constant of motion, \emph{i.e.}, $\d_t S_{\mathrm{tot}}={ -k_\mathrm{B}}\int\d\Gamma_N \{P_N,H\}=0$, because the Liouville equation~\eqref{eq:liouville} preserves the volume in phase space.
On the other hand, the Boltzmann entropy~\eqref{eq:entropy_kinetic_def} can be related to the macroscopic entropy under certain conditions that we now examine.

To do so, we first express the Shannon entropy~\eqref{eq:shannon_entropy} in terms of the Boltzmann entropy~\eqref{eq:entropy_kinetic_def}:
\begin{equation}
     S_\mathrm{tot}(t) 
     =S_\mathrm{B}(t) -k_\mathrm{B}  D_{KL}\left(P_N(t)\left|\prod_{n=1}^N P_1(t)\right.\right) + k_\mathrm{B}  N (\ln N -1)\,, \label{eq:dkl_decomposition_total_entropy}
\end{equation}
where we used i) multiplication and division by the logarithm of the probability distribution $\prod_{n=1}^N P_1(t)\equiv\prod_{n=1}^N P_1(\gamma_n;t)$ that the system would have if there were no correlations between particles, ii) the fact that particles are identical and iii)
the relative entropy (or Kullback--Leibler divergence) $D_{KL}(p|q)$  between two probability distributions $p$ and $q$
\begin{align}
    D_{KL}(p|q)=\int \d x\, p(x) \ln \frac{p(x)}{q(x)} \geq0\,. \label{eq:dkl_def} 
\end{align}
Equation~\eqref{eq:dkl_decomposition_total_entropy}, together with $\d_tS_\mathrm{tot} = 0$, implies that 
 \begin{align}
     S_\mathrm{tot}(t) = S_\mathrm{tot}(0) =  S_\mathrm{B}(0) + k_\mathrm{B}  N(\ln N-1) \,,
     \label{eq:coarse_grained_entropy_identity_factorized_ic}
 \end{align}
if the system  is initialized in a state of completely uncorrelated particles, \emph{i.e.}, $P_N(\Gamma_N;0) = \prod_{n=1}^N P_1(\gamma_n;0)$. 
Hence, by combining Eqs.~\eqref{eq:coarse_grained_entropy_identity_factorized_ic} and~\eqref{eq:dkl_decomposition_total_entropy},
we obtain that the change of Boltzmann entropy is given by the relative entropy between the $N$-particle description, in terms of the probability density $P_N$, and a factorized description of the $N$-particles, by means of products of one-particle distribution $P_1$ (\emph{i.e.}, the mutual information with respect to the factorized state~\cite{cover1999elements}):
 \begin{align}
     S_\mathrm{B}(t)-S_\mathrm{B}(0) = k_\mathrm{B}  D_{KL}\left(P_N \left| \prod_n^N P_1\right.\right) \geq 0\,.
     \label{eq:ep_as_decorrelation_hamiltonian_level}
 \end{align}
The thermodynamic identity~\eqref{eq:ep_as_decorrelation_hamiltonian_level} represents a (weaker) microscopic analog of the second law of thermodynamics predicting the increase of Boltzmann entropy in isolated systems (like for the macroscopic entropy in Eq.~\eqref{eq:total_entropy_isolated}).
However, Eq.~\eqref{eq:ep_as_decorrelation_hamiltonian_level} is not equivalent to Eq.~\eqref{eq:total_entropy_isolated} as it does not imply the monotone increase of $S_\mathrm{B}$ (\emph{i.e.}, a H-theorem) since recurrences may occur~\cite{esposito2010entropy}.
This becomes evident  when expressing the variation $\Delta S_\mathrm{B}(t)$ during a very short time increment $\Delta t$ via Eq.~\eqref{eq:dkl_decomposition_total_entropy} together with $\d_t S_{\mathrm{tot}}=0$:
\begin{align}
    \Delta t \d_t S_\mathrm{B}(t) &\approx S_\mathrm{B}(t+\Delta t) - S_\mathrm{B}(t) 
    = {k_\mathrm{B} }D_{KL}\left( P_N(t+\Delta t) \left|\prod_n P_1(t+\Delta t) \right.\right) - {k_\mathrm{B} }D_{KL}\left( P_N(t) \left|\prod_n P_1(t) \right.\right)\,,
    \label{eq:microscopic_entropy_increase}
\end{align}
where we did not use any hypothesis on the initial state at time $t$. 
Indeed, according to the thermodynamic identity in  Eq.~\eqref{eq:microscopic_entropy_increase}, a positive rate of change of the Boltzmann entropy, namely, $ \d_t S_\mathrm{B}(t)\geq 0 $ at all times, is granted if the contribution coming from the second relative entropy on the right-hand side of Eq.~\eqref{eq:microscopic_entropy_increase} is always negligible compared to the first one, in agreement with the intuition gained via different approaches~\cite{lanford1975time, spohn2012large}.
This can happen when the stronger requirement of molecular chaos is satisfied, that is when the second relative entropy on the right-hand side of Eq.~\eqref{eq:microscopic_entropy_increase} vanishes at all times.
We can thus conclude that the coarse-grained description alone, in terms of the one-particle distribution function,
is not enough to guarantee irreversible thermodynamic behavior and should be complemented with hypotheses on the dynamics, \textit{e.g.,} molecular chaos.

For isolated systems,
the entropy is the thermodynamic potential that is maximized at equilibrium and
links the microscopic and macroscopic scales via its information-theoretic interpretation 
(see, for instance, Eq.~\eqref{eq:ep_as_decorrelation_hamiltonian_level}).
On the other hand,
non-isolated and non-isothermal systems feature
a different thermodynamic potential liking the microscopic and macroscopic scales via its information-theoretic interpretation.
Hence, 
we define the nonequilibrium Massieu potential~\cite{esposito2010entropy,rao2018conservation} 
at the microscopic scale as
\begin{align}
    \Phi(t) &\equiv   S_\mathrm{B}(t) -   \frac{1}{T} \davg{H_1}(t)\label{eq:massieu_def} \,,
\end{align}
where the average $\davg{...}$ stands for $\davg{...} = \int\d\gamma_1 f_1(\gamma_1) (...)$, in such a way that it gives an extensive contribution, with $T$ an arbitrary temperature.
Its equilibrium counterpart  $\Phi^\mathrm{eq}$  is obtained via Eq.~\eqref{eq:massieu_def}
by using the equilibrium one-particle distribution which reads, 
in the limit of impulsive interactions (\emph{i.e.}, short-range interaction not contributing to the total energy on average),
\begin{align}
     f_1^\mathrm{eq}{ (\gamma_1;T)} = NP_1^\mathrm{eq}{ (\gamma_1;T)} =\exp\left\{ -\frac{1}{k_{\mathrm{B}}T} \left({H_1} { (\gamma_1)}- \mucost_\mathrm{eq}\right)\right\}\,, \label{eq:one_particle_distr_eqm}
\end{align}
where $\mucost_\mathrm{eq}$ is the equilibrium chemical potential which gives the correct normalization.
Note that  i) at equilibrium $P_N^\mathrm{eq}(\Gamma_N; T) = \prod_n P_1^\mathrm{eq}(\gamma_n;T)$
and ii) the limit of impulsive interactions allows us to express the equilibrium one-particle distribution in terms of the $H_1$ only~\cite{kirkwood1946statistical}.

 In the following, we show  that i) $\Phi(t)$ is upper bounded by its equilibrium value $\Phi^\mathrm{eq}$ and
ii) $\Phi(t) - \Phi(0) \geq0$ when the initial state satisfies $P_N({ \Gamma_N};0) = \prod_{n=1}^N P_1(\gamma_n;0)$.
 First, by using Eq.~\eqref{eq:one_particle_distr_eqm} in Eq.~\eqref{eq:massieu_def} and summing and subtracting $f^\mathrm{eq}_b$, we obtain the following thermodynamic identity for the nonequilibrium Massieu potential:
\begin{align}
\begin{split}
    \Phi(t) - \Phi^\mathrm{eq}
    &= - k_\mathrm{B}  \hat{D}_{KL}\left(f_1(t)||f_1^\mathrm{eq}\right)\leq0\,,\label{eq:massieu_identity}
\end{split}
\end{align}
where
\begin{align} \hat{D}_{KL}(f|g)=
\int \d x  \left(f(x) \ln \frac{f(x)}{g(x)} +g(x)-f(x) \right)\geq0
\label{eq:generalized_dkl_def}
\end{align}
is the generalization of the relative entropy for non-normalized, positive functions $f$ and $g$ (also called Shear Lyapunov function in the context of chemical reaction networks~\cite{shear1967analog, higgins1968some}).
Notice that the same result~\eqref{eq:massieu_identity} applies also in the case of a time-dependent Hamiltonian of the type $H_1 = \vec{p}^2/{2} + {\phi(\vec x,t)}$, which we will use in \S~\ref{sec:mesoscopic_boltzmann_open}.

Second, 
by using Eqs.~\eqref{eq:massieu_def} and~\eqref{eq:ep_as_decorrelation_hamiltonian_level} 
together with the conservation of the single-particle energy,
\begin{equation}
\davg{H_1}{(t)}-\davg{H_1}{(0)}=0\,, 
\label{eq:hamiltonian_energy_balance}
\end{equation}
which holds for systems where the resulting force acting on the single particle due to impulsive interactions vanishes in average, \emph{i.e.},
$\int{ \prod_{n=1}^N} \d { \vec{x}_n}\, P_N \nabla_{\vec{x}}\psi(\abs{\vec{x}_1 -\vec{x}_m}))=0$, 
we obtain that the nonequilibrium Massieu potential increases in time:
\begin{align}
     \Phi(t)-\Phi(0) = S_\mathrm{B}(t) - S_\mathrm{B}(0)  \geq 0 \,.
     \label{eq:total_entropy_balance_hamiltonian_level_II}
\end{align}
Note that Eq.~\eqref{eq:total_entropy_balance_hamiltonian_level_II}, as well as Eq.~\eqref{eq:ep_as_decorrelation_hamiltonian_level}, holds
when the system  is initialized in a state of completely uncorrelated particles, \emph{i.e.}, $P_N(\Gamma_N;0) = \prod_{n=1}^N P_1(\gamma_n;0)$.

Finally, we notice that similar  thermodynamic identities have been already observed in different contexts. 
A similar thermodynamic identity to Eq.~\eqref{eq:ep_as_decorrelation_hamiltonian_level} has been already obtained using integral fluctuation relations~\cite{kawai2007dissipation, gaspard2022statistical}.
The difference is that we compare the $N$-particles distribution with a (completely decorrelated) product of one-particle distributions, while they introduce the backward Hamiltonian evolution which enters the integral fluctuation relations.
The thermodynamic identities~\eqref{eq:ep_as_decorrelation_hamiltonian_level}, \eqref{eq:microscopic_entropy_increase}, \eqref{eq:massieu_identity} are also similar to results obtained for systems coupled to equilibrium baths~\cite{esposito2010entropy}, but here there is no distinction between degrees of freedom of the system and of the baths.
The identities presented in this Section rely on coarse-graining the many degrees of freedom in favor of a description in terms of the one-particle distribution and the factorization of the initial distribution.
They are exact results which are consistent with the thermodynamic intuition, but do not lead to a full irreversible thermodynamics with a positive EPR.
In fact, the microscopic dynamics may not be chaotic enough or may involve too few particles to obtain thermodynamic behavior \cite{gaspard1998chaos, chakraborti2022entropy}.
A fully consistent thermodynamic description can be obtained at the mesoscopic level by using the assumption of molecular chaos, as we proceed to do in \S~\ref{sec:mesoscopic_boltzmann}.

\section{Mesoscopic Scale: Boltzmann equation}
\label{sec:mesoscopic_boltzmann}
The standard approach to introduce irreversibility  in the evolution of the one-particle distribution function relies on the crucial assumption of molecular chaos~\cite{kardar2007statistical},
which means that particles are assumed to be uncorrelated before they collide~\cite{spohn2012large}.
The resulting mesoscopic evolution is the Boltzmann equation.
We now review the dynamics and thermodynamics of the Boltzmann equation both in general and after applying the Chapman--Enskog expansion to obtain the hydrodynamic equations.

\subsection{Dynamics}
\label{p:boltzmann_general_dynamics}
The Boltzmann equation  
\begin{align}
    \partial_t f_1 + \{f_1,H_1\} =  C(f_1,f_1)\, \label{eq:boltzmann}
\end{align}
is a closed dynamical equation for the one-particle distribution function $f_1$  
resulting from truncating the BBGKY hierarchy~\cite{dgm1962nonequilibrium, kardar2007statistical, gaspard2022statistical} at the first equation~\eqref{eq:one-particle_ev} and 
approximating the collision integral involving the two-particles distribution function $f_2$ in terms of the product of two one-particle distribution functions times a scattering rate $W$:
\begin{align}
    C(f_1,f_1) = \int 
    {\d \vec{p}_2} 
    {\d\vec{p}_1' }
    {\d \vec{p}_2' }
    \, \left( W(\vec{p},\vec{p}_2|\vec{p}_1',\vec{p}_2')f_1(\vec{x},\vec{p}_1') f_1(\vec{x},\vec{p}_2')  -  W(\vec{p}_1',\vec{p}_2'|\vec{p},\vec{p}_2)  f_1(\vec{x},\vec{p})f_1(\vec{x},\vec{p}_2) \right)\,,\label{eq:collision}
\end{align}
Because of the underlying Hamiltonian dynamics~\cite{van1992stochastic}, the scattering rate $W$ must satisfy the time-reversal symmetry
\begin{align}
     W(\vec{p}_1,\vec{p}_2|\vec{p}_1',\vec{p}_2') =  W(-\vec{p}_1',-\vec{p}_2'|-\vec{p}_1,-\vec{p}_2)\,,\label{eq:symmetry_time_rev}
\end{align}
and  be invariant under a generic rotation  which implies invariance under inversion of the coordinate reference frame~$(\vec{x},\vec{p})\mapsto (-\vec{x},-\vec{p})$
\begin{align}
     W(\vec{p}_1,\vec{p}_2|\vec{p}_1',\vec{p}_2') =  W(-\vec{p}_1,-\vec{p}_2|-\vec{p}_1',-\vec{p}_2')\,.\label{eq:symmetry_inversion}
\end{align}
Combining Eqs.~\eqref{eq:symmetry_time_rev} and~\eqref{eq:symmetry_inversion} leads to the microscopic reversibility condition:
\begin{align}
W(\vec{p}_1,\vec{p}_2|\vec{p}_1',\vec{p}_2')=W(\vec{p}_1',\vec{p}_2'|\vec{p}_1,\vec{p}_2)\,. \label{eq:microreversibility}
\end{align}

In this framework, the so-called summational invariant functions  $A(\vec{x},\vec{p})$ 
represent
quantities satisfying  the conservation law 
\begin{equation}
    A(\vec{x},\vec{p}_1) +A(\vec{x},\vec{p}_2) - A(\vec{x},\vec{p}_1') -A(\vec{x},\vec{p}_2')=0\,,
    \label{eq:conservation_law_summational}
\end{equation}
when two particles collide in $\vec{x}$ and the momenta $\vec{p}_1$, $\vec{p}_2$, $\vec{p}_1'$ and $\vec{p}_2'$ satisfy the equation of motion, \emph{i.e.}, $W(\vec{p}_1',\vec{p}_2'|\vec{p}_1,\vec{p}_2)\neq 0$.
Their integral over the momentum is zero when multiplied by the collision integral~\cite{dgm1962nonequilibrium}:
\begin{align}
    \int \d \vec{p} \,A(\vec{x}, \vec{p})  C(f_1,f_1)  =0\,.
    \label{eq:summational_invariant}
\end{align}
This can be proved by multiplying the conservation law~\eqref{eq:conservation_law_summational} by the integrand of $C(f_1,f_1)$, integrating over all the momenta and using the symmetry~\eqref{eq:microreversibility} and indistinguishability of particles to obtain four times the integral in Eq.~\eqref{eq:summational_invariant}.
Functions $A(\vec{x})$ of the position only
  and constant functions are always summational invariants.
Moreover, since the scattering rate $W$ derives from an underlying Hamiltonian dynamics, also the kinetic energy and the momentum are summational invariants: 
$(\vec p_1)^2+(\vec p_2)^2-(\vec p_1')^2-(\vec p_2')^2 = 0$ and
$\vec p_1 + \vec p_2 - \vec p_1'- \vec p_2'=0$.
Note that Eq.~\eqref{eq:summational_invariant} implies that $\int\d\boldsymbol{p}\, C(f_1,f_1) =0$ and, therefore, that the Boltzmann equation~\eqref{eq:boltzmann} conserves the total number of particles.

To describe the behavior of the system in the configuration space, we introduce the (mesoscopic) spatio-temporal fields mass density and velocity as
\begin{subequations}
\begin{align}
   \aden(\vec{x}) & = M \int \frac{\d \vec{p}}{(2\pi \hbar)^{3}}\, f_1(\vec{x},\vec{p})\,,
    \label{eq:mass_density_velocity_boltzmann_defs_d}\\
    \avel(\vec{x}) & = (\aden(\vec{x}))^{-1}\int \frac{\d\vec{p}}{(2\pi \hbar)^{3}}\, \vec{p}\,  f_1(\vec{x},\vec{p}) \,. 
    \label{eq:mass_density_velocity_boltzmann_defs_v}
\end{align}
\label{eq:mass_density_velocity_boltzmann_defs}%
\end{subequations}
Their balance equations are obtained by   
applying the the material derivative $D_t = \partial_t + \avel\cdot\nabla_{\vec x} $
and
using the Boltzmann equation~\eqref{eq:boltzmann} when evaluating $\partial_t f_1$
in Eqs.~\eqref{eq:mass_density_velocity_boltzmann_defs}.
The balance equation for mass density reads
\begin{align}
    D_t \aden = - \aden \nabla_{\vec x} \cdot \avel\,,  \label{eq:mass_balance_boltzmann}
\end{align}
while the balance equation for the average velocity reads
\begin{align}
    \aden D_t \avel = - \nabla_{\vec x} \cdot \apres + \aden \vec{F}\,,
     \label{eq:momentum_balance_boltzmann}%
\end{align}%
where the pressure tensor can be expressed as
\begin{align}
    \apres(\vec{x}) = \frac{1}{M}\int\frac{\d \vec{p}}{(2\pi \hbar)^{3}}\,
    (\vec{p}-M\avel(\vec x)) (\vec{p}-M\avel(\vec x)) f_1(\vec{x},\vec{p})
     \label{eq:pressure_boltzmann}\,,
\end{align}
with $(\vec{p}-M\avel) (\vec{p}-M\avel)$ being the matrix whose entries are $({p_i}-M{\avele{i}}) ({p_j}-M{\avele{j}})$ ($i$ and $j$ labeling here the three spatial components) and $\vec{F} = -\nabla_{\vec x} \phi$ being the external force per unit mass at point $\vec{x}$.
Note that these mesoscopically-derived balance equations~\eqref{eq:mass_balance_boltzmann} and~\eqref{eq:momentum_balance_boltzmann} are in one-to-one correspondence to the macroscopic Navier-Stokes equations~\eqref{eq:NS}.

\subsection{Thermodynamics}
\label{p:boltzmann_general_thermodynamics}
We now introduce the {mesoscopic} definitions, as well as their respective balance equations, of the specific thermodynamic quantities examined at the macroscopic level in \S~\ref{sec:macro_balances}.
The (specific) internal energy is given by
\begin{align}
  \aien(\vec x) &= \frac{(\aden(\vec{x}))^{-1}}{2M} \int \frac{\d \vec{p}}{(2\pi \hbar)^{3}} (\vec{p}-M\avel(\vec x))^2 f_1(\vec{x},\vec{p}) \,, \label{eq:energy_boltzmann_def}
 \end{align}
and its balance equation, obtained using again the Boltzmann equation~\eqref{eq:boltzmann}, reads
\begin{align}
    \aden D_t \aien &= - \nabla_{\vec{x}} \cdot \aheat - \Tr \apres\nabla_{\vec{x}} \avel\,,
     \label{eq:energy_balance_boltzmann}
\end{align}
 where the heat flux is given by
\begin{align}
    \aheat(\vec x) = \frac{1}{2M^2} \int \frac{\d \vec{p}}{(2\pi \hbar)^{3}}\,
    (\vec{p}-M\avel(\vec x))^2 (\vec{p}-M\avel(\vec x)) f_1(\vec{x},\vec{p})  \label{eq:heat_flux_boltzmann}\,.
\end{align}
The specific entropy per unit mass defined, for consistency with Eq.~\eqref{eq:entropy_kinetic_def}, as
\begin{align}
     \aent(\vec{x}) \equiv -k_\mathrm{B} (\aden(\vec{x}))^{-1}\int \frac{\d \vec{p}}{(2\pi \hbar)^{3}}\, f_1(\vec{x},\vec{p}) (\ln f_1(\vec{x},\vec{p}) -1)\,, \label{eq:specific_entropy_boltzmann_def}
\end{align}
 satisfies the balance equation
\begin{align}
    \aden D_t \aent = - \nabla_{\vec{x}} \cdot \aef + \aepr\,. \label{eq:h-theorem-balance}
\end{align}
Here, the entropy flux is defined as
\begin{align}
    \aef(\vec{x}) = -k_\mathrm{B} \int \frac{\d \vec{p}}{(2\pi \hbar)^{3}} \,
    \left(\frac{\vec{p}}{M}-\avel(\vec x)\right) f_1(\vec{x},\vec{p}) \ln f_1(\vec{x},\vec{p}) \;.\label{eq:entropy_flux_boltzmann}
\end{align}
It is worth noting that, unlike the corresponding macroscopic fluxes in Eq.~\eqref{eq:entropy_balance}, the mesoscopic entropy flux~\eqref{eq:entropy_flux_boltzmann} is not proportional to the mesoscopic heat flux~\eqref{eq:heat_flux_boltzmann}. 
This difference emerges from the absence, in general, of the local equilibrium condition. 
We will verify in \S~\ref{sec:thermo_local_eqm} that if $f_1$ satisfies the local equilibrium, the proportionality between $\aef$ and $\aheat$ is recovered.
By omitting the spatial dependence for shortness and with the notation $\d^4 p=(2\pi\hbar)^{-3}\prod_{i=1,2} \d \vec{p}_i\d \vec{p}_i'$, the (local) EPR $\aepr$ reads
\begin{align}
    \aepr(\vec x)
           = \frac{k_\mathrm{B} }{4} \int \d^4p \, W(\vec{p}_1,\vec{p}_2|\vec{p}_1',\vec{p}_2') (f_1(\vec{p}_1') f_1(\vec{p}_2')  -  f_1(\vec{p}_1)f_1(\vec{p}_2)) \ln \frac{W(\vec{p}_1,\vec{p}_2|\vec{p}_1',\vec{p}_2')f_1(\vec{p}_1') f_1(\vec{p}_2')}{W(\vec{p}_1',\vec{p}_2'|\vec{p}_1,\vec{p}_2)f_1(\vec{p}_1)f_1(\vec{p}_2)} \geq 0\,, \label{eq:epr_boltzmann}
\end{align}
where the inequality follows from microscopic reversibility~\eqref{eq:microreversibility}. 
Note that the balance equation~\eqref{eq:h-theorem-balance} implies that for an isolated system
\begin{align}
    \d_t \ament =\int \d \vec{x}\, \aden(\vec{x}) \, D_t\aent(\vec{x}) = 
    \int \d \vec{x}\, \aepr(\vec{x}) \geq 0\,, \label{eq:h-theorem-inequality} 
\end{align}
namely, the (global) Boltzmann's entropy $\ament$ defined in Eq.~\eqref{eq:entropy_kinetic_def} monotonously increases in time according to the Boltzmann equation, in contrast to what we obtained in the microscopic theory (see Eq.~\eqref{eq:ep_as_decorrelation_hamiltonian_level}).
This is an alternative statement of Boltzmann's H-theorem~\cite{dgm1962nonequilibrium}.

The theory developed here is the mesoscopic analog of the macroscopic thermodynamics discussed in \S~\ref{sec:macro_balances}, but they are still not equivalent:
the crucial local equilibrium condition~\eqref{eq:local_eq}, or equivalently, the Euler relation~\eqref{eq:local_extensivity}, has not been derived yet.
To recover them, we need to introduce perturbative solutions of the Boltzmann equation as we now do in \S~\ref{sec:mesoscopic_chapman_enskog}.

\subsection{Linking the Mesoscopic and Macroscopic Description via the Chapman--Enskog Expansion}
\label{sec:mesoscopic_chapman_enskog}

Many perturbative solutions of the Boltzmann equation have been derived~\cite{dgm1962nonequilibrium, cercignani1988boltzmann}.
Since we are interested in
the dynamics of the spatio-temporal fields accounted by the balance equations~\eqref{eq:mass_balance_boltzmann},~\eqref{eq:momentum_balance_boltzmann},~\eqref{eq:energy_balance_boltzmann},~\eqref{eq:h-theorem-balance}, 
we apply the Chapman--Enskog scheme, \emph{i.e.}, a type of multiple-scales perturbation theory~\cite{van1987chapman}.

\subsubsection{Zeroth-order of the Chapman--Enskog expansion.}
\label{sec:par_zeroCE}

We start by identifying two (temporal and spatial) scales of the Boltzmann dynamics.
We introduce a macroscopic length scale~$L$ on which the external potential $\phi$ varies, and the associated macroscopic timescale $\tau=L/v_m$, with $v_m$ the typical molecular velocity.
The second timescale corresponds to the mean free time between collisions~$\tau_f$ and is related to the mean free path~$l=v_m \tau_f$. 
The Knudsen number is defined as the ratio between these two scales,
\begin{align}
   \varepsilon =  l/L = \tau_f / \tau \,,\label{eq:Knudsen}
\end{align}
and plays the role of the small parameter in the Chapman--Enskog perturbative scheme.
Indeed, by using these length and time scales to define adimensional variables (marked by the primes),
\begin{align}
    t = \tau t'  \,,\qquad 
    \vec{x} = L \vec{x}' \,,\qquad 
    \vec{p} = M v_m \vec{p}' \,,\qquad
    \vec{F}=\frac{{M}L}{\tau^2}\vec{F}'\,,\qquad
    C= \frac{C'}{\tau_f}\,, \qquad
    \label{eq:adimensional_coords} 
\end{align}
the Boltzmann equation~\eqref{eq:boltzmann} becomes (removing the primes for compactness)
\begin{align}
     \partial_t f_1  + \, \vec{p}\cdot\nabla_{\vec{x}} f_1+\vec{F}\cdot\nabla_{\vec{p}}f_1 =  \varepsilon^{-1} C(f_1,f_1)\,. \label{eq:boltzmann_adimensional}
\end{align}
 We are interested in the limit  $\varepsilon\to 0$, in which the collision term dominates the evolution of the one-particle distribution.
 
We thus consider the expansion
\begin{align}
    f_1 =  f^{(0)} + \varepsilon f^{(1)} + \varepsilon^2 f^{(2)} +\dots \,,\label{eq:expansion_f_1}
\end{align}
and substitute it in Eq.~\eqref{eq:boltzmann_adimensional}.
The zeroth-order solution is given by any one-particle distribution function for which the collision integral of the Boltzmann equation vanishes:
\begin{align}
    C(f^{(0)},f^{(0)})&=0\,. \label{eq:ce_0th_boltzmann} 
\end{align}
To identify $f^{(0)}$,
we notice that a solution for which the EPR in Eq.~\eqref{eq:epr_boltzmann} vanishes must also satisfy Eq.~\eqref{eq:ce_0th_boltzmann}.
Thus, we look for solutions of $\aepr = 0$ which can be rewritten in the form~\eqref{eq:summational_invariant} for the quantity $A=\ln f^{(0)}$ by using 
the symmetry in Eq.~\eqref{eq:microreversibility}
and the indistinguishability of the particles.
This implies that $\ln f^{(0)}$ must be a function of summational invariants, like mass, momentum and kinetic energy.
We can thus verify that the 
local Maxwell--Boltzmann distribution
\begin{equation}
    f^\mathrm{MB}(\vec{x},\vec{p}) =  \exp\left\{-\frac{1}{T(\vec{x})} \left(\frac{\left( \vec{p} - \vel(\vec{x})\right)^2}{2} -\mu(\vec{x})\right)\right\}\,,\label{eq:local_maxwell_boltzmann}
\end{equation}
with, in general, $\vec{x}$-dependent and $t$-dependent temperature $T(\vec{x})$, velocity $\vel(\vec{x})$, chemical potential of the ideal gas
\begin{equation}
     \mu(\vec{x}) = T(\vec{x})  \ln \left(z(\vec{x})\right) - \frac{3}{2} T(\vec{x}) \ln \frac{T(\vec{x})}{2\pi \hbar^2} \,,\label{eq:chemical_potential_ideal_adimensional}
\end{equation}
and density $z(\vec{x})$ (satisfying $\int\d\vec{x}\, z(\vec{x})=N$), solves Eq.~\eqref{eq:ce_0th_boltzmann}, namely, $f^{(0)} = f^\mathrm{MB}$.
Here all the quantities in Eqs.~\eqref{eq:local_maxwell_boltzmann} and~\eqref{eq:chemical_potential_ideal_adimensional} are written using the scaling defined in Eq.~\eqref{eq:adimensional_coords} (in particular, energy, temperature and $\hbar$ are expressed in ${Mv_m^2}$, ${Mv_m^2}/{k_\mathrm{B} }$ and $MLv_m$ units, respectively).
At this stage, the temperature $T(\vec{x})$, 
the velocity $\vel(\vec{x})$ 
and the density $z(\vec x)$ in Eq.~\eqref{eq:local_maxwell_boltzmann} are arbitrary parameters of the Maxwell--Boltzmann distribution, but they will be determined in the following using the balance equations. 
For convenience, we recast the expansion in Eq.~\eqref{eq:expansion_f_1} truncated at the first order using the Maxwell--Boltzmann distribution as
\begin{align}
    f_1(\vec x, \vec p) =  f^\mathrm{MB}(\vec x, \vec p) (1 + \varepsilon \chi(\vec x, \vec p)) + O(\varepsilon^2)\,,
     \label{eq:chapman_enskog_expansion}
\end{align}
namely, defining $\chi(\vec x, \vec p)\equiv f^{(1)}(\vec x, \vec p)/f^\mathrm{MB}(\vec x, \vec p)$.

Now we examine how the parameters of the Maxwell--Boltzmann distribution~\eqref{eq:local_maxwell_boltzmann}, that is the zeroth-order solution of \eqref{eq:boltzmann_adimensional}, are related to the (mesoscopic) spatio-temporal fields mass density $\aden(\vec x)$ and velocity $\avel (\vec x)$ introduced in \S~\ref{p:boltzmann_general_dynamics}.
By using Eq.~\eqref{eq:local_maxwell_boltzmann} in Eqs.~\eqref{eq:mass_density_velocity_boltzmann_defs_d} and~\eqref{eq:mass_density_velocity_boltzmann_defs_v}, we obtain
\begin{subequations}
\begin{align}
     \aden(\vec{x}) &=  \int \frac{\d \vec{p}}{(2\pi\hbar)^3}\, f^\mathrm{MB}(\vec{x},\vec{p}) = z(\vec x)\,, \label{eq:ce_0th_mass}\\
     \aden(\vec{x})\avel(\vec{x}) &=  \int \frac{\d \vec{p}}{(2\pi\hbar)^3} \,\vec{p}f^\mathrm{MB}(\vec{x},\vec{p}) = z(\vec{x}) \vel (\vec{x})\,,\label{eq:ce_0th_momentum}
\end{align}
\end{subequations}
and, consequently, the density $z(\vec x)$ and velocity $\vel (\vec x)$ are determined by the balance equations~\eqref{eq:mass_balance_boltzmann} and~\eqref{eq:momentum_balance_boltzmann} which can be rewritten now as
\begin{subequations}
\begin{align}
    D_t z &= - z \nabla \cdot\vel \,, \label{eq:zeroCE_mass} \\
    z D_t \vel &= - \nabla p + z \vec{F}\,,\label{eq:ce_0th_momentum_CL}
\end{align}\label{eq:zeroCE_NK}
\end{subequations}
by solving the integral in Eq.~\eqref{eq:pressure_boltzmann} using Eq.~\eqref{eq:local_maxwell_boltzmann} which leads to a diagonal pressure tensor,
\begin{equation}
    \aprese{ij}(\vec{x})=p(\vec{x})\delta_{ij}=z(\vec x)T(\vec{x})\delta_{ij}\,.
    \label{eq:pressur_tensor_MB}
\end{equation}
On the one hand, the zeroth-order Chapman--Enskog expansion
recovers the macroscopic mass balance equation (namely, Eq.~\eqref{eq:zeroCE_mass} coincides with Eq.~\eqref{eq:mass_conservation_lagrangian}).
On the other hand, it describes only fluids without viscous friction since a diagonal pressure tensor enters the zeroth order momentum balance equation~\eqref{eq:ce_0th_momentum_CL} unlike in Eq.~\eqref{eq:momentum_conservation}.

Second, we focus on the thermodynamic quantities introduced in \S~\ref{p:boltzmann_general_thermodynamics},
as well as their balance equations.
By using the Maxwell--Boltzmann distribution~\eqref{eq:local_maxwell_boltzmann},
the (specific) internal energy~\eqref{eq:energy_boltzmann_def} becomes
\begin{equation}
    \aden(\vec{x}) \aien(\vec{x})
    =\frac{1}{2} \int \frac{\d \vec{p}}{(2\pi\hbar)^3}\, (\vec{p} -  {\vel})^2 {f^\mathrm{MB}(\vec{x},\vec{p})}
    =\frac{3}{2}z(\vec x) T(\vec{x})\,, \label{eq:mb_internal}
\end{equation}
while its balance equation~\eqref{eq:energy_balance_boltzmann} reads
\begin{align}
      D_t T= -\frac{2}{3} T \nabla_{\vec{x}} \cdot \vel\,,  \label{eq:ce_0th_energy_CL}
\end{align}
since the heat flux vanishes, \emph{i.e.}, $\aheat(\vec x) = 0$.
The balance equation~\eqref{eq:ce_0th_energy_CL} implies that the temperature $T(\vec x)$ changes only due to the transport caused by the velocity field.
Analogously, the balance equation~\eqref{eq:h-theorem-balance} for the local entropy becomes
\begin{align}
    D_t \aent = 0\,,
    \label{eq:constant_entropy_zeroth_MB}
\end{align}
since the Maxwell--Boltzmann distribution leads to a vanishing entropy flux $\aef(\vec x) = 0$
and entropy producation rate $\aepr=0$.
This physically means that, at the zeroth-order of Chapman--Enskog expansion, the evolution of a fluid is isentropic along the flow lines.

Finally, by using the using the Maxwell--Boltzmann distribution~\eqref{eq:local_maxwell_boltzmann} in Eq.~\eqref{eq:specific_entropy_boltzmann_def} together with Eq.~\eqref{eq:mb_internal} and $p(\vec{x})=z(\vec x)T(\vec{x})$, we can write the specific entropy as
\begin{equation}
    T(\vec x)\aent(\vec{x}) = \aien(\vec{x}) + \frac{p(\vec x)}{z(\vec x)} - \mu(\vec x)\,,\label{eq:local_equilibrium_mb}
\end{equation}
which represents the  mesoscopic  formulation of the Euler relation~\eqref{eq:local_extensivity}. 
Note that, even if the zeroth-order of Chapman--Enskog expansion is consistent with the Euler relation~\eqref{eq:local_extensivity}, or equivalently with the local equilibrium condition~\eqref{eq:local_eq}, this consistency is a trivial one.
In fact, the dissipative effects characterizing the macroscopic thermodynamics and responsible for the variation of the local entropy are not recovered (see Eq.~\eqref{eq:constant_entropy_zeroth_MB}).
We thus move to the next order of the Chapman--Enskog expansion.

\subsubsection{First-order of the Chapman--Enskog expansion.}

The first-order equation of the Chapman--Enskog expansion of the (adimensional) Boltzmann equation~\eqref{eq:boltzmann_adimensional} can be written as
\begin{align}
\begin{split}
    {D_t} f^\mathrm{MB}  &+ (\vec{p}-\vel) \cdot \nabla_{\vec{x}} f^\mathrm{MB} + \vec{F}\cdot \nabla_{\vec{p}} f^\mathrm{MB} =\\
    &=\int  \d \vec{p}_2 \d\vec{p}_1' \d \vec{p}_2' \, f^{\mathrm{MB}}(\vec{x},\vec{p})f^{\mathrm{MB}}(\vec{x},\vec{p}_2) W(\vec{p}_1',\vec{p}_2'|\vec{p},\vec{p}_2) 
    \left( \chi(\vec{p}_1')+\chi(\vec{p}_2')-\chi(\vec{p})-\chi(\vec{p}_2)  \right)
    \,,
    \end{split}
    \label{eq:chapman_enskog_boltzmann_I}
\end{align}
by using Eq.~\eqref{eq:chapman_enskog_expansion}, 
the material derivative $D_t \equiv \partial_t + \vel\cdot \nabla_{\vec{x}}$, 
$C(f^\mathrm{MB}, f^\mathrm{MB})=0$,
and $f^{\mathrm{MB}}(\vec{p}_1')f^{\mathrm{MB}}(\vec{p}_2') = f^{\mathrm{MB}}(\vec{p}_1)f^{\mathrm{MB}}(\vec{p}_2)$ because of the conservation of the kinetic energy and momentum.
The first order correction $\chi$ must not contribute to the value of the spatio-temporal fields mass $\aden(\vec x)$, momentum $\avel (\vec x)$ and internal energy $\aien(\vec x)$ since they are already recovered from the the parameters $z(\vec x)$, $\vel(\vec x)$ and $T(\vec x)$ of the zeroth-order solution~\eqref{eq:local_maxwell_boltzmann} according to Eqs.~\eqref{eq:ce_0th_mass},~\eqref{eq:ce_0th_momentum} and~\eqref{eq:mb_internal}, respectively.
This implies that 
\begin{subequations}
\begin{align}
         & \int \frac{\d \vec{p}}{(2\pi\hbar)^3}\, \chi(\vec{x},\vec{p})f^\mathrm{MB}(\vec{x},\vec{p})=0 \label{eq:constraint_chi_I} \,,\\
    &\int \frac{\d \vec{p}}{(2\pi\hbar)^3} \,\vec{p}\,\chi(\vec{x},\vec{p})f^\mathrm{MB}(\vec{x},\vec{p})=0 \label{eq:constraint_chi_II}\,, \\
  & \int \frac{\d \vec{p}}{(2\pi\hbar)^3}\, (\vec{p} -  {\vel})^2 \chi(\vec{x},\vec{p})f^\mathrm{MB}(\vec{x},\vec{p})=0 \label{eq:constraint_chi_III}\,.
\end{align}
\end{subequations}
However, the correction $\chi$ should affect the balance equations of momentum, internal energy and entropy via the non-diagonal elements of the pressure tensor and the heat flux and, furthermore, lead to a nonvanishing entropy flux and EPR. 

To determine $\chi$, we start by expressing the terms on left-hand side of Eq.~\eqref{eq:chapman_enskog_boltzmann_I} using Eq.~\eqref{eq:local_maxwell_boltzmann} together with
Eqs.~\eqref{eq:zeroCE_mass},~\eqref{eq:ce_0th_momentum_CL} and~\eqref{eq:ce_0th_energy_CL}.
We obtain
\begin{subequations}
\begin{align}
     D_t \ln f^\mathrm{MB}
    &=\left( - \frac{2}{3} \nabla_{\vec{x}}\cdot \vel \right) \frac{(\vec{p}-{\vel})^2}{{2}T}  + \frac{\vec{p}-{\vel}}{T}\cdot \left( -\nabla_{\vec{x}} T - \frac{T}{z}\nabla_{\vec{x}}z + \vec{F}\right)\,,\\
    { \nabla_{x_i}\ln f^\mathrm{MB}}
    &=
    { \nabla_{x_i} T \left( \frac{(\vec{p}-{\vel})^2}{2T^2} -\frac{3}{2T} \right) + \frac{\sum_{j}({p}_j-{\vele_j})\partial_{x_i}\vele_j}{T} +\frac{1}{z}\nabla_{x_i}z}  \,,\\
    \nabla_{\vec{p}}\ln f^\mathrm{MB}&= - { \frac{1}{T}} \left(\vec{p}-{\vel}\right)\,,
\end{align}
\end{subequations}
with $i$ and $j$ labeling the three spatial components.
By using the strain tensor $\epsilon_{ij}=\frac{1}{2}(\partial_{x_i}{\vele}_j+\partial_{x_j} {\vele}_i)$, the left-hand side of Eq.~\eqref{eq:chapman_enskog_boltzmann_I} becomes
\begin{align}
\begin{split}
   &  D_t  f^\mathrm{MB}  + (\vec{p}-{\vel})  \cdot \nabla_{\vec{x}} f^\mathrm{MB} + \vec{F} \cdot \nabla_{\vec{p}} f^\mathrm{MB}=\\
       &=  f^\mathrm{MB} \left\{ \frac{1}{T}\sum_{ij} \left( (p_i-{\vele}_i)(p_j-{\vele}_j) - \frac{{1}}{3}(\vec{p}-{\vel})^2 \delta_{ij}\right)  \epsilon_{ij} +\frac{1}{T} \left(\frac{(\vec{p}-{\vel})^2}{2T} - \frac{5}{2}\right) (\vec{p}-{\vel})\cdot \nabla_{\vec{x}} T \right\} \,. \label{eq:c-e_derivatives_result}
\end{split}
\end{align}
Comparing this expression with Eq.~\eqref{eq:chapman_enskog_boltzmann_I} suggests that $\chi$ is a linear function of the temperature $\nabla_{\vec{x}}\vec{T}$ and velocity $\{\partial_{x_j}{\vele}_i\}$ gradients~\cite{dgm1962nonequilibrium}.
To determine its expression, one could introduce an expansion in appropriate orthogonal polynomials~\cite{chapman1990mathematical}.
However, for the sake of simplicity, we will employ the relaxation-time approximation or single-collision-time approximation~\cite{kardar2007statistical, lundstrom2002fundamentals}.
This approximation corresponds to use the properties of the Maxwell--Boltzmann distribution $f^\mathrm{MB}$ to obtain the following estimate of the  collision operator  on the right-hand side of Eq.~\eqref{eq:chapman_enskog_boltzmann_I},
\begin{small}
\begin{align}
\begin{split}
    C(f,f) &\simeq  \int \d\vec{p}_2\d\vec{p}_1'\d\vec{p}_2'\, W(\vec{p},\vec{p}_2|\vec{p}_1',\vec{p}_2') f^\mathrm{MB}(\vec{p}_1')f^\mathrm{MB}(\vec{p}_2')\left( \chi(\vec{p}_1')+\chi(\vec{p}_2')\right) + \\
    &\quad- f^\mathrm{MB}(\vec{p})\chi(\vec{p})\underbrace{\int\d\vec{p}_2 \d\vec{p}_1'\d\vec{p}_2'\, f^\mathrm{MB}(\vec{p}_2)W(\vec{p},\vec{p}_2|\vec{p}_1',\vec{p}_2')\left(1 + \frac{\chi(\vec{p}_2)}{\chi(\vec{p})}\right) }_{1/\tau_R} \approx -\tau_R^{-1} f^\mathrm{MB}(\vec{p}) \chi(\vec{p})\,.\label{eq:rtapprox}
\end{split}
\end{align}
\end{small}
Here, we assumed that before a collision takes place the distribution is the equilibrium one, \emph{i.e.}, $\chi(\vec{p}'_1)=\chi(\vec{p}'_2)\approx0$.
Furthermore, we assumed that the factor $f^{\mathrm{MB}}(\vec{p}_2)W(\vec{p},\vec{p}_2|\vec{p}_1',\vec{p}_2')$ decays fast enough as a function of the difference $\vec{p}_2-\vec{p}$ that only the zeroth-order contribution in the expansion around $\vec{p}$, namely $\chi(\vec{p}_2)\simeq \chi(\vec{p})$, can be retained.
Therefore, the resulting scattering timescale $\tau_R$ does not depend on $\chi$ or on the gradients of temperature or velocity.

Thus, by using the relaxation time approximation~\eqref{eq:rtapprox} together with Eq.~\eqref{eq:c-e_derivatives_result} in Eq.~\eqref{eq:chapman_enskog_boltzmann_I}, we conclude that
\begin{align}
    \chi(\vec{p}) &= -\frac{\tau_R}{ T} \left\{\sum_{ij} \left((p_i-{\vele}_i)(p_j-{\vele}_j) - \frac{{1}}{3}(\vec{p}-{\vel})^2 \delta_{ij}\right) \epsilon_{ij}+ \left(\frac{(\vec{p}-{\vel})^2}{2T} - \frac{5}{2}\right) (\vec{p}-{\vel})\cdot \nabla_{\vec{x}} T \right\}\,. \label{eq:chi_single_collision}
\end{align}
The relaxation time approximation is only appropriate for modeling near-equilibrium systems~\cite{lundstrom2002fundamentals}. A more rigorous treatment leads to the emergence of two different scattering timescales associated to viscous forces, accounted by the first term on the right-hand side of Eq.~\eqref{eq:chi_single_collision}, 
and thermal diffusion, accounted by the second term on the right-hand side of Eq.~\eqref{eq:chi_single_collision}.
This discrepancy becomes important in determining the quantitative values of the pressure tensor~\eqref{eq:pressure_boltzmann} and the heat flux~\eqref{eq:heat_flux_boltzmann},
but it is not fundamental for our qualitative discussion of thermodynamics.
For this reason, we will further simplify the expressions assuming $\tau_R=1$ in the following.

Notice that $\chi$ given in Eq.~\eqref{eq:chi_single_collision} satisfies the constraints in Eqs.~\eqref{eq:constraint_chi_I},~\eqref{eq:constraint_chi_II} and~\eqref{eq:constraint_chi_III} as one can verify using the properties of the Gaussian distribution:
\begin{subequations}
\begin{equation}
    \avg{(p_i-\vele_i) (p_j-\vele_j) }_\mathrm{MB} =  z(\vec x) T(\vec x)\delta_{ij} \,, \label{eq:gaussian_second_centered_moment}
    \end{equation}
    \begin{equation}
\avg{ (p_i-\vele_i) (p_j-\vele_j) (p_k-\vele_k) (p_l-\vele_l)}_\mathrm{MB} =z(\vec x) (T(\vec x))^2(\delta_{ij}\delta_{kl}+\delta_{ik}\delta_{jl}+\delta_{il}\delta_{jk})\,.\label{eq:gaussian_fourth_centered_moment}
    \end{equation}
\end{subequations}
In particular, Eq.~\eqref{eq:constraint_chi_II} is equivalent to 
$\int {\d \vec{p}}\, (\vec{p} -  {\vel}) \chi(\vec{x},\vec{p})f^\mathrm{MB}(\vec{x},\vec{p})=0$
(by using Eq.~\eqref{eq:constraint_chi_I}),
which is verified by $\chi$ given in Eq.~\eqref{eq:chi_single_collision}  because 
i) the terms multiplied by the strain tensor $\epsilon_{ij}$ independently vanish by symmetry,
while ii) the terms multiplied by the temperature gradient $\nabla_{\vec{x}}T$ sum up to zero after integration.

We have now all the elements to assess the consistency of the first order solution~\eqref{eq:chapman_enskog_expansion} (with $\chi$ given in Eq.~\eqref{eq:chi_single_collision}) with the macroscopic theory, as originally done by Prigogine~\cite{prigogine1949domaine}.

First, we consider the balance equations for the (mesoscopic) spatio-temporal fields mass $\aden(\vec x)$, momentum $\avel (\vec x)$ and internal energy $\aien(\vec x)$ given in
Eqs.~\eqref{eq:mass_balance_boltzmann},~\eqref{eq:momentum_balance_boltzmann} and~\eqref{eq:energy_balance_boltzmann}.
These equations have the same form as the macroscopic balance equations~\eqref{eq:mass_conservation_lagrangian},~\eqref{eq:momentum_conservation} and~\eqref{eq:internal_energy_balance_lagrangian}  using the zeroth-order solution.
Therefore, we only need to use the correction $\chi$ to evaluate the changes in the pressure tensor $\apres$ in Eq.~\eqref{eq:pressure_boltzmann} and the heat flux $\aheat$ in Eq.~\eqref{eq:heat_flux_boltzmann} which become equal to
\begin{subequations}
\begin{align}
    \aprese{ij}(\vec x) &=p(\vec{x}) \delta_{ij} + \apreseoff{ij}(\vec x) = \left(p+ \frac{2}{3}\varepsilon zT \sum_k\epsilon_{kk}\right)\delta_{ij} -2\varepsilon zT\epsilon_{ij} \,, \label{eq:chi_P}\\
    \aheat (\vec x)&=- \frac{5}{2}\varepsilon zT\nabla_{\vec{x}} T \,. \label{eq:chi_Jq}
\end{align}
\label{eq:chapman_enskog_fluxes_result}%
\end{subequations}
Unlike the zeroth-order expression \eqref{eq:pressur_tensor_MB}, 
Eqs.~\eqref{eq:chi_P} and~\eqref{eq:chi_Jq} account for the additional viscous part of the tensor pressure as well as a nonzero  thermal conductivity.
Furthermore, Eqs.~\eqref{eq:chi_P} and~\eqref{eq:chi_Jq} show a linear relation between $\apreseoff{}$ and $\aheat$, on the one hand, and their conjugate thermodynamic forces $\{\epsilon_{ij}\}$ and $\nabla_{\vec{x}} T$, on the other hand. 
This means that the first-order of the Chapman--Enskog expansion describes fluids in the linear regime, namely, close to equilibrium.

Second, the local thermodynamic equilibrium is left unaffected by the inclusion of the first Chapman--Enskog correction:
substituting the first order solution~\eqref{eq:chapman_enskog_expansion} (with $\chi$ in Eq.~\eqref{eq:chi_single_collision}) in Eq.~\eqref{eq:specific_entropy_boltzmann_def}, we obtain again Eq.~\eqref{eq:local_equilibrium_mb}, as a consequence of the constraints in Eqs.~\eqref{eq:constraint_chi_I},~\eqref{eq:constraint_chi_II} and~\eqref{eq:constraint_chi_III}.
On the other hand, the first order solution leads to a nonvanishing EPR which reads
\begin{align}
    \aepr= \frac{\varepsilon }{4} \int \d^{4}p\, W(\vec{p}_{1}',\vec{p}_{2}'|\vec{p}_{1},\vec{p}_{2})f^{\mathrm{MB}}(\vec{p}_{1})f^{\mathrm{MB}}(\vec{p}_{2})
  \left(\chi(\vec{p}_{1}')+ \chi(\vec{p}_{2}')-\chi(\vec{p}_{1})-\chi(\vec{p}_{2})\right)^2 = \dot{\sigma}_\mathrm{macro}+ O(\varepsilon^2), \label{eq:ce_epr_matching}
\end{align}
where the $\dot{\sigma}_\mathrm{macro}$ is the one computed in the linear regime of irreversible thermodynamics, \emph{i.e.}, after linearization of $\apreseoff{}$ and $\aheat$ in terms of $(\epsilon_{ij})$ and $\nabla_{\vec{x}}T$ as given in Eq.~\eqref{eq:epr_linear_regime}.

Third, we verify that in the presence of local equilibrium the entropy and heat flux are proportional to each other.
By substitutiong the Chapman--Enskog solution~\eqref{eq:chapman_enskog_expansion} in Eq.~\eqref{eq:entropy_flux_boltzmann} and using the expression of the Maxwell--Boltzmann and the property~\eqref{eq:constraint_chi_II}, we find
\begin{equation}
   \aef = \underbrace{-\int \frac{\d \vec{p}}{(2\pi\hbar)^3}\left( \vec{p}-\avg{\vec{v}}\right) f^\mathrm{MB}(\ln f^{\mathrm{MB}}+\varepsilon 
   \chi) }_{=0} - \varepsilon \int \frac{\d \vec{p}}{(2\pi\hbar)^3}\left( \vec{p}-\avg{\vec{v}}\right) f^\mathrm{MB}\chi \ln f^{\mathrm{MB}} +O(\varepsilon^2) = \frac{1}{T}\aheat +O(\varepsilon^2)\,.\label{eq:entropy_flow_proportionality_local_eq}
\end{equation}

The Chapman--Enskog procedure can in principle be continued to obtain higher order corrections in the Knudsen number.
However, the resulting Burnett and super-Burnett equations display unphysical oscillations or other artificial behavior near the boundary~\cite{cercignani1988boltzmann, struchtrup2007h}.
Furthermore, by considering the higher perturbative orders one looses the crucial property of local equilibrium~\cite{prigogine1949domaine, dgm1962nonequilibrium}.
Therefore, the higher orders of this approximation have little significance for thermodynamics and will not be considered in the following.

\subsection{Thermodynamics of the Transient Relaxation to Local Equilibrium}
\label{sec:thermo_local_eqm}

The Chapman--Enskog solution~\eqref{eq:chapman_enskog_expansion} describes only a restricted class of solutions of the Boltzmann equation~\eqref{eq:boltzmann_adimensional} close to the equilibrium for which the EPR~\eqref{eq:epr_boltzmann} coincides with the macroscopic one in the linear regime according to Eq.~\eqref{eq:ce_epr_matching}.
However, the Chapman--Enskog perturbation series does not describe the relaxation of an arbitrary initial condition towards the distribution~\eqref{eq:chapman_enskog_expansion}.
{
The study of the relaxation process of an arbitrary initial condition to the Chapman--Enskog solution is as difficult as the original Boltzmann equation.}
Therefore, we employ a perturbative treatment following Grad's method ~\cite{grad1963asymptotic, cercignani1988boltzmann}.
In this way, we are able to identify the contribution of this relaxation process to the EPR in Eq.~\eqref{eq:epr_boltzmann}.

We consider an initial condition $f_1(0)$ for the one-particle distribution close to the near-equilibrium solution~\eqref{eq:chapman_enskog_expansion}.
Then, we introduce two separate timescales.
One timescale is associated with the (slow) dynamics of the Chapman--Enskog solution~\eqref{eq:chapman_enskog_expansion}.
The other one, $\tau={\varepsilon^{-1}}t$, is related to a fast relaxation process, and is relevant only over times comparable to the Knudsen number~\eqref{eq:Knudsen}.
To account for both timescales, we add a  correction term to the Chapman--Enskog solution~\eqref{eq:chapman_enskog_expansion} and write
\begin{align}
 f_1(t,\tau) =  f_1^{\mathrm{CE}}(t)+\varepsilon f_1^{\mathrm{rel}}(\tau)   \,, \label{eq:boltzmann_transient_multiple_scale}
\end{align}
with the condition that $ f_1^{\mathrm{rel}}(\tau)$ decays to zero exponentially fast for $\tau\gg 1$.
By substituting Eq.~\eqref{eq:boltzmann_transient_multiple_scale} in the Boltzmann equation~\eqref{eq:boltzmann_adimensional}, we obtain
\begin{align}
\begin{split}
        &\varepsilon[(\partial_t+\vec{p}\cdot\nabla_{\vec{x}}+\vec{F}\cdot\nabla_{\vec{p}})f_1^{\mathrm{CE}}(t)] -C(f_1^{\mathrm{CE}},f_1^{\mathrm{CE}})=\\
        &\,= -\varepsilon^2[(\partial_t+\vec{p}\cdot\nabla_{\vec{x}}+\vec{F}\cdot\nabla_{\vec{p}})f_1^{\mathrm{rel}}(\tau)]  + \varepsilon\left(C(f_1^{\mathrm{rel}},f_1^{\mathrm{CE}})+C(f_1^{\mathrm{CE}},f_1^{\mathrm{rel}})\right) +\varepsilon^2 C(f_1^{\mathrm{rel}},f_1^{\mathrm{rel}})\,.
\end{split}
\label{eq:boltzmann_expanded_relaxation_I}
\end{align}
By introducing $\partial_\tau = \varepsilon\partial_t$, 
keeping only terms linear in $\varepsilon$, 
and using the fact that $f_1^{\mathrm{CE}}$ makes the left-hand side of Eq.~\eqref{eq:boltzmann_expanded_relaxation_I} vanish, being a solution of Eq.~\eqref{eq:boltzmann_adimensional} up to order $O(\varepsilon)$, 
we obtain
\begin{align}
    \partial_\tau f_1^{\mathrm{rel}} &= L\left(\frac{f_1^{\mathrm{rel}}}{f^{\mathrm{MB}}}\right)\,,\label{eq:initial_relaxation_dynamics}
\end{align}
where we introduced the linearized collision operator (acting on a generic distribution function $g$)
\begin{align}
    L(g) \equiv \int \d\vec{p}_1' \d \vec{p}_2 \d \vec{p}_2' \, W(\vec{p},\vec{p}_2|\vec{p}_1',\vec{p}_2')f^{\mathrm{MB}}(\vec{p}_1')f^{\mathrm{MB}}(\vec{p}_2')\left(g(\vec{p}_1')+ g(\vec{p}_2') - g(\vec{p})-g(\vec{p}_2)\right)\,,\label{eq:collision_linearized}
\end{align}
with $f^{\mathrm{MB}}$ the Maxwell--Boltzmann distribution~\eqref{eq:local_maxwell_boltzmann} satisfying $f^{\mathrm{MB}}(\vec{p}_1')f^{\mathrm{MB}}(\vec{p}_2') = f^{\mathrm{MB}}(\vec{p}_1)f^{\mathrm{MB}}(\vec{p}_2)$ because of the conservation of the kinetic energy.
The dynamics of $ f^{\mathrm{rel}}_1(\tau)$  is obtained by solving Eq.~\eqref{eq:initial_relaxation_dynamics}, and we refer to the original article~\cite{grad1963asymptotic} for the quantitative theory.
Here, we focus on the entropy production~\eqref{eq:epr_boltzmann} for distributions of the type in Eq.~\eqref{eq:boltzmann_transient_multiple_scale}.
The result in the dominant order in $\varepsilon$ is given by
\begin{equation}
  \dot{\sigma}_{\mathrm{B}} \simeq
  \dot{\sigma}_{\mathrm{macro}}+ \dot{\sigma}_{\mathrm{rel}} + \dot{\sigma}_\mathrm{cross} \,, \label{eq:epr_fast_relaxation_decoupled}
\end{equation}
where $\dot{\sigma}_{\mathrm{macro}}$ is given in Eq.~\eqref{eq:ce_epr_matching},
\begin{align}
    \dot{\sigma}_{\mathrm{rel}} = \frac{\varepsilon }{4} \int \d^{4}p\, W(\vec{p}_{1}',\vec{p}_{2}'|\vec{p}_{1},\vec{p}_{2})f^{\mathrm{MB}}(\vec{p}_{1})f^{\mathrm{MB}}(\vec{p}_{2})
  \left(
   \frac{f^{\mathrm{rel}}_1(\vec{p}_1')}{f^{\mathrm{MB}}(\vec{p}_{1}')}+\frac{ f^{\mathrm{rel}}_1(\vec{p}_{2}')}{f^{ \mathrm{MB}}(\vec{p}_{2}')}-\frac{ f^{\mathrm{rel}}_1(\vec{p}_{1})}{f^{\mathrm{MB}}(\vec{p}_{1})}-\frac{ f^{\mathrm{rel}}_1(\vec{p}_{2})}{f^{\mathrm{MB}}(\vec{p}_{2})} \right)^{2}\label{eq:epr_relaxation_def}
\end{align}
is the EPR of  the relaxation to the Chapman--Enskog distribution (the local equilibrium distribution) and becomes negligible for long times $\tau\gg 1$.
It is important to note, however, that for short times $\tau \lesssim 1$, the two contributions are of the same order in $\varepsilon$.
The third term in Eq.~\eqref{eq:epr_fast_relaxation_decoupled},
\begin{small}
    \begin{equation}
    \dot{\sigma}_\mathrm{cross} =\frac{\varepsilon }{2} \int \d^{4}p\, Wf^{\mathrm{MB}}(\vec{p}_{1})f^{\mathrm{MB}}(\vec{p}_{2})
  \left(
  \chi(\vec{p}_{1}')+ \chi(\vec{p}_{2}')-\chi(\vec{p}_{1})-\chi(\vec{p}_{2})
  \right)
  \left(\frac{f^{\mathrm{rel}}_1(\vec{p}_1')}{f^{\mathrm{MB}}(\vec{p}_{1}')}+\frac{ f^{\mathrm{rel}}_1(\vec{p}_{2}')}{f^{ \mathrm{MB}}(\vec{p}_{2}')}-\frac{ f^{\mathrm{rel}}_1(\vec{p}_{1})}{f^{\mathrm{MB}}(\vec{p}_{1})}-\frac{ f^{\mathrm{rel}}_1(\vec{p}_{2})}{f^{\mathrm{MB}}(\vec{p}_{2})}\right) \,,
\end{equation}
\end{small}
does not have, in general, a definite sign.
This means that while the total EPR in Eq.~\eqref{eq:epr_fast_relaxation_decoupled} is positive definite, it is not necessarily larger than the one due to the macroscopic processes, \emph{i.e.}, $\dot{\sigma}_{\mathrm{macro}}$.

\section{Mesoscopic Scale: generalized Boltzmann equation with external reservoirs}
\label{sec:mesoscopic_boltzmann_open}

The microscopic and mesoscopic theory
discussed in \S\S~\ref{sec:microscopic_hamiltonian} and~\ref{sec:mesoscopic_boltzmann}, respectively, are derived starting from an isolated system setup.
Therefore, they do not explicitly account for what happens at the system's boundary which is crucial in open systems, as it can lead to thermodynamic inconsistencies when arbitrary conditions are fixed at the boundary~\cite{gallavotti2002foundations, cercignani1988boltzmann, gaspard2022statistical}.
By taking inspiration from both old and recent works in nonequilibrium thermodynamics and ST~\cite{bergmann1955new, lebowitz1957irreversible, van2015stochastic, horowitz2016work}, we develop here a model that includes the boundary directly in the Boltzmann equation and examine the corresponding thermodynamics.
Note that throughout this Section we will use the adimensional variables introduced in Eq.~\eqref{eq:adimensional_coords}.

\subsection{Dynamics}
The boundary $\partial \Omega$ is split into different reservoirs, or baths, $\partial \Omega = \cup_{\bath} \partial \Omega_{\bath}$ like in \S~\ref{par:general_EPR}.
We assume that
each reservoir $b$ is composed of non-interacting particles always in equilibrium at the temperature $T_b$ and its one-particle distribution function is given by
\begin{align} 
    {f_1^{\bath}}(\vec{X},\vec{P}) =  \frac{N_b}{Z(T_{b})}\exp\left\{-\frac{1}{T_{\bath}} \left(\frac{(\vec{P}-\vec{v}_b)^2}{2M_b} + \mathcal{U}_b(\vec{X})\right)\right\} \,.\label{eq:bath_eqm_distribution}
\end{align}
Here, $Z(T_{b})$ is the partition function, 
$N_b$ is the total number of particles,
$\vec{v}_b$ is the (externally controlled) velocity of the reservoir, 
$\mathcal{U}_b(\vec{X})$ is a (arbitrary) potential that confines the reservoir's particles to $\partial\Omega_{\bath}$,
\emph{i.e.}, $N_b=\int_{\partial \Omega_{\bath}} \d\vec{X}  \int \d\vec{P}\,  f_1^{\bath}(\vec{X},\vec{P}) /(2\pi\hbar)^3$.
For simplicity, we further assume that the particles of the reservoir have the same mass of those of the system: $M_b=M $.

In the limit of dilute fluids, collisions are rare and can only take place either between pairs of particles of the system or between one particle of the system and one particle of a single bath.
Thus, we modify the Boltzmann equation~\eqref{eq:boltzmann_adimensional} by adding collision terms corresponding to each reservoir:
\begin{align}
  \partial_{t}f_1 + \vec{p}\cdot \nabla_{{\vec{x}}}f_1 + \vec{F}\cdot \nabla_{\vec{p}}f_1 = \varepsilon^{-1}C(f_1,f_1)+{\varepsilon^{-1}}\sum_{b=1}^{B}  C_{\bath}(f_1)\,, \label{eq:boltzmann_with_reservoirs}
\end{align}
where $C(f_{1},f_{1})$ is given in Eq.~\eqref{eq:collision}, while the linear collision term due to the reservoirs reads
\begin{align}
    C_{\bath}(f_{1})=& \int \d\vec{p}'\d \vec{P}\d\vec{P}'\, 
    (W_{\bath}(\vec{p},\vec{P}|\vec{p}',\vec{P}')f_{1}(\vec{p}')f_1^{\bath}(\vec{P}')- 
    W_{\bath}(\vec{p}',\vec{P}'|\vec{p},\vec{P})f_{1}(\vec{p})f_1^{\bath}(\vec{P}))  \,,  \label{eq:collision_bath}
\end{align}
with the scattering rate $W_{\bath}$ satisfying Eq.~\eqref{eq:microreversibility} like $W$.
Note that Eq.~\eqref{eq:boltzmann_with_reservoirs} generalizes previous formulations of linear Boltzmann equations discussed in the framework of ST~\cite{van2015stochastic,horowitz2016work} by including the nonlinear collision integral~$C(f_1,f_{1})$.

The balance equations
for the (mesoscopic) mass density and velocity 
defined in Eqs.~\eqref{eq:mass_density_velocity_boltzmann_defs} 
according to the generalized Boltzmann equation~\eqref{eq:boltzmann_with_reservoirs}
(which are analogous to Eqs.~\eqref{eq:mass_balance_boltzmann} and~\eqref{eq:momentum_balance_boltzmann}) read:
\begin{subequations}
    \begin{align}
    \partial_t\avg{\rho} &= -\nabla_{\vec{x}}\cdot\left(  \avg{\rho}\avg{\vec{v}}\right)  \,,\label{eq:open_boltzmann_mass_balance}\\
        \avg{\rho}\partial_t\avg{\vec{v}}&=-\avg{\rho}\avg{\vec{v}}\cdot\nabla_{\vec{x}}\avg{\vec{v}}-\nabla_{\vec{x}}\cdot \avg{P}+ \avg{\rho}\vec{F}+
        \varepsilon^{-1}{\sum_\bath} \int\frac{\d \vec{p}}{(2\pi\hbar)^3}\, \vec{p}\,C_\bath(f_1)\,,\label{eq:open_boltzmann_momentum_balance}
    \end{align}%
\label{eq:open_boltzmann_navier_stokes}%
\end{subequations}
where the pressure tensor $\avg{P}(\vec{x})$ is defined in Eq.~\eqref{eq:pressure_boltzmann},
the force can be written as $\vec{F}=\vec{F}_{\mathrm{int}}+\vec{F}_{\mathrm{w}}$ (see \S~\ref{sec:macro_energy_balances}),
and we used Eq.~\eqref{eq:boltzmann_with_reservoirs} together with the property of summational invariance Eq.~\eqref{eq:summational_invariant}.
{Notice that 
i) $\avg{\vec{v}}$ must satisfy the boundary conditions $\avg{\vec{v}} |_{\partial \Omega_b}= \vec{v}_\bath$, 
and ii) $\vec{F}_{\mathrm{w}}$ as well as $C_b(f_1)$ are nonvanishing only on the boundary $\partial \Omega$.}

\subsection{Thermodynamics}

We now derive the thermodynamic description corresponding to the generalized Boltzmann equation~\eqref{eq:boltzmann_with_reservoirs}.

\subsubsection{First and Second Law.\label{generalized_boltzmann_1and2_law}} 

The global momentum, the kinetic and potential energy balances corresponding to Eqs.~\eqref{eq:open_boltzmann_navier_stokes} are
\begin{subequations}
\begin{align}
        \d_t\davg{\vec{p}}&=\davg{\vec F}+
        \varepsilon^{-1}{\sum_\bath} \int_{\Omega(t)}\d\vec{x} \int\frac{\d \vec{p}}{(2\pi\hbar)^3}\, \vec{p}\,C_\bath(f_1)\,,\label{eq:open_boltzmann_momentum_integrated}\\
        \frac{1}{2}\d_t \davg{\vec{p}^2} &= \davg{\vec{p}\cdot \vec{F} }
       +\varepsilon^{-1}{\sum_{b}}\int_{\Omega(t)}\d\vec{x} \int\frac{\d \vec{p}}{(2\pi\hbar)^3}\, \frac{\vec{p}^2 }{2}  \,      C_\bath(f_1)\,,\label{eq:open_boltzmann_kin_en_balance}\\
       \d_t \davg{\phi}
       &= -\davg{\vec{p}\cdot\vec{F}} + {\sum_b  \vec{v}_b\cdot\int_{\partial\Omega_b}\d\vec{x} \avg{\rho}\vec{F}_{\mathrm{w}}}\,,\label{eq:open_boltzmann_pot_en_balance}
\end{align}%
\label{eq:open_boltzmann_global_balances}%
\end{subequations}
respectively.
In Eq.~\eqref{eq:open_boltzmann_pot_en_balance},
we used that boundary is split in different reservoirs moving at different velocities changing the confining potential $\phi_{\mathrm{w}}$ in such a way that 
$\int_{\Omega(t)}\d\vec{x}\, \partial_t\phi=
\int_{\partial\Omega(t)}\d\vec{x}\, \partial_t\phi_{\mathrm{w}} = \sum_b\vec{v}_b\cdot\int_{\partial\Omega_{b}(t)}\d\vec{x}\,\vec{F}_{\mathrm{w}}$.

We now use Eqs.~\eqref{eq:open_boltzmann_navier_stokes} and ~\eqref{eq:open_boltzmann_kin_en_balance} to express the balance for the (global) internal energy ${U} =\davg{(\vec{p}-{\avg{\vec{v}}})^2}/2$ in terms of the {power} $\dot{W}$ and the heat flows at the boundaries $\dot{Q}_\bath$ with each reservoir $\bath$:
\begin{subequations}
\begin{align}
   \d_t U & = \frac{1}{2}\d_t\davg{\vec{p}^2}-\frac{1}{2} \int_{\Omega(t)}\d\vec{x} \avg{\vec{v}}^2 \partial_t\avg{\rho} - \int_{\Omega(t)}\d\vec{x} \avg{\rho}\avg{\vec{v}}\cdot\partial_t\avg{\vec{v}}  \\
     &=\underbrace{-\int_{\Omega(t)}\d\vec{x}  \Tr \left(\avg{P}\nabla_{\vec{x}} \avg{\vec{v}}\right)}_{\equiv \dot{W}  }    +{\sum_\bath} \underbrace{\varepsilon^{-1}\int_{\Omega{(t)}}\d\vec{x} \int\frac{\d \vec{p}}{(2\pi\hbar)^3} \left(\frac{\vec{p}^2}{2}-\avg{\vec{v}}\cdot\vec{p}\right)C_\bath(f_1)}_{\equiv\dot{Q}_b}\,.\label{eq:final_first_law_generalized_boltzmann}
\end{align}
\end{subequations}
Equation~\eqref{eq:final_first_law_generalized_boltzmann} is the global formulation of the first law for the generalized Boltzmann equation~\eqref{eq:boltzmann_with_reservoirs}.

Before turning to the entropy balance, we consider the equilibrium distribution corresponding to the $b$-th reservoir, 
\begin{align}
    f_\bath^{\mathrm{eq}}(\vec{x},\vec{p}) =  \exp\left\{ - \frac{1}{T_{b}} \left( \frac{\left(\vec{p}-\vec{v}_b\right)^2}{2} - \mu^b_{\mathrm{eq}}(\vec x) \right)\right\}\,, \label{eq:final_eqm_distr}
\end{align}
with a time-dependent chemical potential $\mu^b_\mathrm{eq}$, 
that becomes the equilibrium solution $f^\mathrm{eq}_B$ of the generalized Boltzmann equation~\eqref{eq:boltzmann_with_reservoirs}  when all reservoirs have the same temperature $T_\bath = T_B$ and velocity $\vec{v}_\bath = \vec{v}_B$. 
Indeed, $f^\mathrm{eq}_B$ satisfies 
\begin{equation}
    D_t {f_B^{\mathrm{eq}}}(\vec{x},\vec{p}) = 0\,,\label{eq:open_boltzmann_eqm_material_derivative}
\end{equation}
where $ D_t = (\partial_t+\vec{v}_B\cdot\nabla_{\vec{x}})$,
and the equilibrium chemical potential $\mu_{\mathrm{eq}}$ satisfies $\nabla_{\vec x}\mu_{\mathrm{eq}}(\vec x) = \vec F(\vec x)$ and therefore reads
\begin{equation}
    \mu_{\mathrm{eq}}(\vec x,t) = -\phi(\vec x,t) + \mucost_{\mathrm{eq}}\,,
    \label{eq:eq_chempot}
\end{equation}
at every time $t$ and with $\mucost_{\mathrm{eq}}$ a constant.
This can be easily proven by direct substitution and using momentum conservation in the form 
${f_B^{{\mathrm{eq}}}}(\vec{p}){f_B^{{\mathrm{eq}}}}(\vec{p}_2) = {f_B^{{\mathrm{eq}}}}(\vec{p}'){f_B^{{\mathrm{eq}}}}(\vec{p}_2')$ 
as well as ${f_1^{b}}(\vec{P}){f_B^{{\mathrm{eq}}}}(\vec{p}) = {f_1^{b}}(\vec{P}'){f_B^{{\mathrm{eq}}}}(\vec{p'})$ when $T_\bath = T_{B}$ and $\vec{v}_b=\vec{v}_B$ for every reservoir.
Equation~\eqref{eq:final_eqm_distr},
together with $\int \d\vec{p}\,C_\bath(f_1)=0$, and the fact that the bath distribution $f^b_1(\vec{P})$ is nonvanishing only on the region ${ \partial}\Omega_b$ where $\avg{\vec{v}}=\vec{v}_b$,
allows us to write the heat flux (defined in Eq.~\eqref{eq:final_first_law_generalized_boltzmann}) as 
\begin{align}
   \dot{Q}_\bath = - \varepsilon^{-1} T_\bath \int_{\Omega(t)}\d\vec{x} \int \frac{\d\vec{p}}{(2\pi\hbar)^3}\, \ln f_\bath^\mathrm{eq}\, C_\bath(f_1) \,.\label{eq:heat_fluxes_equivalence}
\end{align}
Therefore, we obtain the second law (the global entropy balance) in the form of
\begin{small}
    \begin{subequations}
\begin{align}
   \d_t S_{\mathrm{B}} 
   &= -\d_t\int_{\Omega(t)}\d\vec{x} \int\frac{\d \vec{p}}{(2\pi\hbar)^3}
   (f_1\ln f_1 - f_1)\\
   &= -\varepsilon^{-1} \int_{\Omega(t)}\d\vec{x} \int\frac{\d \vec{p}}{(2\pi\hbar)^3} \left(C(f_1,f_1) \ln f_1  + { \sum_\bath} C_\bath(f_1 )  \ln \frac{f_1}{f^{{\mathrm{eq}}}_\bath} + {\sum_\bath} C_\bath(f_1)\ln f^{{\mathrm{eq}}}_\bath  \right) \\
   &=  \dot{\Sigma}_\mathrm{int} + {\sum_\bath} \left( \dot{\Sigma}_\bath + \frac{\dot{Q}_\bath}{T_\bath}\right)   \,.\label{eq:entropy_balance_generalized_boltzmann_open}%
\end{align}%
\end{subequations}%
\end{small}%
Equation~\eqref{eq:entropy_balance_generalized_boltzmann_open} follows from Eq.~\eqref{eq:reynolds}, 
the summational invariance~\eqref{eq:summational_invariant},
summing and subtracting $D_tf_1\ln f_\bath^{{{\mathrm{eq}}}}$, 
and identifying the EPR inside the system
\begin{align}
    \dot{\Sigma}_{\mathrm{int}} 
    =&-\varepsilon^{-1}\int_{\Omega(t)} \d \vec{x}\int\frac{\d \vec{p}}{(2\pi\hbar)^3} \, C(f_1,f_1)\ln f_1 \,,\label{eq:epr_inner_open_boltzmann}
\end{align}
and on the boundary
\begin{align}
    \dot{\Sigma}_\bath= &- \varepsilon^{-1}\int_{\Omega(t)} \d \vec{x}\int\frac{\d \vec{p}}{(2\pi\hbar)^3} \, C_{\bath}(f_1)\ln\frac{f_1}{f_\bath^{\mathrm{eq}}}\,,\label{eq:epr_boundary_open_boltzmann}
\end{align}
which are separately nonnegative as we now show.
Indeed, by using the indistinguishability of the particles and the symmetry~\eqref{eq:microreversibility},
Eq.~\eqref{eq:epr_inner_open_boltzmann} can be rewritten as
\begin{align}
    \dot{\Sigma}_{\mathrm{int}} 
    =& \frac{\varepsilon^{-1}}{4}\int_{\Omega(t)} \d \vec{x}\int \frac{\d \vec{p}\d \vec{p}_1'\d \vec{p}_2\d \vec{p}_2'}{(2\pi\hbar)^3} \,W(\vec{p},\vec{p}_2|\vec{p}_1',\vec{p}_2')(f_1(\vec{p}_1')f_1(\vec{p}_2')-f_1(\vec{p})f_1(\vec{p}_2))\ln\frac{f_1(\vec{p}_1')f_1(\vec{p}_2')}{f_1(\vec{p})f_1(\vec{p}_2)}\geq 0\,.\label{eq:open_boltzmann_inner_epr_inequality}
\end{align}
This approach cannot be applied to Eq.~\eqref{eq:epr_boundary_open_boltzmann} since the particles of the system and the particles of the reservoirs are not indistinguishable.
However, by using i) the explicit expression of $f_1^\bath$ and $f_\bath^{\mathrm{eq}}$ in Eqs.~\eqref{eq:bath_eqm_distribution} and~\eqref{eq:final_eqm_distr}, respectively, together with 
ii) the conservation of the kinetic energy in the form, \emph{i.e.}, $(\vec{p}-\vec{v}_{b})^2 - (\vec{p}'-\vec{v}_{b})^2 = (\vec{P}'-\vec{v}_{b})^2 - (\vec{P}-\vec{v}_{b})^2$, we obtain
\begin{subequations}
\begin{align}
    \dot{\Sigma}_\bath
    =& \frac{\varepsilon^{-1}}{2}\int_{\Omega(t)} \d \vec{x}\int \frac{\d \vec{p}\d \vec{P}\d \vec{p}'\d \vec{P}'}{(2\pi\hbar)^3} \,W_{\bath}(\vec{p},\vec{P}|\vec{p}',\vec{P}')(f_1(\vec{p}')f_1^{\bath}(\vec{P}')-f_1(\vec{p})f_1^{\bath}(\vec{P}))\ln\frac{f_1(\vec{p}')f_\bath^{\mathrm{eq}}(\vec{p})}{f_1(\vec{p})f_\bath^{\mathrm{eq}}(\vec{p}')} \,,\\
    =& \frac{\varepsilon^{-1}}{2}\int_{\Omega(t)} \d \vec{x}\int \frac{\d \vec{p}\d \vec{P}\d \vec{p}'\d \vec{P}'}{(2\pi\hbar)^3} \,W_{\bath}(\vec{p},\vec{P}|\vec{p}',\vec{P}')(f_1(\vec{p}')f_1^{\bath}(\vec{P}')-f_1(\vec{p})f_1^{\bath}(\vec{P}))\ln\frac{f_1(\vec{p}')f_1^{\bath}(\vec{P}')}{f_1(\vec{p})f_1^{\bath}(\vec{P})}\geq 0\,.
\end{align}\label{eq:epr_boundary_open_boltzmann_bis}
\end{subequations}

\subsubsection{Global thermodynamic potential.}
\label{sec:open_boltzmann_global_potential}

We now show that, when the system is
i) closed in a domain $\Omega(t)$ with the boundary moving at constant and uniform velocity ${\vec{v}_{B}}$,
and ii) in contact with a single external reservoir (identified by $b=B$) with temperature $T_{B}$, the dynamical equation~\eqref{eq:boltzmann_with_reservoirs} satisfies the analogous of a H-theorem (\emph{i.e.}, it admits a Lyapunov function) ruling the relaxation to equilibrium. 
To do so, we rely on the nonequilibrium Massieu potential 
$\Phi = S_\mathrm{B}(t) -  \davg{((\vec{p}{ -\vec{v}_B})^2 / 2 + \phi)}/T_{B}$ (defined in general in Eq.~\eqref{eq:massieu_def}), and we proceed in three steps.

First, as proven in Eq.~\eqref{eq:massieu_identity}, $\Phi$ is upper bounded by its equilibrium value $\Phi^\mathrm{eq}$ when the time-dependent Hamiltonian $H_1 = (\vec{p}{ -\vec{v}_B})^2/{2} + {\phi(\vec x,t)}$ results from all reservoirs moving with the same uniform velocity $\vec{v}_B$ 
and all reservoirs having the same temperature~$T_B$.

Second,
for the general case of multiple reservoirs $b=1,...,B$ with different velocities and temperatures, the time-derivative $\d_t\Phi$ reads
\begin{subequations}
    \begin{align}
    \d_t \Phi 
    &= \d_t  S_{\mathrm{B}} -  \frac{1}{2T_B}\d_t \davg{\vec{p}^2} { - \frac{1}{T_B} \d_t \davg{\vec{p}}\cdot\vec{v}_B}- \frac{1}{T_B} \d_t \davg{\phi}\\
  &= 
    \d_t S_\mathrm{B} 
 -  \sum_b \frac{\dot{Q}_b}{{T_B}}
 {-\frac{1}{T_B} \underbrace{\left\{\sum_b \int_{\Omega(t)}\d\vec{x}\, (\vec{v}_b{ -\vec{v}_B}) \cdot \left(\avg{\rho} \vec{F}_{\mathrm{w}}+\varepsilon^{-1}\int \frac{\d\vec{p}}{{ (2\pi\hbar)^3}} \,\vec{p}\,C_b(f_1)\right) {  -\davg{\vec{F}_{\text{int}}}\cdot\vec{v}_B}\right\}}_{\equiv \dot{W}_{\mathrm{mech}}}} \label{eq:open_boltzmann_massieu_partial}\\ 
    &= \dot{\Sigma}_{\mathrm{int}} + \sum_b  \dot{\Sigma}_{b} 
    +\sum_b \left( \frac{1}{T_b} - \frac{1}{T_B} \right)\dot{Q}_b -\frac{\dot{W}_{\mathrm{mech}}}{T_B}\,,
\label{eq:open_boltzmann_massieu_final}
\end{align}
\end{subequations}
by using Eqs.~\eqref{eq:open_boltzmann_global_balances},~\eqref{eq:entropy_balance_generalized_boltzmann_open} and~\eqref{eq:final_first_law_generalized_boltzmann} (with $\avg{\vec{v}}|_{\partial\Omega_b}=\vec{v}_b$), {  $\vec{F} = \vec{F}_\text{w} + \vec{F}_{\text{int}}$},
and summing and subtracting $({\varepsilon^{-1}}/{T_B})  \sum_b \int_{\Omega(t)}\d\vec{x}\int {\d\vec{p}}\,\vec{v}_b\cdot \vec{p}\,C_b(f_1)/{(2\pi\hbar)^3}$.
The mechanical power $\dot{W}_{\mathrm{mech}}$ generalizes the power $\dot{W}$ featuring the first law~\eqref{eq:final_first_law_generalized_boltzmann}. 
If the boundary is at rest, \emph{i.e.}, $\vec{v}_b = 0$ $\forall b$,  
the mechanical power vanishes,  $\dot{W}_{\mathrm{mech}}=0$.

Third, for the case of a single external temperature $T_B$ and velocity $\vec{v}_b=0$ $\forall b$ on the boundary, Eq.~\eqref{eq:open_boltzmann_massieu_final} becomes
\begin{align}
    \d_t \Phi =\dot{\Sigma}_{{ \text{int}}} +\sum_b \dot{\Sigma}_b\geq 0\,,
\end{align}
and, consequently, (minus) the nonequilibrium Massieu potential $\Phi$ plays the role of a Lyapunov function for the relaxation to equilibrium of the dynamics~\eqref{eq:boltzmann_with_reservoirs}. 
Note that the system relaxes to equilibrium also when the boundaries move with a single velocity $\vec{v}_b = \vec{v}_B$ $\forall b$
and the center of mass (CM) of the system moves at the same velocity,
i.e., $\vec{v}_{\text{CM}}=\vec{v}_{B}$.
Indeed, in this case the 
the mechanical power $\dot{W}_{\mathrm{mech}}$ in Eq.~\eqref{eq:open_boltzmann_massieu_partial} reduces to the rate of change of the potential energy of the center of mass, namely $\dot{W}_{\mathrm{mech}}=m\d_t \phi_{\text{int}}(\vec{x}_{\text{CM}})$ (see Eq.~\eqref{eq:meso_E_cm_rate_velocity_control}), 
provided that the internal potential energy $\phi_{\text{int}}$ is linear in $\vec{x}$,
namely $\phi_{\text{int}}((\vec{x})=\vec{g}\cdot\vec{x}$ 
(as it is for the gravitational potential), 
and the thermodynamic potential becomes $\Phi' =  S_\text{B}- ( \davg{((\vec{p}-\vec{v}_B)^2 / 2 + \phi)}-m \phi_{\text{int}}(\vec{x}_{\text{CM}}))/T_B$, satisfying $\d_t \Phi' =\dot{\Sigma}_{{ \text{int}}} +\sum_b \dot{\Sigma}_b\geq 0$.

Finally, we recast Eq.~\eqref{eq:open_boltzmann_massieu_final} in the form of a decomposition of the EPR. 
By isolating the total EPR, we obtain
\begin{equation}
    \dot{\Sigma}_\mathrm{tot}\equiv\dot{\Sigma}_\mathrm{int} +\sum_{\bath=1}^{{{B}}}\dot{\Sigma}_\bath = \d_t \Phi +\frac{\dot{W}_{\mathrm{mech}}}{T_B}  + \sum_{b} \dot{Q}_\bath \left(\frac{1}{T_B}-\frac{1}{T_\bath}\right)\,,
\end{equation}
that is the mesoscopic analogue of Eq.~\eqref{eq:epr_balance_open_full_td_2} (see also Eq.~\eqref{eq:epr_closed_velocity_II}).
Indeed, it features the time derivative of a thermodynamic potential $\d_t\Phi$;
a mechanical power contribution $\dot{W}_\mathrm{mech}$ due to the  velocity externally imposed on the boundary; 
a flux-force contribution due to the presence of heat fluxes $\dot{Q}_\bath$ exchanged with the environment and triggered by the presence of global differences in temperature.
For clarity of presentation, we did not consider the possible time-dependence in externally controlled parameters $\vec{v}_b$ or $T_b$, and therefore the corresponding terms in Eq.~\eqref{eq:epr_balance_open_full_td_2} do not appear here. 
Therefore, we can identify the nonequilibrium Massieu potential defined at the mesoscopic scale with the one featuring in Eq.~\eqref{eq:epr_balance_open_full_td_2}, \emph{i.e.}, $\Phi = Y$, if the one-particle distribution $f_1$ satisfies the local equilibrium.
This point is further examined in the following Section.

\subsubsection{Local thermodynamic potential.}

We now identify the local thermodynamic potential $\tpotmeso$  (\emph{i.e.}, the local version of the global potential $\Phi$), upper bounded by its equilibrium value $\tpotmeso_\mathrm{eq}$,  corresponding to the global macroscopic thermodynamic potential $\tpotmacro$ given in Eq.~\eqref{eq:macro_termo_pot},
and which governs the relaxation to equilibrium when  all the reservoirs have the same velocity $\vec{v}_b=\vec{v}_B$ and temperature $T_b=T_B$.
We start by observing that 
\begin{small}
    \begin{equation}
    0\leq\int \frac{\d \vec{p}}{(2\pi \hbar)^3}  \left( f_1(\vec{x},\vec{p}) \ln \frac{f_1(\vec{x},\vec{p})}{f_{{B}}^{\mathrm{eq}}(\vec{x},\vec{p})} +f_B^\mathrm{eq}(\vec{x},\vec{p})-f_1(\vec{x},\vec{p}) \right) = 
    -\aden \aent + \aden_\mathrm{eq}+\frac{\aden}{T_{B}}
    \left(\frac{(\avel - \vec v_{B} )^2}{2} + \phi+ \aien {-} \mucost_\mathrm{eq} \right)
    \,, \label{eq:local_variational_principle}%
\end{equation}%
\end{small}%
by using Eqs.~\eqref{eq:mass_density_velocity_boltzmann_defs},~\eqref{eq:energy_boltzmann_def},~\eqref{eq:specific_entropy_boltzmann_def},~\eqref{eq:final_eqm_distr},~\eqref{eq:eq_chempot} and $(\vec p - \vec v_{B})^2 = (\vec p - \avel)^2 +2\vec p\cdot (\avel - \vec v_{B} ) { -} \avel^2+(\vec v_{B})^2$.
We then identify the local total energy in the frame comoving with $\partial\Omega_b$ as
\begin{equation}
    \aten \equiv \frac{(\avel - \vec v_{B} )^2}{2} + \phi+ \aien\,,
\end{equation}
and recall that the pressure tensor satisfies Eq.~\eqref{eq:pressur_tensor_MB} when $f_1$ is a Maxwell--Boltzmann distribution like $f_B^\mathrm{eq}$ in Eq.~\eqref{eq:final_eqm_distr} yielding
\begin{equation}
    \aden_\mathrm{eq} = \frac{p_{B}}{T_{B}}\,,
\end{equation}
where $p_{B}$ is the pressure that the system would reach at equilibrium.
This allows us to write the mescoscopic thermodynamic potential as (minus) the right-hand side of Eq.~\eqref{eq:local_variational_principle} divided by the local density $\aden$:
\begin{equation}
  \tpotmeso = \aent -\frac{\aten + p_{B}\aden^{-1}}{T_{B}}{+}\frac{\mucost_\mathrm{eq}}{T_{B}}\,,
  \label{eq:local_potential}
\end{equation}
with $\aden^{-1}$ being the specific volume as in \S~\ref{sec:local_eq}. 
Note that $\tpotmeso_\mathrm{eq}=0$ because of the local equilibrium~\eqref{eq:local_equilibrium_mb} and that $\tpotmeso - \tpotmeso_\mathrm{eq}\leq 0$ because of Eq.~\eqref{eq:local_variational_principle}.
Furthermore, $\tpotmeso$ is the local, mesoscopic analog of the global, macroscopic potential $\tpotmacro$, and the exact identity $\int_{\Omega} \d\vec{x} \avg{\rho} \avg{y} = Y$ is recovered when {  (i) $\vec{v}_B=\vec{0}$ and (ii)} $f_1$ satisfies the local equilibrium, as it is the case for the Chapman--Enskog solution~\eqref{eq:chapman_enskog_expansion} that turns the mesoscopic specific entropy $\aent$ into the macroscopic specific entropy.  
Notice that the local thermodynamic potential~\eqref{eq:local_potential} has the same form as the macroscopic one~\eqref{eq:macro_termo_pot}, the only difference being the term $\mucost_\mathrm{eq}/ T_{B}$, which emerges because the system is locally open (\emph{i.e.}, it exchanges matter with its surroundings) as already observed in Ref.~\cite{avanzini2019thermodynamics}.
This same fact also prevents the existence of a local version of a H-theorem for $\avg{y}$.

\section{Stochastic thermodynamics for viscous fluids}
\label{sec:stochastic_thermo}

By building on Eq.~\eqref{eq:boltzmann_with_reservoirs}, we now develop a linear theory for the one-particle distribution function, endowed with the property of local detailed balance featuring an explicit non-conservative contribution, which is the form usually assumed in Stochastic Thermodynamics~\cite{katz1983phase, seifert2012stochastic, rao2018conservation}.
This theory further admits a Chapman--Enskog solution identical to the one obtained in \S~\ref{sec:mesoscopic_chapman_enskog} and therefore gives rise to the hydrodynamic equations presented in \S~\ref{sec:macro_balances}.

\subsection{Dynamics and Local Detailed Balance.}
\label{sec:ldb_viscous}

We consider a diluted suspension of $N_\alpha$ particles $\alpha$ in a weakly out-of-equilibrium fluid made up of $N_\beta$ particles $\beta$, with respective molecular masses are $M_\alpha$ and $M_\beta$. 
Since $N_\beta \gg N_\alpha$, the fluid of particles~$\beta$ is not affected by the presence of particles~$\alpha$.
The one-particle distribution, $f_{1\beta}$, is thus well described by the Chapman-Enskog form $f_{1\beta}= f_\beta^{\mathrm{MB}} (1+{ \varepsilon} \chi_\beta)$ 
with $f_{\beta}^{\mathrm{MB}}$ given in Eq.~\eqref{eq:local_maxwell_boltzmann} and $\chi_\beta$ given in Eq.~\eqref{eq:chi_single_collision}, 
except for a trivial change in normalization from $N$ to $N_\beta$. Also, $\langle\vec{v}_\beta\rangle= \vec{v}$, where $\vec{v}$ is the velocity of the fluid solving the macroscopic Navier-Stokes equations.
Since the particles $\alpha$ are diluted, the collisions among them are neglected, while their collisions with the fluid of particles $\beta$ and with the reservoirs constituting the boundaries are retained.
Therefore, dynamics of the one-particle distribution $f_{1\alpha}$ is given by \eqref{eq:boltzmann_with_reservoirs}, where the collision integral $C(f_{1\alpha}, f_{1\alpha})$ is neglected, a collision integral $C(f_{1\alpha}\,, f_{1\beta})$ is added and the collision integral $C_{\bath}(f_{1\alpha})$ is retained.

First, we consider the interaction with the boundary $\partial\Omega_{\bath}$
and we rewrite the collision integral~\eqref{eq:collision_bath} as 
\begin{align}
    C_{\bath}(f_{1\alpha}) = \int \d \vec{p}_{\alpha}' R_{\bath}(\vec{p}_{\alpha}|\vec{p}_{\alpha}') f_{1\alpha}(\vec{p}_{\alpha}') - f_{1\alpha}(\vec{p}_{\alpha}) \int \d \vec{p}_{\alpha}'  R_{\bath}(\vec{p}_{\alpha}'|\vec{p}_{\alpha})\,, \label{eq:collision_bath_linear}
\end{align}
where we introduced the transition rates
\begin{align}
    R_{\bath}(\vec{p}_{\alpha}|\vec{p}_{\alpha}') &\equiv \int \d \vec{P} \d \vec{P}'\, W_{\bath}(\vec{p}_{\alpha},\vec{P}|\vec{p}_{\alpha}',\vec{P}') f_{\bath}(\vec{P}')\,.\label{eq:mod_boltzmann_trans_rate_bath}
\end{align}
By using the one-particle distribution function~\eqref{eq:bath_eqm_distribution},
we prove that the transition rates satisfy the local detailed balance condition~\cite{bergmann1955new, katz1983phase, van2015stochastic, horowitz2016work, maes2021local, falasco2021local}
\begin{align}
    \frac{R_{\bath}(\vec{p}_{\alpha}|\vec{p}_{\alpha}')}{R_{\bath}(\vec{p}_{\alpha}'|\vec{p}_{\alpha})} = \exp\left\{-\frac{\Delta k_{\bath}(\vec{p}_{\alpha}|\vec{p}_{\alpha}')}{T_{\bath}} \right\} \,,\label{eq:ldb_no_force}
\end{align}
with
\begin{align}
    \Delta k_{\bath}(\vec{p}_{\alpha}|\vec{p}'_{\alpha}) \equiv \frac{1}{2}((\vec{p}_{\alpha}-\vec{v}_b)^2 - (\vec{p}'_{\alpha}-\vec{v}_b)^2) = \frac{1}{2} \left[ (\vec{p}_\alpha)^2-(\vec{p}_{\alpha}')^2\right] + \vec{v}_b \cdot \left( \vec{p}_\alpha - \vec{p}_{\alpha}' \right)
    \label{eq:energy_cons_ldb_res}
\end{align}
the variation of the kinetic energy of the system (measured with respect to the wall velocity $\vec{v}_\bath$) due to the transition $\vec{p}_{\alpha}'\mapsto \vec{p}_{\alpha}$ 
satisfying $ \Delta k_{\bath}(\vec{p}_{\alpha}|\vec{p}_{\alpha}') = ((\vec{P}'-\vec{v}_b)^2 - (\vec{P}-\vec{v}_b)^2)/2$
because of the conservation of the kinetic energy.
Note that Eq.~\eqref{eq:ldb_no_force} means that the local detailed balance condition holds for each reservoir individually.
Furthermore, 
the change in kinetic energy in Eq.~\eqref{eq:energy_cons_ldb_res} is expressed as the sum of two terms of the same order in the Knudsen number $\varepsilon$ but with different thermodynamic meaning.
The first term  after the second equality in Eq.~\eqref{eq:energy_cons_ldb_res} 
plays the role of the change in a thermodynamic potential.
The second term after the second equality in Eq.~\eqref{eq:energy_cons_ldb_res} is a nonconservative term that emerges from the collisions with the reservoir moving at the  velocity $\vec{v}_b$.
Indeed, it is the same as in the local detailed balance condition obtained in Ref.~\cite{horowitz2016work}.

Second, we consider the dynamics far from the boundary and we write the  the collision integral $C(f_{1\alpha}, f_{1\beta})$ as
\begin{equation}
    C_\varepsilon(f_{1\alpha}) = \int\d \vec{p}_{\alpha}'\, R_{\varepsilon}(\vec{p}_{\alpha}|\vec{p}_{\alpha}') f_{1\alpha}(\vec{p}_{\alpha}') - f_{1\alpha}(\vec{p}_{\alpha}) \int\d \vec{p}_{\alpha}'\, R_{\varepsilon}(\vec{p}_{\alpha}'|\vec{p}_{\alpha}) \,,
    \label{eq:linear_collision_bulk}
\end{equation}
with rates
\begin{align}
    R_{\varepsilon}(\vec{p}_{{\alpha}}|\vec{p}_{\alpha}') &\equiv  \int \d \vec{p}_{\beta} \d \vec{p}_{\beta}'\, W(\vec{p}_\alpha,\vec{p}_\beta|\vec{p}_\alpha',\vec{p}_\beta') {f_\beta^{\mathrm{MB}}(\vec{p}_\beta')(1+\varepsilon \chi_\beta(\vec{p}'_\beta))} \,,\label{eq:mod_boltzmann_trans_rate_gas}
\end{align}
since we assumed that $f_{1\beta}= f_\beta^{\mathrm{MB}} (1+\epsilon \chi_\beta)$.
Note that Eqs.~\eqref{eq:collision_bath_linear}, \eqref{eq:mod_boltzmann_trans_rate_bath}, \eqref{eq:linear_collision_bulk} and \eqref{eq:mod_boltzmann_trans_rate_gas} define a linear evolution equation for $f_{1\alpha}$.
To verify whether the rates~\eqref{eq:mod_boltzmann_trans_rate_gas} satisfy the local detailed balance condition or not, we expand the ratio $R_{\varepsilon}(\vec{p}_{\alpha}|\vec{p}_{\alpha}')/ R_{\varepsilon}(\vec{p}_{\alpha}'|\vec{p}_{\alpha})$ to first order in $\varepsilon$ 
and we obtain
\begin{equation}
    \frac{R_{\varepsilon}(\vec{p}_{\alpha}|\vec{p}_{\alpha}')}{R_{\varepsilon}(\vec{p}_{\alpha}'|\vec{p}_{\alpha})} = \exp\left\{-\frac{\Delta k(\vec{p}_{\alpha}|\vec{p}_{\alpha}')}{T(\vec x)}\right\}
    \left\{1+\frac{\varepsilon}{T(\vec x)}\frac{\int \d \vec{p}_\beta\d \vec{p}_\beta' W(\vec{p}_{\alpha},\vec{p}_\beta|\vec{p}_{\alpha}', \vec{p}_\beta')  f_\beta^{\mathrm{MB}}(\vec{p}_\beta') w(\vec{p}_\beta', \vec{p}_\beta)}{\int \d \vec{p}_\beta\d \vec{p}_\beta' W(\vec{p}_{\alpha},\vec{p}_\beta|\vec{p}_\alpha', \vec{p}_{\beta}')  f_\beta^{\mathrm{MB}}(\vec{p}_\beta') }\right\}\,,
    \label{eq:ldb_ce_I}
\end{equation}
where we used the notations
\begin{subequations}
\begin{align}
\Delta k(\vec{p}_\alpha|\vec{p}_\alpha') &\equiv  \frac{1}{2}((\vec{p}_{\alpha}-\vec{v})^2 - (\vec{p}'_{\alpha}-\vec{v})^2) = \frac{1}{2} \left[ (\vec{p}_{\alpha})^2 - (\vec{p}_{\alpha}')^2\right] + \vec{v} \cdot\left( \vec{p}_{\alpha}' - \vec{p}_{\alpha}\right) \,,    \label{eq:deltaK} \\
    \frac{w(\vec{p}_\beta', \vec p_\beta)}{T(\vec{x})}&\equiv\chi_\beta(\vec{p}_\beta') -\chi_\beta(\vec{p}_\beta).
    \label{eq:chi_reversed} 
\end{align} 
\end{subequations}
Equation~\eqref{eq:ldb_ce_I} does not correspond to the local detailed balance condition.
Indeed, on the one hand, the most general formulation of the local detailed balance~\cite{rao2018conservation} for a transition $e$ reads $R({e})/R({-e})=\exp\{-(\Delta_e \Phi - F(e)Q(e))/T_e\}$, where
$R({e})$ (resp. $R({-e})$) is the forward (resp. backward) rate,
$\Delta_e\Phi$ is the change of the thermodynamic potential, as in the local detailed balance~\eqref{eq:ldb_no_force} for collisions at the boundary,
while $F(e)Q(e)$ is the nonconservative work that 
depends only on the transition and is given by the product between a thermodynamic force $F(e)$ and the  conjugate quantity~$Q(e)$.
On the other hand, the term of order $\varepsilon$ in Eq.~\eqref{eq:ldb_ce_I} does not depend only on the change in momentum, which identifies a transition in our case. 
However, Eq.~\eqref{eq:ldb_ce_I} becomes equivalent to the local detailed balance condition in the small variance limit $ M_\alpha\vec{v}^2 \gg k_\mathrm{B}T$ where the Maxwell--Boltzmann distribution $f_\beta^{\mathrm{MB}}(\vec{p}_\beta')$ becomes sharply peaked around $\vec{v}$ (see Eq.~\eqref{eq:local_maxwell_boltzmann}).
This, together with the momentum conservation $\vec{p}_\beta' = \vec{p}_\beta + \Delta \vec{p}_\alpha $ 
(with $\Delta \vec{p}_\alpha = \vec {p}_\alpha- \vec{p}_\alpha'$), leads to 
\begin{equation}
    \frac{R_{\varepsilon}(\vec{p}_\alpha|\vec{p}_\alpha')}{R_{\varepsilon}(\vec{p}_\alpha'|\vec{p}_\alpha)} = \exp\left\{-\frac{\Delta k(\vec{p}_\alpha|\vec{p}_\alpha')-\varepsilon w(\vec{v},\vec{v} - \Delta \vec{p}_\alpha )}{T(\vec x)}\right\}\,,\label{eq:ldb_ce_II}
\end{equation}
using $e^{\varepsilon a} \simeq 1+\varepsilon a$.
Equation~\eqref{eq:ldb_ce_II} is the local detailed balance for the collisions between particles of type $\alpha$ and $\beta$, and features two contributions with a different scaling in the Knudsen number $\varepsilon$.
The zeroth order contribution, $\Delta k(\vec{p}_\alpha|\vec{p}_\alpha')$, 
can be split into two terms (see Eq.~\eqref{eq:deltaK}) with different thermodynamic meanings, as for $\Delta k_b(\vec{p}_\alpha|\vec{p}_{\alpha}')$ in  Eq.~\eqref{eq:energy_cons_ldb_res}.
The first term after the second equality in Eq.~\eqref{eq:deltaK} 
plays the role of the change in a thermodynamic potential.
The second term after the second equality in Eq.~\eqref{eq:deltaK} is a nonconservative term that emerges from the collisions with the particles $\beta$ moving at the average velocity $\vec{v}$.
It generalizes Eq.~\eqref{eq:energy_cons_ldb_res} to 
reservoirs
with spatially nonuniform temperature and velocity.
The first order contribution, $w(\vec{v},\vec{v}-\Delta\vec{p}_\alpha)$,
is an additional nonconservative term due to temperature and velocity gradients.
Indeed, by using Eqs.~\eqref{eq:chi_reversed} and~\eqref{eq:chi_single_collision},  
$w(\vec{v},\vec{v}-\Delta\vec{p}_\alpha)$ becomes
\begin{equation}
    w(\vec{v},\vec{v} -\Delta \vec p_\alpha)= - T\chi_\beta(\vec{v} -\Delta \vec p_\alpha) = 
 \left\{\sum_{ij} \left( (\Delta p_{\alpha,i})(\Delta p_{\alpha,j}) - \frac{{1}}{3}(\Delta \vec{p}_\alpha)^2 \delta_{ij}\right) \epsilon_{ij}{-} \left(\frac{(\Delta\vec{p}_\alpha)^2}{2T} - \frac{5}{2}\right) (\Delta\vec{p}_\alpha)\cdot \nabla_{\vec{x}} T \right\}\,.
\end{equation}
The thermodynamic forces are the local strain tensor $\epsilon_{ij}$ and the local temperature gradient $\nabla_{\vec{x}}T$, while the conjugate quantities are $\{\Delta p_{\alpha,i}\}$, which are conserved in the corresponding thermodynamically isolated system.
The thermodynamic forces can be determined by solving the macroscopic equations for $\vec{v}$ and $T$ and vanish when temperature and velocity become uniform in space.
The necessity of a small noise approximation (like the small variance limit leading to Eq.~\eqref{eq:ldb_ce_II}) when deriving the local detailed balance condition as a coarse-graining of an underlying nonconservative dynamics has been already observed in the context of overdamped diffusion in a potential landscape plus a small nonconservative force~\cite{falasco2021local}.

Therefore, by using the rates in Eqs.~\eqref{eq:mod_boltzmann_trans_rate_bath} and~\eqref{eq:mod_boltzmann_trans_rate_gas} satisfying the local detailed balance conditions in Eqs.~\eqref{eq:ldb_no_force} and~\eqref{eq:ldb_ce_II}, the generalized Boltzmann equation~\eqref{eq:boltzmann_with_reservoirs} for the type $\alpha$ particles  becomes the following linear equation (dropping the $\alpha$ subscript when redundant from now on):\begin{equation}
    \varepsilon (\partial_t + {\vec{p}}\cdot\nabla_{\vec{x}} + \vec{F}\cdot \nabla_{\vec{p}})f_{1\alpha} = \int \d \vec{p}' \left[ R_\varepsilon(\vec{p}|\vec{p}')f_{1\alpha}(\vec{p}') -  R_\varepsilon(\vec{p}'|\vec{p})f_{1\alpha}(\vec{p}) \right] + \sum_b \int \d \vec{p}' \left[ R_\bath(\vec{p}|\vec{p}')f_{1\alpha}(\vec{p}') -  R_\bath(\vec{p}'|\vec{p})f_{1\alpha}(\vec{p}) \right]\,. \label{eq:generalized_master_eq}
\end{equation}

\subsection{Chapman--Enskog Expansion}
\label{sec:ce_alpha_part}
We now show that Eq.~\eqref{eq:generalized_master_eq} admits a Chapman–Enskog solution and, consequently, it has a macroscopic hydrodynamic limit.
To do so, we apply the Chapman--Enskog procedure of \S~\ref{sec:mesoscopic_chapman_enskog} to Eq.~\eqref{eq:generalized_master_eq} in the small variance limit in which the local detailed balance \eqref{eq:ldb_ce_II} holds.
By inserting $f_{1\alpha} = f_{1\alpha}^{(0)}+{ \varepsilon} f_{1\alpha}^{(1)}+O({ \varepsilon}^2)$ into Eq.~\eqref{eq:generalized_master_eq}, zeroth-order in $\varepsilon$ reads
\begin{align}
    \int \d \vec{p}' \left[R_0(\vec{p}|\vec{p}')f_{1\alpha}^{(0)}(\vec{p}')  - R_0(\vec{p}'|\vec{p})f_{1\alpha}^{(0)}(\vec{p})  \right] + \sum_b \int \d \vec{p}' \left[ R_\bath(\vec{p}|\vec{p}')f_{1\alpha}^{(0)}(\vec{p}')  -  R_\bath(\vec{p}'|\vec{p})f_{1\alpha}^{(0)}(\vec{p}) \right] =0\,. \label{eq:master_eq_ce_0}
\end{align}
We now assume that the two terms in Eq.~\eqref{eq:master_eq_ce_0} vanish separately and then check the consistency of the result.
Note that the first term must vanish in every point of the domain $\Omega$, while the second only on the boundary $\partial \Omega$ because of the definition of $R_\bath$.
By using the local detailed balance~\eqref{eq:ldb_ce_II} with $\varepsilon = 0$,
we find that the first term is zero if $f_{1\alpha}^{(0)}$ is a Maxwell--Boltzmann distribution~\eqref{eq:local_maxwell_boltzmann}
with the temperature and mesoscopic velocity field $\vec v$ 
 solving the macroscopic Navier-Stokes equations determined by the type $\beta$ particles.
By using the local detailed balance~\eqref{eq:ldb_no_force},
we find that the Maxwell--Boltzmann distribution makes also the second term in Eq.~\eqref{eq:master_eq_ce_0} vanish when the temperature and mesoscopic velocity field assume the values imposed by the reservoirs on the boundary, \textit{i.e.}, $T_b$ and $\vec{v}_b$.

We now examine the first-order correction in $\varepsilon$ of Eq.~\eqref{eq:generalized_master_eq}. 
The first-order correction of the left-hand side of Eq.~\eqref{eq:generalized_master_eq} is the same as the one obtained in Eq.~\eqref{eq:c-e_derivatives_result} when applying the Chapman--Enskog procedure to the original Boltzmann equation~\eqref{eq:boltzmann_adimensional}.
To determine the first-order correction of the right-hand side of Eq.~\eqref{eq:generalized_master_eq}, we use $f_{1\alpha}=  f_{\alpha}^{\mathrm{MB}}(1 +{ \varepsilon} \chi_\alpha) +O({ \varepsilon}^2)$ together with Eqs.~\eqref{eq:master_eq_ce_0} and the local detailed balance conditions~\eqref{eq:ldb_no_force} and~\eqref{eq:ldb_ce_II} obtaining
\begin{subequations}
\begin{align}
    & R_\varepsilon(\vec{p}|\vec{p}')f_{1\alpha}(\vec{p}') - R_\varepsilon(\vec{p}'|\vec{p}) f_{1\alpha}(\vec{p}) = \varepsilon\big[
    R_0(\vec{p}|\vec{p}')f_{\alpha}^\mathrm{MB}(\vec{p}')\frac{w(\vec{v},\vec{v}-\Delta\vec{p})}{T} 
    - R_0(\vec{p}'|\vec{p})f_{\alpha}^\mathrm{MB}(\vec{p})(\chi_\alpha(\vec{p}) -\chi_\alpha(\vec{p}'))  \big] + O(\varepsilon^2)\,\label{eq:collision_rates_alpha_CE_I}\\
    & R_\bath(\vec{p}|\vec{p}') f_{1\alpha}(\vec{p}')- R_\bath(\vec{p}'|\vec{p})  f_{1\alpha}(\vec{p}) =
     \varepsilon R_\bath(\vec{p}'|\vec{p})f_{\alpha}^\mathrm{MB}(\vec{p})\left[\chi_{\alpha}(\vec{p}')- \chi_\alpha(\vec{p})   \right] + O(\varepsilon^2)\,.\label{eq:collision_rates_alpha_CE_II}
\end{align}\label{eq:collision_rates_alpha_CE}%
\end{subequations}%
We now apply again the relaxation time approximation~\eqref{eq:rtapprox}.
By assuming that i) the perturbation is negligible before a collision, \emph{i.e.}, $\chi_\alpha(\vec{p}'_1)\approx0$ 
and ii) $R_0(\vec{p}'|\vec{p})$ decays fast enough with $\Delta\vec{p}$ implying  $w(\vec{v},\vec{v}-\Delta\vec{p})=O(\abs{\Delta \vec{p}}^2)$, 
Eq.~\eqref{eq:collision_rates_alpha_CE_I} leads to
\begin{align}
    \int \d \vec{p}' \left\{ 
    R_0(\vec{p}|\vec{p}')f_\alpha^\mathrm{MB}(\vec{p}')\frac{w(\avg{\vec{v}},\avg{\vec{v}}-\Delta\vec{p})}{T} 
    - R_0(\vec{p}'|\vec{p})f_\alpha^\mathrm{MB}(\vec{p})\left[\chi_\alpha(\vec{p}) -\chi_\alpha(\vec{p}')  \right] \right\}\approx - {(\tau_R')^{-1}}{f_\alpha^\mathrm{MB}(\vec{p})\chi_\alpha(\vec{p})}\,, \label{eq:rtapprox_mastereq_bulk}
\end{align}
where $(\tau_R')^{-1} \equiv \int\d\vec p'\,R_0(\vec{p}'|\vec{p})$ is $\vec p$-independent for the same reasons as in Eq.~\eqref{eq:rtapprox}.
By applying the same reasoning to Eq.~\eqref{eq:collision_rates_alpha_CE_II}, 
we obtain
\begin{align}
 \sum_{\bath} \int \d \vec{p}' R_\bath(\vec{p}'|\vec{p})f_\alpha^\mathrm{MB}(\vec{p})\left[\chi_\alpha(\vec{p}')- \chi_\alpha(\vec{p})   \right] \approx -  f_\alpha^\mathrm{MB}(\vec{p})\chi_\alpha(\vec{p}) \sum_\bath \tau^{-1}_\bath\,,\label{eq:rtapprox_mastereq_bath}
\end{align}
with $(\tau_b)^{-1} \equiv \int\d\vec p'\,R_b(\vec{p}'|\vec{p})$.
By combining the left-hand side of Eq.~\eqref{eq:c-e_derivatives_result} with Eqs.~\eqref{eq:rtapprox_mastereq_bulk}, and~\eqref{eq:rtapprox_mastereq_bath},
we verify that the expression in Eq.~\eqref{eq:chi_single_collision}, with the relaxation time $\tau_R^{-1} = ((\tau_R')^{-1}+\sum_b \tau_b^{-1})$, is the first-order correction in the Chapman-Enskog solution of Eq.~\eqref{eq:generalized_master_eq}.
Notice that by following  the same approach, one can also show that $f_{1\beta}$ is the Chapman-Enskog solution of Eq.~\eqref{eq:boltzmann_with_reservoirs} corresponding to the values of temperature  and velocity imposed at the boundaries.

The corresponding dynamical equations for $\rho_\alpha$ and ${\langle}\vec{v}_\alpha{\rangle}=\vec{v}$ are, therefore, obtained substituting the Chapman-Enskog expansion of $f_{1\alpha}$ into the balance equations for mass and momentum, Eqs.~\eqref{eq:mass_density_velocity_boltzmann_defs}, and result in the macroscopic Eqs.~\eqref{eq:NS} (with $\rho=\rho_\alpha$).
Notice that the pressure tensor $P$ and heat flux $\vec{J}_q$ coincide with those in Eqs.~\eqref{eq:chapman_enskog_fluxes_result}, except for a different relaxation time taking into account the effect of the boundary.

\subsection{Thermodynamics}
We now formulate the thermodynamics, namely, the first and second law, corresponding to the Chapman-Enskog solution of Eq.~\eqref{eq:generalized_master_eq} 
(ensuring in particular that  $\langle \vec{v}_{\alpha}\rangle = \vec v$) by following the same steps as in \S~\ref{generalized_boltzmann_1and2_law}. 
We start by writing the balance equations for the velocity and the kinetic energy 
\begin{subequations}
\begin{align}
    \langle\rho_\alpha\rangle\partial_t \langle \vec v_\alpha\rangle &=
    -\avg{\rho}\avg{\vec{v}}\cdot\nabla_{\vec{x}}\avg{\vec{v}}-\nabla_{\vec{x}}\cdot \avg{P}+ \avg{\rho}\vec{F}+
        \varepsilon^{-1} \int_{\Omega(t)}\d\vec{x} \int\frac{\d \vec{p}}{(2\pi\hbar)^3}\, {\vec{p}}\,C_\varepsilon(f_{1\alpha})
        +\varepsilon^{-1}{\sum_\bath} \int\frac{\d \vec{p}}{(2\pi\hbar)^3}\, \vec{p}\,C_\bath(f_1)\label{eq:open_boltzmann_vel_balance_II}\\
        \frac{1}{2}\d_t \davg{\vec{p}^2} &= \davg{\vec{p}\cdot \vec{F} } +
        \varepsilon^{-1} \int_{\Omega(t)}\d\vec{x} \int\frac{\d \vec{p}}{(2\pi\hbar)^3}\, \frac{\vec{p}^2}{2}\,C_\varepsilon(f_{1\alpha})+
        \varepsilon^{-1}{\sum_{b}}\int_{\Omega(t)}\d\vec{x} \int\frac{\d \vec{p}}{(2\pi\hbar)^3}\, \frac{\vec{p}^2}{2}   \,   C_\bath(f_{1\alpha})\,,\label{eq:open_boltzmann_kin_en_balance_II}
\end{align}
\label{eq:open_boltzmann_st_balances}%
\end{subequations}%
 respectively.
The difference between Eqs.~\eqref{eq:open_boltzmann_vel_balance_II} and~\eqref{eq:open_boltzmann_kin_en_balance_II}, on the one hand, and Eqs.~\eqref{eq:open_boltzmann_momentum_balance} and~\eqref{eq:open_boltzmann_kin_en_balance}, on the other, results from $\vec p^2$ and $\vec p$ not being summational invariants under the collision operator $C_\varepsilon(f_{1\alpha})$ in Eq.~\eqref{eq:linear_collision_bulk}.
Hence, Eqs.~\eqref{eq:open_boltzmann_st_balances} lead to the following first law
\begin{align}
\begin{split}
   \d_t U &= \frac{1}{2} \d_t\davg{(\vec{p}-{\langle}\vec{v_{\alpha}}{\rangle})^2}=\frac{1}{2}\d_t\davg{\vec{p}^2}-\frac{1}{2} \d_t \int_{\Omega(t)}\d\vec{x} \, {\langle}\rho_\alpha{\rangle} { \langle}\vec{v}_{\alpha}{\rangle}^2  \\
   &= \dot{W} + {\sum_\bath}  \dot{Q}_\bath + \dot{Q}_\varepsilon\,,
   \label{eq:open_boltzmann_first_law_with_diss_fluxes}
\end{split}
\end{align}
where the $\dot{Q}_\bath $ and $\dot{W}$ are defined like in Eq.~\eqref{eq:final_first_law_generalized_boltzmann}, 
while $Q_\epsilon$ reads
\begin{align}
     \dot{Q}_\varepsilon &\equiv \varepsilon^{-1}\int_{\Omega{(t)}}\d\vec{x} \int\frac{\d \vec{p}}{(2\pi\hbar)^3} \left(\frac{\vec{p}^2}{2} - \vec{p}\cdot{\langle}\vec{v}_{\alpha}{\rangle}^2\right)C_\varepsilon(f_{1\alpha})\,.
     \label{eq:heat_inside_II}
\end{align}
Like in \S~\ref{generalized_boltzmann_1and2_law}, the heat flow in the equilibrium reservoirs can also be written as $\dot Q_\bath = - \varepsilon^{-1} T_\bath \int_{\Omega(t)}\d\vec{x} \int \d\vec{p} \,\ln (f_\bath^\mathrm{eq})\, C_\bath(f_{1\alpha})/(2\pi\hbar)^3$ with $f_\bath^\mathrm{eq}$ given in Eq.~\eqref{eq:final_eqm_distr}, and analogously the heat exchanged with the particles of type $\beta$ can be written as $\dot Q_\varepsilon = \int_{\Omega(t)}\d\vec{x}\,  \dot{q}_\varepsilon(\vec{x}) $ with the local (scalar) heat flow defined as $ \dot{q}_\varepsilon(\vec{x}) \equiv - \varepsilon^{-1}    {T(\vec x)} \int \d\vec{p}\,  \ln {(f_\alpha^\mathrm{MB})}\, C_\varepsilon(f_{1\alpha})/(2\pi\hbar)^3$.
Note that Eq.~\eqref{eq:collision_rates_alpha_CE_I} implies that $\dot{q}_\varepsilon(\vec{x})$ is of order zero in $\varepsilon$ when the local equilibrium holds, namely, $f_{1\alpha}=f_{1\alpha}^\mathrm{CE}$.

We now turn to the second law.
By using Eq.~\eqref{eq:generalized_master_eq}, the global entropy balance for the particles of type $\alpha$ now reads 
\begin{small}
\begin{subequations}
\begin{align}
   \d_t S 
   &= -\d_t\int_{\Omega(t)}\d\vec{x} \int\frac{\d \vec{p}}{(2\pi\hbar)^3}\,
   (f_{1\alpha}\ln f_{1\alpha} - f_{1\alpha})\\
   &= 
   -\varepsilon^{-1} \int_{\Omega(t)}\d\vec{x} \int\frac{\d \vec{p}}{(2\pi\hbar)^3}\, C_\varepsilon(f_{1\alpha}) \ln {f_{1\alpha}} +  \sum_\bath \left( \dot{\Sigma}_\bath + \frac{\dot{Q}_\bath}{T_\bath}\right) \label{eq:linearized_boltzmann_integral_to_split} \\
   &=  \dot{\Sigma}_\mathrm{int}  + \ef
   + \sum_\bath \left( \dot{\Sigma}_\bath + \frac{\dot{Q}_\bath}{T_\bath}\right) \,, \label{eq:open_boltzmann_second_law_with_diss_fluxes}
\end{align}
\end{subequations}
\end{small}
where $\dot{\Sigma}_\bath$ is the EPR on the boundary
in Eqs.~\eqref{eq:epr_boundary_open_boltzmann} and~\eqref{eq:epr_boundary_open_boltzmann_bis}. 
We split the integral in Eq.~\eqref{eq:linearized_boltzmann_integral_to_split} into 
\begin{equation}
    \dot{\Sigma}_{\mathrm{int}} 
    =\frac{\varepsilon^{-1}}{2}\int_{\Omega(t)} \d \vec{x}\int\frac{\d \vec{p}\d \vec{p}'}{(2\pi\hbar)^3} \big[R_\varepsilon(\vec{p}|\vec{p}')f_{1\alpha}(\vec{p}') - R_\varepsilon(\vec{p}'|\vec{p}) f_{1\alpha}(\vec{p})\big]
    \ln\frac{R_\varepsilon(\vec{p}|\vec{p}')f_{1\alpha}(\vec{p}')}{ R_\varepsilon(\vec{p}'|\vec{p}) f_{1\alpha}(\vec{p})}\geq0\,  \,,\label{eq:epr_inner_open_boltzmann_II}
\end{equation}
and 
\begin{equation}
    \ef 
    =\frac{\varepsilon^{-1}}{2}\int_{\Omega(t)}\d\vec{x} \int \frac{\d\vec{p}\d\vec{p}'}{(2\pi\hbar)^3}\,
    \big[R_\varepsilon(\vec{p}|\vec{p}')f_{1\alpha}(\vec{p}') - R_\varepsilon(\vec{p}'|\vec{p}) f_{1\alpha}(\vec{p})\big]
    \ln\frac{ R_\varepsilon(\vec{p}'|\vec{p})}{R_\varepsilon(\vec{p}|\vec{p}')}\, ,\label{eq:linear_boltzmann_entropy_flow}
\end{equation}
corresponding to the EPR and the entropy flow inside the system, respectively.
 By introducing the local entropy flow $\dot{s}_{ \varepsilon}(\vec{x}) $ satisfying $\ef 
    = \int_{\Omega(t)} \d\vec{x}\, \dot{s}_{ \varepsilon}(\vec{x})$ and using 
the local detailed balance condition \eqref{eq:ldb_ce_II}, we obtain 
\begin{align}
\begin{split}
     \dot{s}_{ \varepsilon}(\vec{x}) 
     &=\frac{\dot{q}_\varepsilon(\vec{x})}{T(\vec{x})} + \frac{1}{2T(\vec{x})}\int \frac{\d\vec{p}\d\vec{p}'}{(2\pi\hbar)^3}\,  \big[R_\varepsilon(\vec{p}|\vec{p}')f_{1\alpha}(\vec{p}') - R_\varepsilon(\vec{p}'|\vec{p}) f_{1\alpha}(\vec{p})\big] w(\vec{v},\vec{v}-\Delta \vec{p})\,.\label{eq:efdensity}
     \end{split}
\end{align}
Note that Eq.~\eqref{eq:efdensity} together with Eq.~\eqref{eq:collision_rates_alpha_CE_I} implies that $\dot{s}_\epsilon(\vec{x})$ is given by the sum of two contributions with a different scaling in the Knudsen number.
The zeroth order contribution accounts for the local heat flow ${\dot{q}_\varepsilon(\vec{x})}$. At this order, the local equilibrium relation $\dot{s}_\epsilon(\vec{x})= \dot{q}_\varepsilon(\vec{x})/T(\vec{x})$ holds.
The first order contribution emerges instead from the nonconservative work (resulting from the irreversible momentum exchanges between $\alpha$ and $\beta$ particles)
and breaks the local equilibrium even if, at the same time, local detailed balance is obeyed at the mesoscopic scale and the macroscopic dynamics is the solution to the hydrodynamic equations.
The local equilibrium is broken also in the framework of the so-called Extended Irreversible Thermodynamics where the nonconservative contributions to the balance equation for the entropy are included on phenomenological grounds or derived from Grad's thirteen-moment approximation~\cite{jou1996extended}. 

We conclude by expressing the entropy production~\eqref{eq:epr_inner_open_boltzmann_II} in the first order in $\varepsilon$ by using Eq.~\eqref{eq:collision_rates_alpha_CE_I}, the local detailed balance condition~\eqref{eq:ldb_ce_II} and $R_0(\vec{p}|\vec{p}')f_\alpha^\mathrm{MB}(\vec{p}') = R_0(\vec{p}'|\vec{p})f_\alpha^\mathrm{MB}(\vec{p})$. This leads to
\begin{equation}
     \dot{\Sigma}_{\mathrm{int}} 
    =\frac{\varepsilon}{2}\int_{\Omega(t)} \d \vec{x}\int\frac{\d \vec{p}\d \vec{p}'}{(2\pi\hbar)^3}  R_0(\vec{p}'|\vec{p})f_\alpha^\mathrm{MB}(\vec{p}) \left[ 
    \frac{w(\vec{v},\vec{v}-\Delta\vec{p})}{T} 
    +\chi_\alpha(\vec{p}') -\chi_\alpha(\vec{p})\right]^2\,,
\end{equation}
which emphasizes the analogy with the expression obtained in ST (in the corresponding linear regime~\cite{forastiere2022linear}).
This expression confirms that the physical meaning of $w(\vec{v},\vec{v}-\Delta\vec{p})$ is
that of a nonconservative work rate done by the local thermodynamic forces $\nabla_{\vec{x}}T$ and $(\epsilon_{ij})$.
On the other hand, $\varepsilon (\chi_\alpha(\vec{p}')-\chi_\alpha(\vec{p})) \simeq
\ln \left[{f_{1\alpha}(\vec{p})} {f^\mathrm{MB}_{1\alpha}(\vec{p})} /{f_{1\alpha}(\vec{p}')} {f^{\mathrm{MB}}_{1\alpha}(\vec{p}')}\right]$ is the information-theoretic contribution to the EPR due to the fact that the transition $\vec{p}\mapsto\vec{p}'$ occurs   out-of-equilibrium.

\section{Conclusions and perspectives}
\label{sec:conclusions}

Our work systematically analyzed how to construct nonequilibrium thermodynamics for systems described by hydrodynamics at the macroscopic level, starting from the microscopic scale.
Emphasis has been placed on the boundary conditions which play a crucial role in determining the right thermodynamic potentials and the thermodynamic forces.

Several further developments can be envisioned.
The description we provide here is based on the Boltzmann equation and allows for relatively explicit calculations, but the argument used to truncate the BBGKY hierarchy holds only for very rarefied gases.
An alternative approach based on an integral fluctuation theorem has recently been developed~\cite{sasa2014derivation,  mabillard2020microscopic, saito2021microscopic, gaspard2022statistical}.
It would be interesting to systematically investigate the thermodynamic implications of this approach employing techniques from large deviations theory~\cite{bouchet2020boltzmann}, which recover irreversibility at the mesoscopic level by focusing on the typical microscopic evolution~\cite{chakraborti2022entropy}.

Our microscopic approach can be extended to include the effect of chemical reactions, where macroscopic nonequilibrium thermodynamics is well established~\cite{prigogine1947etude, dgm1962nonequilibrium, landau2013statistical, hill2005free} and continues to receive attention~\cite{rao2016nonequilibrium,avanzini2019thermodynamics, avanzini2021nonequilibrium,avanzini2022}.
Recent formulations of nonequilibrium thermodynamics for systems undergoing chemical reactions ~\cite{falasco2018information, avanzini2019thermodynamics} are currently assuming the absence of macroscopic motion of the fluid and isothermal conditions.
The effects of viscous flows and heat exchanges are thus neglected but could be accounted by such extensions.

\section*{Acknowledgments}
This research was supported by project ChemComplex
(C21/MS/16356329, FA and ME) funded by FNR (Luxembourg), European Research Council, project NanoThermo (ERC-2015-CoG Agreement No.~681456, DF and ME), and Università degli Studi di Padova (DF with the grant 'BAIE BIRD2021 01', and FA with the grant 'P-DiSC\#BIRD2023-UNIPD').

\section*{Data availability statement}
No new data were created or analysed in this study.

\appendix
 
\section{Reynolds' transport theorem }
\label{app:reynolds}

We summarize here a formal proof of Reynolds' transport theorem.
To compute integrals over the time-dependent domain~$\omega(t)$ of~$\mathbb{R}^3$ with a smooth boundary we introduce the indicator function $I({\pos},\omega(t))$.
Thus,
\begin{align}
    \d_t \int_{\omega(t)} \d {\pos}\, f({\pos},t) = \d_t \int_{\mathbb{R}^3} \d {\pos}\, f({\pos},t)I({\pos},\omega(t)) = \int_{\omega(t)} \d {\pos}\, \partial_t f({\pos},t)  +  \int_{\mathbb{R}^3} \d{\pos}\,f({\pos},t) \partial_t I({\pos},\omega(t))\,. \label{eq:reynolds_I}
\end{align}
The boundary $\partial \omega(t)$ admits a smooth parametrization by means of two functions $m_i(t,\{x_j\}_{j\neq i})$ and $M_i(t,\{x_j\}_{j\neq i})$ giving respectively the minimum and maximum value that the $i$-th coordinate can take for a given value of the other two coordinates $\{x_j\}_{j\neq i}$.
This means that ${\pos}=(\dots, x_i,\dots)\in \omega$ if and only if all its coordinates satisfy $m_i(t,\{x_j\}_{j\neq i})\leq x_i \leq M_i(t,\{x_j\}_{j\neq i})$.
The indicator function can therefore be  written as
\begin{align}
    I({\pos},\omega(t)) = \prod_{i=1}^3 \bigg( \theta(x_i - m_i(t,\{x_j\}_{j\neq i})) + \theta(M_i(t,\{x_j\}_{j\neq i}) - x_i)-1 \bigg)\,.
\end{align}
Its derivative with respect to time reads 
\begin{align}
    \partial_t I({\pos},\omega(t)) &=  \sum_{i=1}^3 \bigg(-\delta(x_i - m_i(t,\{x_j\}_{j\neq i})) \; \partial_t m_i(t,\{x_j\}_{j\neq i}) + \delta(M_i(t,\{x_j\}_{j\neq i}) - x_i) \; \partial_t M_i(t,\{x_j\}_{j\neq i})\bigg) I({\pos},i,\omega(t))\,,
\end{align}
where
\begin{equation}
    I({\pos},i,\omega(t)) = 
    \prod_{k\neq i}^3 \bigg(\theta(x_k - m_k(t,\{x_j\}_{j\neq k})) + \theta(M_k(t,\{x_j\}_{j\neq k}) - x_k)-1\bigg)\,,
\end{equation}
is a restricted indicator function that ignores the $i$-th coordinate.
The $i$-th components of the velocity field $\vec{v}_b$ on the boundary are given by $ v_{b,i}^{+}(\{x\}_{j\neq i})= \partial_t M_i(t,\{x_j\}_{j\neq i})$   and $v_{b,i}^{-}(\{x\}_{j\neq i})= -\partial_t m_i(t,\{x_j\}_{j\neq i})$.
This means that the second integral in~\eqref{eq:reynolds_I} is restricted to the boundary by the $\delta$-function and can be recast as
\begin{align}
    \int_{\mathbb{R}^3}\d{\pos}\, f(\vec{x},t) \partial_t I(\vec{x},\omega(t)) = \int_{\partial \omega(t)} f(\vec{x},t) \vec{v}_b \cdot \d \vec{n}  \,,
\end{align}
where $\d \vec{n}$ is an oriented surface element pointing outwards.
The integral can be transformed back to an integral over $\omega(t)$ using Stokes' theorem, obtaining the well-known formula
\begin{align}
    \d_t \int_{\omega(t)} \d {\pos}\, f({\pos},t) = \int_{\omega(t)} \d {\pos}\, \left(  \partial_t f({\pos},t) +\nabla_{\pos}\cdot (f({\pos},t) \vec{v})\right) =\int_{\omega(t)} \d \vec{x}\, (D_t f (\pos,t)+ f (\pos,t)(\nabla_{\vec{x}}\cdot{\vec{v}})) \,, \label{eq:reynolds}
\end{align}
with $\vec{v}$ being the velocity field on the domain $\omega(t)$.

\section{Minimum entropy production theorem}
\label{app:mep}

Equation~\eqref{eq:epr_linear_regime} can be expressed as in Eq.~\eqref{eq:epr_ss_decomposition} if
\begin{equation}
    \int_{\Omega} \d \vec{x} \begin{pmatrix}\partial_{x_i} 1/T^{ss}\\  \epsilon_{ij}^{ss} \end{pmatrix}\tr \begin{pmatrix}
        \kappa \delta_{ij} & 0\\
        0 & L_{(ij),(i'j')}\\
    \end{pmatrix}
\begin{pmatrix}\partial_{x_j} (1/T-1/T^{ss})\\ \epsilon_{i'j'} -\epsilon_{i'j'}^{ss}\end{pmatrix} 
\simeq
\int_{\Omega} \d \vec{x} \begin{pmatrix}\partial_{x_i} T^{ss}\\  \epsilon_{ij}^{ss} \end{pmatrix}\tr \begin{pmatrix}        (T^\mathrm{eq})^{-4} \kappa \delta_{ij} & 0\\
        0 & L_{(ij),(i'j')}\\
    \end{pmatrix}
\begin{pmatrix}\partial_{x_j} (T-T^{ss})\\ \epsilon_{i'j'} -\epsilon_{i'j'}^{ss}\end{pmatrix}
    = 0  \,,\label{eq:app_orthogonality_condition}
\end{equation}
by consider a near-equilibrium condition for which $T=T^\mathrm{eq}+\delta T$ and, therefore,
$\partial_{x_j} (1/T)$ (as well as $\partial_{x_j} (1/T^{ss})$) 
can be written as $\partial_{x_j} (1/T) = {
-(T^\mathrm{eq})^{-2}\partial_{x_j} T + {O(\delta T^2)}}$,
 by using
$\partial_{x_i}T^{\mathrm{eq}}=0$
or, equivalently,
$\partial_{x_i}T=\partial_{x_i}\delta T$.
A set of sufficient conditions for Eq.~\eqref{eq:app_orthogonality_condition} to hold is given by
\begin{subequations}
\begin{align}
    &\int_\Omega \d\vec{x}\, \nabla_{\vec{x}}
    T^\mathrm{ss}
\cdot\nabla_{\vec{x}}(T-T^\mathrm{ss}) =0\,,\label{eq:app_orthogonality_condition_2a}\\
    & \sum_{ij}\int_{\Omega} \d\vec{x}\, (  \partial_{x_i} v_i^{\mathrm{ss}} ) \, \partial_{x_j}(v_j -v_j^\mathrm{ss})=0\,,\label{eq:app_orthogonality_condition_2b}\\
    &\sum_{ij} \int_{\Omega} \d\vec{x}\,  (  \partial_{x_i} v_j^{\mathrm{ss}} ) \, \partial_{x_i}(v_j -v_j^\mathrm{ss})  =0\,,\label{eq:app_orthogonality_condition_2c}\\
    &\sum_{ij} \int_{\Omega} \d\vec{x}\,   (    \partial_{x_i} v_j^{\mathrm{ss}} )\,  \partial_{x_j}(v_i -v_i^\mathrm{ss}) =0\,,\label{eq:app_orthogonality_condition_2d}
\end{align}   
\label{eq:app_orthogonality_condition_2}%
\end{subequations}
where we used
$L_{(ij),(i'j')} \equiv (\lambda \delta_{ij} \delta_{i'j'} + 2 \rho_{\mathrm{eq}} \nu  \delta_{ii'} \delta_{jj'})/{T_{\mathrm{eq}}}$ and 
$\epsilon_{ij}=(\partial_{x_i} v_j + \partial_{x_j} v_i)/2$. 

In the following, we show that Eq.~\eqref{eq:app_orthogonality_condition_2} holds
near equilibrium 
(with $\rho = \rho_{\mathrm{eq}}+\delta \rho$, 
$\vec{v}=\vec{v}^\mathrm{eq}+\delta \vec{v}$, 
$T=T^\mathrm{eq}+\delta T$, 
and $p=p_{\mathrm{eq}} + \delta p$)
for incompressible fluids, \emph{i.e.},
\begin{align}
    \nabla_{\vec{x}}\cdot\vec{v} = 0 \, , \label{eq:incompressibility}
\end{align}
whose internal energy reads
\begin{equation}
    u=c_V T\,,
\end{equation}
where $c_V$ is the heat capacity (per unit volume, computed at constant 
volume)~\cite{landau2013statistical, de2006hydrodynamic}.
Note that the approximation of incompressible fluids is justified when 
i) the Mach number $\textrm{Ma}$ is negligible, \emph{i.e.}, $\textrm{Ma}={v_0}/{c}\ll1$ 
(where $v_0$ is  the characteristic velocity  of the flow 
and $c$ the speed of sound in the fluid),
ii) fast transient phenomena, like pressure waves, are absent and  iii) thermal expansion can be ignored \cite{batchelor1967introduction}.
The dependence of $\rho$ on the temperature and pressure gradients is therefore assumed to be negligible, \emph{i.e.}, $\delta \rho=0$. 

We employ a perturbative analysis for the balance equations of mass, momentum and energy 
(in Eqs.~\eqref{eq:mass_conservation_lagrangian}, \eqref{eq:momentum_conservation} and \eqref{eq:internal_energy_balance_lagrangian}, respectively) near equilibrium.
First, since the EPR~\eqref{eq:epr_local_defs} vanishes everywhere at equilibrium, $T^\mathrm{eq}$ and $\vec{v}^\mathrm{eq}$ satisfy 
\begin{subequations}
    \begin{align}
    \nabla_{\vec{x}}T^\mathrm{eq}=0\,,\\
    \nabla_{\vec{x}}\vec{v}^\mathrm{eq}=0\,.
\end{align}
\label{eq:eqm_def_T_v}%
\end{subequations}
In general, $\vec{v}^\mathrm{eq}\neq 0$, but
Hence, the balance equations~\eqref{eq:mass_conservation_lagrangian}, \eqref{eq:momentum_conservation} and~\eqref{eq:internal_energy_balance_lagrangian}
at the zeroth order read
\begin{subequations}
\begin{align}  
    \nabla_{\vec{x}} \rho^\mathrm{eq} & = 0 \label{eq:eqm_unif_density}\,,\\
    -\nabla_{\vec{x}}p^\mathrm{eq} +\rho^\mathrm{eq}\vec{F} & =0\,,\label{eq:eqm_momentum_balance}\\
    \partial_t T^\mathrm{eq}&=0\,.\label{eq:eqm_energy_balance}
\end{align}
    \label{eq:eqm_balances}%
\end{subequations}%
Second, the balance equations, \eqref{eq:momentum_conservation} and~\eqref{eq:internal_energy_balance_lagrangian}
at the first order in $\delta \vec{v}$, $\delta T$ and $\delta p$ read
\begin{subequations}
    \begin{align}
    \rho^\mathrm{eq} \partial_t \delta \vec{v}
    &=- \nabla_{\vec{x}}\delta p + \nu  \rho^\mathrm{eq}\nabla_{\vec{x}}^2 \delta \vec{v} \,,\label{eq:linearized_momentum_balance}\\    
    \rho^\mathrm{eq} c_V{(T^{\text{eq}})^2} \partial_t \delta T   
    &=\kappa \nabla_{\vec{x}}^2  \delta T \,,\label{eq:fourier_eq}
\end{align}%
\label{eq:transport_eqs_linearized}%
\end{subequations}%
while the balance equation~\eqref{eq:mass_conservation_lagrangian} vanishes at the first order because $\delta\rho = 0$.
At steady state, Eq.~\eqref{eq:transport_eqs_linearized} becomes
\begin{subequations}
    \begin{align}
        - \nabla_{\vec{x}} 
    \delta p^\mathrm{ss} + \nu  \rho^\mathrm{eq} \nabla_{\vec{x}}^2  \vec{v}^\mathrm{ss} &=0\,,\label{eq:stationary_vector_helmoltz_v}\\
     \nabla_{\vec{x}}^2   T^\mathrm{ss} &= 0 \,.\label{eq:stationary_fourier_eq}
\end{align}
\label{eq:stationary_eqs_linearized}%
\end{subequations}
Notice that for the temperature and the velocity, we have $\nabla_{\vec{x}}\delta T=\nabla_{\vec{x}}T$ and $\nabla_{\vec{x}}\delta\vec{v}=\nabla_{\vec{x}}\vec{v}$ 
as a consequence of Eq.~\eqref{eq:eqm_def_T_v}.

Third, we show that Eq.~\eqref{eq:stationary_eqs_linearized} implies Eq.~\eqref{eq:app_orthogonality_condition_2}.
Indeed, Eq.~\eqref{eq:app_orthogonality_condition_2a} reads
\begin{align}
     \int_{\Omega} \d \vec{x}\, \nabla_{\vec{x}}T^{\mathrm{ss}}\cdot \nabla_{\vec{x}}(T-T^\mathrm{ss}) &= \int_{\partial\Omega} \d \vec{n}\cdot \nabla_{\vec{x}}T^\mathrm{ss} (T-T^\mathrm{ss}) - \int_{\Omega} \d \vec{x}\, \nabla_{\vec{x}}^2 T^{\mathrm{ss}} (T-T^\mathrm{ss}) =0\,,
\end{align}
where the first term the vanishes because $T=T^\mathrm{ss}$ on the boundary, 
while the second term vanishes because of Eq. \eqref{eq:stationary_fourier_eq}.
Equation~\eqref{eq:app_orthogonality_condition_2b} holds because of incompressibility~\eqref{eq:incompressibility}.
Equation~\eqref{eq:app_orthogonality_condition_2c} reads
\begin{subequations}
    \begin{align}
    \sum_{ij} \int_{\Omega} \d \vec{x} \partial_{x_i} v^{\mathrm{ss}}_j  \partial_{x_i}({v}_j-{v}^\mathrm{ss}_j) &= \sum_{j} \int_{\partial \Omega} \d n \cdot (\nabla_{\vec{x}} v_j^\mathrm{ss})   (v_j-{v}_j^\mathrm{ss}) - \sum_j \int_{\Omega} \d \vec{x}  \,\big(\sum_i\partial_{x_i}^2 v_j^{\mathrm{ss}} \big) ({v}_j-v_j^\mathrm{ss})\,,\\
    &= -\frac{1}{{\nu\rho^{\mathrm{eq}}}} \sum_j \int_{\Omega} \d \vec{x}  \,(\partial_{x_j}{ 
    \delta p^\mathrm{ss}} ) ({v}_j-v_j^\mathrm{ss})\,,\\
    &= - \frac{1}{{\nu\rho^{\mathrm{eq}}}}  \int_{\partial\Omega} \d \vec{n} \cdot( { 
    \delta p^\mathrm{ss}}  (\vec{v}-\vec{v}^\mathrm{ss}))+ \frac{1}{{\nu\rho^{\mathrm{eq}}}} \int_{\Omega} \d \vec{x}  \, { 
    \delta p^\mathrm{ss}}   \nabla_{\vec{x}}\cdot (\vec{v}-\vec{v}^\mathrm{ss})=0\,,
\end{align}
\end{subequations}
where we used $\vec{v}= \vec{v}^\mathrm{ss}$ on the boundaries as well as Eqs.~\eqref{eq:incompressibility} and~\eqref{eq:stationary_vector_helmoltz_v}.
Equation~\eqref{eq:app_orthogonality_condition_2d} reads
\begin{subequations}
    \begin{align}
    \sum_{ij} \int_{\Omega} \d \vec{x} \partial_{x_i} v^{\mathrm{ss}}_j  \partial_{x_j}({v}_i-{v}^\mathrm{ss}_i) &= \sum_{ij} \int_{\partial \Omega} \d n_j \left(  \partial_{x_i} v^{\mathrm{ss}}_j  ({v}_i-{v}^\mathrm{ss}_i)\right) - \sum_{i} \int_{\Omega} \d \vec{x} \, \partial_{x_i}\big(\sum_j \partial_{x_j} v^{\mathrm{ss}}_j \big) ({v}_i-{v}^\mathrm{ss}_i)
    =0\,,
\end{align}
\end{subequations}
where the first term vanishes because $\vec{v}= \vec{v}^\mathrm{ss}$ on the boundaries, while the second vanishes because of Eq.~\eqref{eq:incompressibility}.

Finally, we prove that $\dot{\Sigma}$ acts as a Lyapunov function for incompressible fluids close to equilibrium.
To do so, we rewrite the EPR~\eqref{eq:epr_linear_regime} in terms of $\nabla_{\vec{x}}\vec{v}$ instead of the strain tensor $\epsilon$:
\begin{equation}
    \dot{\Sigma} 
    = \int_{\Omega} \d \vec{x} 
    \begin{pmatrix}
    \nabla_{\vec{x}}T\\
    \nabla_{\vec{x}}\vec{v}
    \end{pmatrix}\tr 
    \mathbb{O}'
    \begin{pmatrix}
    \nabla_{\vec{x}}T\\
    \nabla_{\vec{x}}\vec{v}
    \end{pmatrix}\,,
\end{equation}
where 
$\mathbb{O}'=\textrm{diag}({ \kappa(T^\mathrm{eq})^{-4}},L')$ 
and $L'_{(i,j),(i',j')}=(\lambda \delta_{ij}\delta_{i'j'} +  \nu \rho^\mathrm{eq}(\delta_{ii'}\delta_{jj'} + \delta_{ij'}\delta_{ji'}))/T_{\mathrm{eq}}$.
Its time derivative reads
\begin{equation}
    \d_t \dot{\Sigma} 
    =
    2 \int_{\Omega} \d \vec{x} 
    \begin{pmatrix}
    \nabla_{\vec{x}}T\\
    \nabla_{\vec{x}}\vec{v}
    \end{pmatrix}\tr 
    \mathbb{O}'
    \begin{pmatrix}
    \nabla_{\vec{x}}\partial_t T\\
    \nabla_{\vec{x}}\partial_t \vec{v}
    \end{pmatrix}
    = - 2 \int_{\Omega} \d \vec{x} 
    \begin{pmatrix}
    \nabla^2_{\vec{x}} T\\
    \nabla^2_{\vec{x}} \vec{v}
    \end{pmatrix}\tr 
    \mathbb{M}
    \begin{pmatrix}
    \nabla^2_{\vec{x}} T\\
    \nabla^2_{\vec{x}} \vec{v} - (\nu \rho^\text{eq})^{-1}\nabla_{\vec{x}}\delta p
    \end{pmatrix}\,,
    \label{eq:epr_as_lyapunov_I}
\end{equation}
by using $L'_{(i,j),(i',j')} = L'_{(i',j'),(i,j)}$ (in the first equality), 
together with i) Eq.~\eqref{eq:transport_eqs_linearized} 
and 
ii) $\nabla_{\vec{x}}\delta T=\nabla_{\vec{x}}T$ and $\nabla_{\vec{x}}\delta\vec{v}=\nabla_{\vec{x}}\vec{v}$ 
because of Eq.~\eqref{eq:eqm_def_T_v}
(in the second equality).
In particular, for the second equality,  
we integrate by parts using that 
$\partial_{t}v_i = 0$ and $\partial_{t}T = 0$ 
on the boundary together with incompressibility~\eqref{eq:incompressibility}, 
and we define the $4\times 4$ matrix 
$\mathbb{M}{ \equiv}
\textrm{diag}({ \kappa}^2{(\rho^{\mathrm{eq}}c_V)^{-1}}{(T^\mathrm{eq})^{-6}},
\nu^2\rho^\mathrm{eq}{ (T^\mathrm{eq})^{-1}})$.
Due to incompressibility~\eqref{eq:incompressibility},
\begin{equation}
     \int_{\Omega}\d \vec{x}\,  \nabla^2_{\vec{x}} \vec{v}\cdot \nabla_{\vec{x}}\delta p =0\,. \label{eq:vanishing_term_I}
    \end{equation}
Indeed, by expanding $\delta p$ and $\vec{v}$ on a basis of plane waves
\begin{subequations}
    \begin{align}
    \delta p(\vec{x})  = \int \d\vec{k}\, \e^{\i \vec{k}\cdot\vec{x}} {\delta \hat{\vec{p}}}(\vec{k})\,,\\
    \vec{v}(\vec{x}) = \int \d\vec{k}\,\e^{\i \vec{k}\cdot\vec{x}}\hat{\vec{v}}(\vec{k})\,,
    \end{align}
\label{eq:fourier_decomposition}
\end{subequations}
incompressibility~\eqref{eq:incompressibility} implies 
\begin{align}
    \vec{k} \cdot\hat{\vec{v}}(\vec{k}) = 0\,,\label{eq:incompressibility_fourier}
\end{align}
and, therefore,  
\begin{equation}
    \int_{\Omega}\d\vec{x}\, \nabla^2_{\vec{x}} \vec{v}\cdot \nabla_{\vec{x}}\delta p = {- \i} \int_{\Omega}\d\vec{x}  \int \d \vec{k} \d \vec{k}' \vec{k}^2\, (\vec{k}' \cdot \hat{\vec{v}}(\vec{k})) \delta \hat{p} (\vec{k}') \e^{\i (\vec{k}+\vec{k}')\cdot \vec{x}} = {2\pi\i} \int \d \vec{k}\,  \vec{k}^2 (\vec{k} \cdot \hat{\vec{v}}(\vec{k}) )\delta \hat{p}(\vec{k}) = 0\,.\label{eq:pressure_term}
\end{equation}
Thus, $\d_t \dot{\Sigma}$ in Eq.~\eqref{eq:epr_as_lyapunov_I} can be written in the manifestly non-positive form
\begin{equation}
    \d_t \dot{\Sigma}
    = - 2 \int_{\Omega} \d \vec{x} 
    \begin{pmatrix}
    \nabla^2_{\vec{x}} T\\
    \nabla^2_{\vec{x}} \vec{v}
    \end{pmatrix}\tr 
    \mathbb{M}
    \begin{pmatrix}
    \nabla^2_{\vec{x}} T\\
    \nabla^2_{\vec{x}} \vec{v}
    \end{pmatrix}  \leq 0 \,,
    \label{eq:epr_as_lyapunov_II}
\end{equation}
since $\mathbb{M}$ is diagonal with only positive entries.

\section{Energetics in the laboratory and center of mass frames}
\label{app:cm}
We express the macroscopic and mesoscopic 
power contribution in the balance equations~\eqref{eq:epr_closed_velocity_II} (in \S~\ref{sec:closed_velocity_control})
and~\eqref{eq:open_boltzmann_massieu_partial} (in \S~\ref{sec:open_boltzmann_global_potential}), respectively,
using the system's center of mass (CM).

\paragraph{Macroscopic Scale}
At the macroscopic scale, the CM and its velocity are defined as 
\begin{align}
    &\vec{x}_{\text{CM}} \equiv \frac{1}{m}\int_{\Omega(t)} \d \vec{x}\, \vec{x}\rho(\vec{x})\,, 
    &\vec{v}_{\text{CM}} \equiv \frac{1}{m}\int_{\Omega(t)} \d \vec{x}\, \rho(\vec{x})\vec{v}(\vec{x})\,,
    \label{eq:macro_CM_def}
\end{align}
respectively, with $m=\int \d\vec{x}\rho(\vec{x})$ being the total mass.
Their time derivatives (by using  
 Eqs.~\eqref{eq:mass_conservation_lagrangian},~\eqref{eq:momentum_conservation}, and~\eqref{eq:reynolds}) read
\begin{align}
    &\d_t \vec{x}_{\text{CM}} = \vec{v}_{\text{CM}}\,,
    &m \d_t \vec{v}_{\text{CM}} = \int_{\Omega(t)} \d\vec{x}\, \left(- \nabla_{\vec{x}}\cdot P(\vec{x}) + \rho(\vec{x})\vec{F}(\vec{x}) \right) \,. \label{eq:cm_balance}
\end{align}
By assuming that the internal potential energy $\phi_{\text{int}}$ is linear in $\vec{x}$,
namely $\phi_{\text{int}}(\vec{x})=\vec{g}\cdot\vec{x}$ 
(as it is for the gravitational potential) 
and that the velocity $\vec{v}_{B}$ is uniform on the boundary,
the power contribution in Eq.~\eqref{eq:epr_closed_velocity_II} becomes
\begin{align}
    \dot{W}=\int_{\Omega(t)} \d\vec{x}\, \vec{v}_{B}\cdot\left(  \rho(\vec{x})\vec{F}_\text{w}(\vec{x})- \nabla_{\vec{x}}\cdot P \right)
    = \d_t E_{CM} +  m \left(\vec{v}_B-\vec{v}_{\text{CM}}\right) \cdot \left(\d_t \vec{v}_{\text{CM}} + \vec{g} \right)\,. 
    \label{eq:macro_E_cm_rate_velocity_control}
\end{align}
where we used $\vec{F} = \vec{F}_\text{w} + \vec{F}_{\text{int}}$, Eq.~\eqref{eq:cm_balance},
and we introduced the total mechanical energy of the CM,
i.e., $E_{\text{CM}}\equiv m\vec{v}_{\text{CM}}^2/2 + m\phi_{\text{int}}(\vec{x}_{\text{CM}})$.

\paragraph{Mesoscopic Scale}
At the mesoscopic scale, the CM and its velocity are defined (in the units of Eq.~\eqref{eq:adimensional_coords}) as 
\begin{align}
    &\vec{x}_{\text{CM}} \equiv \frac{1}{m}\int_{\Omega(t)} \d \vec{x}\, \vec{x}\aden(\vec{x})\,, 
    &\vec{v}_{\text{CM}} \equiv \frac{1}{m}\int_{\Omega(t)} \d \vec{x}\, {\aden(\vec{x})}\avel(\vec{x})\,,
    \label{eq:meso_CM_def}
\end{align}
respectively, by using the fields mass density~\eqref{eq:mass_density_velocity_boltzmann_defs_d} and velocity~\eqref{eq:mass_density_velocity_boltzmann_defs_v}
and with $m=\int \d\vec{x}\aden(\vec{x})$.
When we explicitly account for external reservoirs on the boundary (as done in \S~\ref{sec:mesoscopic_boltzmann_open}),
their  time derivatives read
\begin{align}
    &\d_t \vec{x}_{\text{CM}} = \vec{v}_{\text{CM}}\,,
    &m \d_t \vec{v}_{\text{CM}} = \int_{\Omega(t)} \d\vec{x}\, \left(\avg{\rho} \vec{F}+\varepsilon^{-1}\sum_b\int \frac{\d\vec{p}}{{(2\pi\hbar)^3}}\,\vec{p}\,C_b(f_1) \right) \,. \label{eq:meso_cm_balance}
\end{align}
by using  
Eqs.~\eqref{eq:open_boltzmann_mass_balance},~\eqref{eq:open_boltzmann_momentum_balance}, and
$\int_{\Omega(t)}\d\vec{x}\nabla_{\vec{x}}\cdot \avg{P} = 0$.
By assuming that the internal potential energy $\phi_{\text{int}}$ is linear in $\vec{x}$,
namely $\phi_{\text{int}}(\vec{x})=\vec{g}\cdot\vec{x}$ 
(as it is for the gravitational potential),
and that all reservoirs move at the same speed $\vec{v}_{B}$,
the power contribution in Eq.~\eqref{eq:open_boltzmann_massieu_partial} becomes
\begin{align}
    \dot{W}_{\mathrm{mech}}= -\davg{\vec{F}_{\text{int}}}\cdot\vec{v}_B
    = m\d_t \phi_{\text{int}}(\vec{x}_{\text{CM}}) +  m \left(\vec{v}_B-\vec{v}_{\text{CM}}\right) \cdot  \vec{g}\,,\label{eq:meso_E_cm_rate_velocity_control}
\end{align}
where we summed and subtracted the potential energy of the center of mass 
$m\d_t \phi_{\text{int}}(\vec{x}_{\text{CM}}) = m\vec{g}\cdot\vec{v}_{\text{CM}}$.

\bibliography{./bibliography}

\begin{thebibliography}{84}%
\makeatletter
\providecommand \@ifxundefined [1]{%
 \@ifx{#1\undefined}
}%
\providecommand \@ifnum [1]{%
 \ifnum #1\expandafter \@firstoftwo
 \else \expandafter \@secondoftwo
 \fi
}%
\providecommand \@ifx [1]{%
 \ifx #1\expandafter \@firstoftwo
 \else \expandafter \@secondoftwo
 \fi
}%
\providecommand \natexlab [1]{#1}%
\providecommand \enquote  [1]{``#1''}%
\providecommand \bibnamefont  [1]{#1}%
\providecommand \bibfnamefont [1]{#1}%
\providecommand \citenamefont [1]{#1}%
\providecommand \href@noop [0]{\@secondoftwo}%
\providecommand \href [0]{\begingroup \@sanitize@url \@href}%
\providecommand \@href[1]{\@@startlink{#1}\@@href}%
\providecommand \@@href[1]{\endgroup#1\@@endlink}%
\providecommand \@sanitize@url [0]{\catcode `\\12\catcode `\$12\catcode
  `\&12\catcode `\#12\catcode `\^12\catcode `\_12\catcode `\%12\relax}%
\providecommand \@@startlink[1]{}%
\providecommand \@@endlink[0]{}%
\providecommand \url  [0]{\begingroup\@sanitize@url \@url }%
\providecommand \@url [1]{\endgroup\@href {#1}{\urlprefix }}%
\providecommand \urlprefix  [0]{URL }%
\providecommand \Eprint [0]{\href }%
\providecommand \doibase [0]{http://dx.doi.org/}%
\providecommand \selectlanguage [0]{\@gobble}%
\providecommand \bibinfo  [0]{\@secondoftwo}%
\providecommand \bibfield  [0]{\@secondoftwo}%
\providecommand \translation [1]{[#1]}%
\providecommand \BibitemOpen [0]{}%
\providecommand \bibitemStop [0]{}%
\providecommand \bibitemNoStop [0]{.\EOS\space}%
\providecommand \EOS [0]{\spacefactor3000\relax}%
\providecommand \BibitemShut  [1]{\csname bibitem#1\endcsname}%
\let\auto@bib@innerbib\@empty
\bibitem [{\citenamefont {de~Groot}\ and\ \citenamefont
  {Mazur}(1962)}]{dgm1962nonequilibrium}%
  \BibitemOpen
  \bibfield  {author} {\bibinfo {author} {\bibfnamefont {S~R}\ \bibnamefont
  {de~Groot}}\ and\ \bibinfo {author} {\bibfnamefont {P}~\bibnamefont
  {Mazur}},\ }\href@noop {} {\emph {\bibinfo {title} {Non-Equilibrium
  Thermodynamics}}}\ (\bibinfo  {publisher} {North-Holland Publishing Company,
  Amsterdam},\ \bibinfo {year} {1962})\BibitemShut {NoStop}%
\bibitem [{\citenamefont {De~Zarate}\ and\ \citenamefont
  {Sengers}(2006)}]{de2006hydrodynamic}%
  \BibitemOpen
  \bibfield  {author} {\bibinfo {author} {\bibfnamefont {Jose M~Ortiz}\
  \bibnamefont {De~Zarate}}\ and\ \bibinfo {author} {\bibfnamefont {Jan~V}\
  \bibnamefont {Sengers}},\ }\href@noop {} {\emph {\bibinfo {title}
  {Hydrodynamic fluctuations in fluids and fluid mixtures}}}\ (\bibinfo
  {publisher} {Elsevier},\ \bibinfo {year} {2006})\BibitemShut {NoStop}%
\bibitem [{\citenamefont {Phillips}\ \emph {et~al.}(2012)\citenamefont
  {Phillips}, \citenamefont {Kondev}, \citenamefont {Theriot},\ and\
  \citenamefont {Garcia}}]{phillips2012physical}%
  \BibitemOpen
  \bibfield  {author} {\bibinfo {author} {\bibfnamefont {Rob}\ \bibnamefont
  {Phillips}}, \bibinfo {author} {\bibfnamefont {Jane}\ \bibnamefont {Kondev}},
  \bibinfo {author} {\bibfnamefont {Julie}\ \bibnamefont {Theriot}}, \ and\
  \bibinfo {author} {\bibfnamefont {Hernan}\ \bibnamefont {Garcia}},\
  }\href@noop {} {\emph {\bibinfo {title} {Physical biology of the cell}}}\
  (\bibinfo  {publisher} {Garland Science},\ \bibinfo {year}
  {2012})\BibitemShut {NoStop}%
\bibitem [{\citenamefont {Peixoto}\ and\ \citenamefont
  {Oort}(1992)}]{peixoto1992physics}%
  \BibitemOpen
  \bibfield  {author} {\bibinfo {author} {\bibfnamefont {Jos{\'e}~Pinto}\
  \bibnamefont {Peixoto}}\ and\ \bibinfo {author} {\bibfnamefont {Abraham~H}\
  \bibnamefont {Oort}},\ }\href@noop {} {\emph {\bibinfo {title} {Physics of
  climate}}},\ Vol.\ \bibinfo {volume} {520}\ (\bibinfo  {publisher}
  {Springer},\ \bibinfo {year} {1992})\BibitemShut {NoStop}%
\bibitem [{\citenamefont
  {Chandrasekhar}(2013)}]{chandrasekhar2013hydrodynamic}%
  \BibitemOpen
  \bibfield  {author} {\bibinfo {author} {\bibfnamefont {Subrahmanyan}\
  \bibnamefont {Chandrasekhar}},\ }\href@noop {} {\emph {\bibinfo {title}
  {Hydrodynamic and hydromagnetic stability}}}\ (\bibinfo  {publisher} {Courier
  Corporation},\ \bibinfo {year} {2013})\BibitemShut {NoStop}%
\bibitem [{\citenamefont {Chaikin}\ \emph {et~al.}(1995)\citenamefont
  {Chaikin}, \citenamefont {Lubensky},\ and\ \citenamefont
  {Witten}}]{chaikin1995principles}%
  \BibitemOpen
  \bibfield  {author} {\bibinfo {author} {\bibfnamefont {Paul~M}\ \bibnamefont
  {Chaikin}}, \bibinfo {author} {\bibfnamefont {Tom~C}\ \bibnamefont
  {Lubensky}}, \ and\ \bibinfo {author} {\bibfnamefont {Thomas~A}\ \bibnamefont
  {Witten}},\ }\href@noop {} {\emph {\bibinfo {title} {Principles of condensed
  matter physics}}},\ Vol.~\bibinfo {volume} {10}\ (\bibinfo  {publisher}
  {Cambridge university press Cambridge},\ \bibinfo {year} {1995})\BibitemShut
  {NoStop}%
\bibitem [{\citenamefont {Gaspard}(2022)}]{gaspard2022statistical}%
  \BibitemOpen
  \bibfield  {author} {\bibinfo {author} {\bibfnamefont {Pierre}\ \bibnamefont
  {Gaspard}},\ }\href@noop {} {\emph {\bibinfo {title} {The Statistical
  Mechanics of Irreversible Phenomena}}}\ (\bibinfo  {publisher} {Cambridge
  University Press},\ \bibinfo {year} {2022})\BibitemShut {NoStop}%
\bibitem [{\citenamefont {Gallavotti}(2002)}]{gallavotti2002foundations}%
  \BibitemOpen
  \bibfield  {author} {\bibinfo {author} {\bibfnamefont {Giovanni}\
  \bibnamefont {Gallavotti}},\ }\href@noop {} {\emph {\bibinfo {title}
  {Foundations of fluid dynamics}}},\ Vol.\ \bibinfo {volume} {172}\ (\bibinfo
  {publisher} {Springer Science \& Business Media},\ \bibinfo {year}
  {2002})\BibitemShut {NoStop}%
\bibitem [{\citenamefont {Chapman}\ and\ \citenamefont
  {Cowling}(1990)}]{chapman1990mathematical}%
  \BibitemOpen
  \bibfield  {author} {\bibinfo {author} {\bibfnamefont {Sydney}\ \bibnamefont
  {Chapman}}\ and\ \bibinfo {author} {\bibfnamefont {Thomas~George}\
  \bibnamefont {Cowling}},\ }\href@noop {} {\emph {\bibinfo {title} {The
  mathematical theory of non-uniform gases: an account of the kinetic theory of
  viscosity, thermal conduction and diffusion in gases}}}\ (\bibinfo
  {publisher} {Cambridge university press},\ \bibinfo {year}
  {1990})\BibitemShut {NoStop}%
\bibitem [{\citenamefont {Lanford}(1975)}]{lanford1975time}%
  \BibitemOpen
  \bibfield  {author} {\bibinfo {author} {\bibfnamefont {Oscar~E}\ \bibnamefont
  {Lanford}},\ }\enquote {\bibinfo {title} {Time evolution of large classical
  systems},}\ \ (\bibinfo  {publisher} {Springer},\ \bibinfo {year} {1975})\
  pp.\ \bibinfo {pages} {1--111}\BibitemShut {NoStop}%
\bibitem [{\citenamefont {Cercignani}(1988)}]{cercignani1988boltzmann}%
  \BibitemOpen
  \bibfield  {author} {\bibinfo {author} {\bibfnamefont {Carlo}\ \bibnamefont
  {Cercignani}},\ }\href@noop {} {\emph {\bibinfo {title} {The Boltzmann
  equation and its applications}}}\ (\bibinfo  {publisher} {Springer},\
  \bibinfo {year} {1988})\BibitemShut {NoStop}%
\bibitem [{\citenamefont {Spohn}(2012)}]{spohn2012large}%
  \BibitemOpen
  \bibfield  {author} {\bibinfo {author} {\bibfnamefont {Herbert}\ \bibnamefont
  {Spohn}},\ }\href@noop {} {\emph {\bibinfo {title} {Large scale dynamics of
  interacting particles}}}\ (\bibinfo  {publisher} {Springer Science \&
  Business Media},\ \bibinfo {year} {2012})\BibitemShut {NoStop}%
\bibitem [{\citenamefont {Gaspard}(1998)}]{gaspard1998chaos}%
  \BibitemOpen
  \bibfield  {author} {\bibinfo {author} {\bibfnamefont {Pierre}\ \bibnamefont
  {Gaspard}},\ }\href@noop {} {\emph {\bibinfo {title} {Chaos, Scattering and
  Statistical Mechanics}}}\ (\bibinfo  {publisher} {Cambridge University
  Press},\ \bibinfo {year} {1998})\BibitemShut {NoStop}%
\bibitem [{\citenamefont {Prigogine}(1949)}]{prigogine1949domaine}%
  \BibitemOpen
  \bibfield  {author} {\bibinfo {author} {\bibfnamefont {Ilya}\ \bibnamefont
  {Prigogine}},\ }\bibfield  {title} {\enquote {\bibinfo {title} {Le domaine de
  validit{\'e} de la thermodynamique des ph{\'e}nom{\`e}nes
  irr{\'e}versibles},}\ }\href
  {https://www.sciencedirect.com/science/article/abs/pii/0031891449900567}
  {\bibfield  {journal} {\bibinfo  {journal} {Physica}\ }\textbf {\bibinfo
  {volume} {15}},\ \bibinfo {pages} {272--284} (\bibinfo {year}
  {1949})}\BibitemShut {NoStop}%
\bibitem [{\citenamefont {Van~Kampen}(1987)}]{van1987chapman}%
  \BibitemOpen
  \bibfield  {author} {\bibinfo {author} {\bibfnamefont {NG}~\bibnamefont
  {Van~Kampen}},\ }\bibfield  {title} {\enquote {\bibinfo {title}
  {Chapman-enskog as an application of the method for eliminating fast
  variables},}\ }\href {https://link.springer.com/article/10.1007/BF01013381}
  {\bibfield  {journal} {\bibinfo  {journal} {Journal of Statistical Physics}\
  }\textbf {\bibinfo {volume} {46}},\ \bibinfo {pages} {709--727} (\bibinfo
  {year} {1987})}\BibitemShut {NoStop}%
\bibitem [{\citenamefont {Jarzynski}(1997)}]{jarzynski1997nonequilibrium}%
  \BibitemOpen
  \bibfield  {author} {\bibinfo {author} {\bibfnamefont {Christopher}\
  \bibnamefont {Jarzynski}},\ }\bibfield  {title} {\enquote {\bibinfo {title}
  {Nonequilibrium equality for free energy differences},}\ }\href
  {https://journals.aps.org/prl/abstract/10.1103/PhysRevLett.78.2690}
  {\bibfield  {journal} {\bibinfo  {journal} {Physical Review Letters}\
  }\textbf {\bibinfo {volume} {78}},\ \bibinfo {pages} {2690} (\bibinfo {year}
  {1997})}\BibitemShut {NoStop}%
\bibitem [{\citenamefont {Seifert}(2012)}]{seifert2012stochastic}%
  \BibitemOpen
  \bibfield  {author} {\bibinfo {author} {\bibfnamefont {Udo}\ \bibnamefont
  {Seifert}},\ }\bibfield  {title} {\enquote {\bibinfo {title} {Stochastic
  thermodynamics, fluctuation theorems and molecular machines},}\ }\href
  {https://iopscience.iop.org/article/10.1088/0034-4885/75/12/126001/meta}
  {\bibfield  {journal} {\bibinfo  {journal} {Reports on Progress in Physics}\
  }\textbf {\bibinfo {volume} {75}},\ \bibinfo {pages} {126001} (\bibinfo
  {year} {2012})}\BibitemShut {NoStop}%
\bibitem [{\citenamefont {Van~den Broeck}\ and\ \citenamefont
  {Esposito}(2015)}]{van2015ensemble}%
  \BibitemOpen
  \bibfield  {author} {\bibinfo {author} {\bibfnamefont {Christian}\
  \bibnamefont {Van~den Broeck}}\ and\ \bibinfo {author} {\bibfnamefont
  {Massimiliano}\ \bibnamefont {Esposito}},\ }\bibfield  {title} {\enquote
  {\bibinfo {title} {Ensemble and trajectory thermodynamics: A brief
  introduction},}\ }\href
  {https://www.sciencedirect.com/science/article/pii/S037843711400346X}
  {\bibfield  {journal} {\bibinfo  {journal} {Physica A: Statistical Mechanics
  and its Applications}\ }\textbf {\bibinfo {volume} {418}},\ \bibinfo {pages}
  {6--16} (\bibinfo {year} {2015})}\BibitemShut {NoStop}%
\bibitem [{\citenamefont {Peliti}\ and\ \citenamefont
  {Pigolotti}(2021)}]{peliti2021stochastic}%
  \BibitemOpen
  \bibfield  {author} {\bibinfo {author} {\bibfnamefont {Luca}\ \bibnamefont
  {Peliti}}\ and\ \bibinfo {author} {\bibfnamefont {Simone}\ \bibnamefont
  {Pigolotti}},\ }\href@noop {} {\emph {\bibinfo {title} {Stochastic
  thermodynamics: an introduction}}}\ (\bibinfo  {publisher} {Princeton
  University Press},\ \bibinfo {year} {2021})\BibitemShut {NoStop}%
\bibitem [{\citenamefont {Strasberg}(2022)}]{strasberg2022quantum}%
  \BibitemOpen
  \bibfield  {author} {\bibinfo {author} {\bibfnamefont {Philipp}\ \bibnamefont
  {Strasberg}},\ }\href@noop {} {\emph {\bibinfo {title} {Quantum Stochastic
  Thermodynamics: Foundations and Selected Applications}}}\ (\bibinfo
  {publisher} {Oxford University Press},\ \bibinfo {year} {2022})\BibitemShut
  {NoStop}%
\bibitem [{\citenamefont {Van~Kampen}(1992)}]{van1992stochastic}%
  \BibitemOpen
  \bibfield  {author} {\bibinfo {author} {\bibfnamefont {Nicolaas~Godfried}\
  \bibnamefont {Van~Kampen}},\ }\href@noop {} {\emph {\bibinfo {title}
  {Stochastic processes in physics and chemistry}}},\ Vol.~\bibinfo {volume}
  {1}\ (\bibinfo  {publisher} {Elsevier},\ \bibinfo {year} {1992})\BibitemShut
  {NoStop}%
\bibitem [{\citenamefont {Breuer}\ \emph {et~al.}(2002)\citenamefont {Breuer},
  \citenamefont {Petruccione} \emph {et~al.}}]{breuer2002theory}%
  \BibitemOpen
  \bibfield  {author} {\bibinfo {author} {\bibfnamefont {Heinz-Peter}\
  \bibnamefont {Breuer}}, \bibinfo {author} {\bibfnamefont {Francesco}\
  \bibnamefont {Petruccione}},  \emph {et~al.},\ }\href@noop {} {\emph
  {\bibinfo {title} {The theory of open quantum systems}}}\ (\bibinfo
  {publisher} {Oxford University Press on Demand},\ \bibinfo {year}
  {2002})\BibitemShut {NoStop}%
\bibitem [{\citenamefont {Jarzynski}(1999)}]{jarzynski1999microscopic}%
  \BibitemOpen
  \bibfield  {author} {\bibinfo {author} {\bibfnamefont {C}~\bibnamefont
  {Jarzynski}},\ }\bibfield  {title} {\enquote {\bibinfo {title} {Microscopic
  analysis of clausius--duhem processes},}\ }\href
  {https://link.springer.com/article/10.1023/A:1004541004050} {\bibfield
  {journal} {\bibinfo  {journal} {Journal of statistical physics}\ }\textbf
  {\bibinfo {volume} {96}},\ \bibinfo {pages} {415--427} (\bibinfo {year}
  {1999})}\BibitemShut {NoStop}%
\bibitem [{\citenamefont {Esposito}\ \emph {et~al.}(2010)\citenamefont
  {Esposito}, \citenamefont {Lindenberg},\ and\ \citenamefont {Van~den
  Broeck}}]{esposito2010entropy}%
  \BibitemOpen
  \bibfield  {author} {\bibinfo {author} {\bibfnamefont {Massimiliano}\
  \bibnamefont {Esposito}}, \bibinfo {author} {\bibfnamefont {Katja}\
  \bibnamefont {Lindenberg}}, \ and\ \bibinfo {author} {\bibfnamefont
  {Christian}\ \bibnamefont {Van~den Broeck}},\ }\bibfield  {title} {\enquote
  {\bibinfo {title} {Entropy production as correlation between system and
  reservoir},}\ }\href
  {https://iopscience.iop.org/article/10.1088/1367-2630/12/1/013013/meta}
  {\bibfield  {journal} {\bibinfo  {journal} {New Journal of Physics}\ }\textbf
  {\bibinfo {volume} {12}},\ \bibinfo {pages} {013013} (\bibinfo {year}
  {2010})}\BibitemShut {NoStop}%
\bibitem [{\citenamefont {Ptaszy{\'n}ski}\ and\ \citenamefont
  {Esposito}(2019)}]{ptaszynski2019entropy}%
  \BibitemOpen
  \bibfield  {author} {\bibinfo {author} {\bibfnamefont {Krzysztof}\
  \bibnamefont {Ptaszy{\'n}ski}}\ and\ \bibinfo {author} {\bibfnamefont
  {Massimiliano}\ \bibnamefont {Esposito}},\ }\bibfield  {title} {\enquote
  {\bibinfo {title} {Entropy production in open systems: The predominant role
  of intraenvironment correlations},}\ }\href
  {https://journals.aps.org/prl/abstract/10.1103/PhysRevLett.123.200603}
  {\bibfield  {journal} {\bibinfo  {journal} {Physical Review Letters}\
  }\textbf {\bibinfo {volume} {123}},\ \bibinfo {pages} {200603} (\bibinfo
  {year} {2019})}\BibitemShut {NoStop}%
\bibitem [{\citenamefont {Soret}\ \emph {et~al.}(2022)\citenamefont {Soret},
  \citenamefont {Cavina},\ and\ \citenamefont
  {Esposito}}]{soret2022thermodynamic}%
  \BibitemOpen
  \bibfield  {author} {\bibinfo {author} {\bibfnamefont {Ariane}\ \bibnamefont
  {Soret}}, \bibinfo {author} {\bibfnamefont {Vasco}\ \bibnamefont {Cavina}}, \
  and\ \bibinfo {author} {\bibfnamefont {Massimiliano}\ \bibnamefont
  {Esposito}},\ }\bibfield  {title} {\enquote {\bibinfo {title} {Thermodynamic
  consistency of quantum master equations},}\ }\href
  {https://journals.aps.org/pra/abstract/10.1103/PhysRevA.106.062209}
  {\bibfield  {journal} {\bibinfo  {journal} {Physical Review A}\ }\textbf
  {\bibinfo {volume} {106}},\ \bibinfo {pages} {062209} (\bibinfo {year}
  {2022})}\BibitemShut {NoStop}%
\bibitem [{\citenamefont {Esposito}\ \emph {et~al.}(2009)\citenamefont
  {Esposito}, \citenamefont {Harbola},\ and\ \citenamefont
  {Mukamel}}]{esposito2009nonequilibrium}%
  \BibitemOpen
  \bibfield  {author} {\bibinfo {author} {\bibfnamefont {Massimiliano}\
  \bibnamefont {Esposito}}, \bibinfo {author} {\bibfnamefont {Upendra}\
  \bibnamefont {Harbola}}, \ and\ \bibinfo {author} {\bibfnamefont {Shaul}\
  \bibnamefont {Mukamel}},\ }\bibfield  {title} {\enquote {\bibinfo {title}
  {Nonequilibrium fluctuations, fluctuation theorems, and counting statistics
  in quantum systems},}\ }\href
  {https://journals.aps.org/rmp/abstract/10.1103/RevModPhys.81.1665} {\bibfield
   {journal} {\bibinfo  {journal} {Review of Modern Physics}\ }\textbf
  {\bibinfo {volume} {81}},\ \bibinfo {pages} {1665} (\bibinfo {year}
  {2009})}\BibitemShut {NoStop}%
\bibitem [{\citenamefont {Esposito}(2012)}]{esposito2012stochastic}%
  \BibitemOpen
  \bibfield  {author} {\bibinfo {author} {\bibfnamefont {Massimiliano}\
  \bibnamefont {Esposito}},\ }\bibfield  {title} {\enquote {\bibinfo {title}
  {Stochastic thermodynamics under coarse graining},}\ }\href
  {https://journals.aps.org/pre/abstract/10.1103/PhysRevE.85.041125} {\bibfield
   {journal} {\bibinfo  {journal} {Physical Review E}\ }\textbf {\bibinfo
  {volume} {85}},\ \bibinfo {pages} {041125} (\bibinfo {year}
  {2012})}\BibitemShut {NoStop}%
\bibitem [{\citenamefont {Falasco}\ and\ \citenamefont
  {Esposito}(2021)}]{falasco2021local}%
  \BibitemOpen
  \bibfield  {author} {\bibinfo {author} {\bibfnamefont {Gianmaria}\
  \bibnamefont {Falasco}}\ and\ \bibinfo {author} {\bibfnamefont
  {Massimiliano}\ \bibnamefont {Esposito}},\ }\bibfield  {title} {\enquote
  {\bibinfo {title} {Local detailed balance across scales: From diffusions to
  jump processes and beyond},}\ }\href
  {https://journals.aps.org/pre/abstract/10.1103/PhysRevE.103.042114}
  {\bibfield  {journal} {\bibinfo  {journal} {Physical Review E}\ }\textbf
  {\bibinfo {volume} {103}},\ \bibinfo {pages} {042114} (\bibinfo {year}
  {2021})}\BibitemShut {NoStop}%
\bibitem [{\citenamefont {Maes}(2021)}]{maes2021local}%
  \BibitemOpen
  \bibfield  {author} {\bibinfo {author} {\bibfnamefont {Christian}\
  \bibnamefont {Maes}},\ }\bibfield  {title} {\enquote {\bibinfo {title} {Local
  detailed balance},}\ }\href
  {https://www.scipost.org/SciPostPhysLectNotes.32?acad_field_slug=mathematics}
  {\bibfield  {journal} {\bibinfo  {journal} {SciPost Physics Lecture Notes}\
  ,\ \bibinfo {pages} {032}} (\bibinfo {year} {2021})}\BibitemShut {NoStop}%
\bibitem [{\citenamefont {Kurchan}(1998)}]{kurchan1998fluctuation}%
  \BibitemOpen
  \bibfield  {author} {\bibinfo {author} {\bibfnamefont {Jorge}\ \bibnamefont
  {Kurchan}},\ }\bibfield  {title} {\enquote {\bibinfo {title} {Fluctuation
  theorem for stochastic dynamics},}\ }\href
  {https://iopscience.iop.org/article/10.1088/0305-4470/31/16/003/meta}
  {\bibfield  {journal} {\bibinfo  {journal} {Journal of Physics A:
  Mathematical and General}\ }\textbf {\bibinfo {volume} {31}},\ \bibinfo
  {pages} {3719} (\bibinfo {year} {1998})}\BibitemShut {NoStop}%
\bibitem [{\citenamefont {Crooks}(1999)}]{crooks1999entropy}%
  \BibitemOpen
  \bibfield  {author} {\bibinfo {author} {\bibfnamefont {Gavin~E}\ \bibnamefont
  {Crooks}},\ }\bibfield  {title} {\enquote {\bibinfo {title} {Entropy
  production fluctuation theorem and the nonequilibrium work relation for free
  energy differences},}\ }\href
  {https://journals.aps.org/pre/abstract/10.1103/PhysRevE.60.2721} {\bibfield
  {journal} {\bibinfo  {journal} {Physical Review E}\ }\textbf {\bibinfo
  {volume} {60}},\ \bibinfo {pages} {2721} (\bibinfo {year}
  {1999})}\BibitemShut {NoStop}%
\bibitem [{\citenamefont {Seifert}(2005)}]{seifert2005entropy}%
  \BibitemOpen
  \bibfield  {author} {\bibinfo {author} {\bibfnamefont {Udo}\ \bibnamefont
  {Seifert}},\ }\bibfield  {title} {\enquote {\bibinfo {title} {Entropy
  production along a stochastic trajectory and an integral fluctuation
  theorem},}\ }\href
  {https://journals.aps.org/prl/abstract/10.1103/PhysRevLett.95.040602}
  {\bibfield  {journal} {\bibinfo  {journal} {Physical Review Letters}\
  }\textbf {\bibinfo {volume} {95}},\ \bibinfo {pages} {040602} (\bibinfo
  {year} {2005})}\BibitemShut {NoStop}%
\bibitem [{\citenamefont {Rao}\ and\ \citenamefont
  {Esposito}(2018{\natexlab{a}})}]{rao2018conservation}%
  \BibitemOpen
  \bibfield  {author} {\bibinfo {author} {\bibfnamefont {Riccardo}\
  \bibnamefont {Rao}}\ and\ \bibinfo {author} {\bibfnamefont {Massimiliano}\
  \bibnamefont {Esposito}},\ }\bibfield  {title} {\enquote {\bibinfo {title}
  {Conservation laws shape dissipation},}\ }\href
  {https://iopscience.iop.org/article/10.1088/1367-2630/aaa15f/meta} {\bibfield
   {journal} {\bibinfo  {journal} {New Journal of Physics}\ }\textbf {\bibinfo
  {volume} {20}},\ \bibinfo {pages} {023007} (\bibinfo {year}
  {2018}{\natexlab{a}})}\BibitemShut {NoStop}%
\bibitem [{\citenamefont {Esposito}\ and\ \citenamefont {Van~den
  Broeck}(2010{\natexlab{a}})}]{esposito2010threefaces}%
  \BibitemOpen
  \bibfield  {author} {\bibinfo {author} {\bibfnamefont {Massimiliano}\
  \bibnamefont {Esposito}}\ and\ \bibinfo {author} {\bibfnamefont {Christian}\
  \bibnamefont {Van~den Broeck}},\ }\bibfield  {title} {\enquote {\bibinfo
  {title} {Three faces of the second law. i. master equation formulation},}\
  }\href {\doibase 10.1103/PhysRevE.82.011143} {\bibfield  {journal} {\bibinfo
  {journal} {Physical Review E}\ }\textbf {\bibinfo {volume} {82}},\ \bibinfo
  {pages} {011143} (\bibinfo {year} {2010}{\natexlab{a}})}\BibitemShut
  {NoStop}%
\bibitem [{\citenamefont {Van~den Broeck}\ and\ \citenamefont
  {Esposito}(2010)}]{van2010threefaces}%
  \BibitemOpen
  \bibfield  {author} {\bibinfo {author} {\bibfnamefont {Christian}\
  \bibnamefont {Van~den Broeck}}\ and\ \bibinfo {author} {\bibfnamefont
  {Massimiliano}\ \bibnamefont {Esposito}},\ }\bibfield  {title} {\enquote
  {\bibinfo {title} {Three faces of the second law. ii. fokker-planck
  formulation},}\ }\href
  {https://journals.aps.org/pre/abstract/10.1103/PhysRevE.82.011144} {\bibfield
   {journal} {\bibinfo  {journal} {Physical Review E}\ }\textbf {\bibinfo
  {volume} {82}},\ \bibinfo {pages} {011144} (\bibinfo {year}
  {2010})}\BibitemShut {NoStop}%
\bibitem [{\citenamefont {Esposito}\ and\ \citenamefont {Van~den
  Broeck}(2010{\natexlab{b}})}]{esposito2010three}%
  \BibitemOpen
  \bibfield  {author} {\bibinfo {author} {\bibfnamefont {Massimiliano}\
  \bibnamefont {Esposito}}\ and\ \bibinfo {author} {\bibfnamefont {Christian}\
  \bibnamefont {Van~den Broeck}},\ }\bibfield  {title} {\enquote {\bibinfo
  {title} {Three detailed fluctuation theorems},}\ }\href
  {https://journals.aps.org/prl/abstract/10.1103/PhysRevLett.104.090601}
  {\bibfield  {journal} {\bibinfo  {journal} {Physical Review Letters}\
  }\textbf {\bibinfo {volume} {104}},\ \bibinfo {pages} {090601} (\bibinfo
  {year} {2010}{\natexlab{b}})}\BibitemShut {NoStop}%
\bibitem [{\citenamefont {Lebowitz}\ and\ \citenamefont
  {Spohn}(1999)}]{lebowitz1999gallavotti}%
  \BibitemOpen
  \bibfield  {author} {\bibinfo {author} {\bibfnamefont {Joel~L}\ \bibnamefont
  {Lebowitz}}\ and\ \bibinfo {author} {\bibfnamefont {Herbert}\ \bibnamefont
  {Spohn}},\ }\bibfield  {title} {\enquote {\bibinfo {title} {A
  gallavotti--cohen-type symmetry in the large deviation functional for
  stochastic dynamics},}\ }\href
  {https://link.springer.com/article/10.1023/A:1004589714161} {\bibfield
  {journal} {\bibinfo  {journal} {Journal of Statistical Physics}\ }\textbf
  {\bibinfo {volume} {95}},\ \bibinfo {pages} {333--365} (\bibinfo {year}
  {1999})}\BibitemShut {NoStop}%
\bibitem [{\citenamefont {Andrieux}\ and\ \citenamefont
  {Gaspard}(2004)}]{andrieux2004fluctuation}%
  \BibitemOpen
  \bibfield  {author} {\bibinfo {author} {\bibfnamefont {David}\ \bibnamefont
  {Andrieux}}\ and\ \bibinfo {author} {\bibfnamefont {Pierre}\ \bibnamefont
  {Gaspard}},\ }\bibfield  {title} {\enquote {\bibinfo {title} {Fluctuation
  theorem and onsager reciprocity relations},}\ }\href
  {https://aip.scitation.org/doi/abs/10.1063/1.1782391} {\bibfield  {journal}
  {\bibinfo  {journal} {The Journal of Chemical Physics}\ }\textbf {\bibinfo
  {volume} {121}},\ \bibinfo {pages} {6167--6174} (\bibinfo {year}
  {2004})}\BibitemShut {NoStop}%
\bibitem [{\citenamefont {Forastiere}\ \emph {et~al.}(2022)\citenamefont
  {Forastiere}, \citenamefont {Rao},\ and\ \citenamefont
  {Esposito}}]{forastiere2022linear}%
  \BibitemOpen
  \bibfield  {author} {\bibinfo {author} {\bibfnamefont {Danilo}\ \bibnamefont
  {Forastiere}}, \bibinfo {author} {\bibfnamefont {Riccardo}\ \bibnamefont
  {Rao}}, \ and\ \bibinfo {author} {\bibfnamefont {Massimiliano}\ \bibnamefont
  {Esposito}},\ }\bibfield  {title} {\enquote {\bibinfo {title} {Linear
  stochastic thermodynamics},}\ }\href
  {https://iopscience.iop.org/article/10.1088/1367-2630/ac836b} {\bibfield
  {journal} {\bibinfo  {journal} {New Journal of Physics}\ }\textbf {\bibinfo
  {volume} {24}},\ \bibinfo {pages} {083021} (\bibinfo {year}
  {2022})}\BibitemShut {NoStop}%
\bibitem [{\citenamefont {Falasco}\ \emph {et~al.}(2018)\citenamefont
  {Falasco}, \citenamefont {Rao},\ and\ \citenamefont
  {Esposito}}]{falasco2018information}%
  \BibitemOpen
  \bibfield  {author} {\bibinfo {author} {\bibfnamefont {Gianmaria}\
  \bibnamefont {Falasco}}, \bibinfo {author} {\bibfnamefont {Riccardo}\
  \bibnamefont {Rao}}, \ and\ \bibinfo {author} {\bibfnamefont {Massimiliano}\
  \bibnamefont {Esposito}},\ }\bibfield  {title} {\enquote {\bibinfo {title}
  {Information thermodynamics of turing patterns},}\ }\href
  {https://journals.aps.org/prl/abstract/10.1103/PhysRevLett.121.108301}
  {\bibfield  {journal} {\bibinfo  {journal} {Physical Review Letters}\
  }\textbf {\bibinfo {volume} {121}},\ \bibinfo {pages} {108301} (\bibinfo
  {year} {2018})}\BibitemShut {NoStop}%
\bibitem [{\citenamefont {Avanzini}\ \emph {et~al.}(2019)\citenamefont
  {Avanzini}, \citenamefont {Falasco},\ and\ \citenamefont
  {Esposito}}]{avanzini2019thermodynamics}%
  \BibitemOpen
  \bibfield  {author} {\bibinfo {author} {\bibfnamefont {Francesco}\
  \bibnamefont {Avanzini}}, \bibinfo {author} {\bibfnamefont {Gianmaria}\
  \bibnamefont {Falasco}}, \ and\ \bibinfo {author} {\bibfnamefont
  {Massimiliano}\ \bibnamefont {Esposito}},\ }\bibfield  {title} {\enquote
  {\bibinfo {title} {Thermodynamics of chemical waves},}\ }\href
  {https://aip.scitation.org/doi/full/10.1063/1.5126528} {\bibfield  {journal}
  {\bibinfo  {journal} {The Journal of Chemical Physics}\ }\textbf {\bibinfo
  {volume} {151}},\ \bibinfo {pages} {234103} (\bibinfo {year}
  {2019})}\BibitemShut {NoStop}%
\bibitem [{\citenamefont {Rao}\ and\ \citenamefont
  {Esposito}(2018{\natexlab{b}})}]{rao2018conservationII}%
  \BibitemOpen
  \bibfield  {author} {\bibinfo {author} {\bibfnamefont {Riccardo}\
  \bibnamefont {Rao}}\ and\ \bibinfo {author} {\bibfnamefont {Massimiliano}\
  \bibnamefont {Esposito}},\ }\bibfield  {title} {\enquote {\bibinfo {title}
  {Conservation laws and work fluctuation relations in chemical reaction
  networks},}\ }\href
  {https://pubs.aip.org/aip/jcp/article/149/24/245101/197785} {\bibfield
  {journal} {\bibinfo  {journal} {The Journal of chemical physics}\ }\textbf
  {\bibinfo {volume} {149}} (\bibinfo {year} {2018}{\natexlab{b}})}\BibitemShut
  {NoStop}%
\bibitem [{\citenamefont {Lazarescu}\ \emph {et~al.}(2019)\citenamefont
  {Lazarescu}, \citenamefont {Cossetto}, \citenamefont {Falasco},\ and\
  \citenamefont {Esposito}}]{lazarescu2019large}%
  \BibitemOpen
  \bibfield  {author} {\bibinfo {author} {\bibfnamefont {Alexandre}\
  \bibnamefont {Lazarescu}}, \bibinfo {author} {\bibfnamefont {Tommaso}\
  \bibnamefont {Cossetto}}, \bibinfo {author} {\bibfnamefont {Gianmaria}\
  \bibnamefont {Falasco}}, \ and\ \bibinfo {author} {\bibfnamefont
  {Massimiliano}\ \bibnamefont {Esposito}},\ }\bibfield  {title} {\enquote
  {\bibinfo {title} {Large deviations and dynamical phase transitions in
  stochastic chemical networks},}\ }\href
  {https://aip.scitation.org/doi/full/10.1063/1.5111110} {\bibfield  {journal}
  {\bibinfo  {journal} {The Journal of Chemical Physics}\ }\textbf {\bibinfo
  {volume} {151}},\ \bibinfo {pages} {064117} (\bibinfo {year}
  {2019})}\BibitemShut {NoStop}%
\bibitem [{\citenamefont {Freitas}\ \emph {et~al.}(2021)\citenamefont
  {Freitas}, \citenamefont {Delvenne},\ and\ \citenamefont
  {Esposito}}]{freitas2021stochastic}%
  \BibitemOpen
  \bibfield  {author} {\bibinfo {author} {\bibfnamefont {Nahuel}\ \bibnamefont
  {Freitas}}, \bibinfo {author} {\bibfnamefont {Jean-Charles}\ \bibnamefont
  {Delvenne}}, \ and\ \bibinfo {author} {\bibfnamefont {Massimiliano}\
  \bibnamefont {Esposito}},\ }\bibfield  {title} {\enquote {\bibinfo {title}
  {Stochastic thermodynamics of nonlinear electronic circuits: A realistic
  framework for computing around k t},}\ }\href
  {https://journals.aps.org/prx/abstract/10.1103/PhysRevX.11.031064} {\bibfield
   {journal} {\bibinfo  {journal} {Physical Review X}\ }\textbf {\bibinfo
  {volume} {11}},\ \bibinfo {pages} {031064} (\bibinfo {year}
  {2021})}\BibitemShut {NoStop}%
\bibitem [{\citenamefont {Freitas}\ and\ \citenamefont
  {Esposito}(2022)}]{freitas2022emergent}%
  \BibitemOpen
  \bibfield  {author} {\bibinfo {author} {\bibfnamefont {Jos{\'e}~Nahuel}\
  \bibnamefont {Freitas}}\ and\ \bibinfo {author} {\bibfnamefont
  {Massimiliano}\ \bibnamefont {Esposito}},\ }\bibfield  {title} {\enquote
  {\bibinfo {title} {Emergent second law for non-equilibrium steady states},}\
  }\href {https://www.nature.com/articles/s41467-022-32700-7} {\bibfield
  {journal} {\bibinfo  {journal} {Nature Communications}\ }\textbf {\bibinfo
  {volume} {13}},\ \bibinfo {pages} {1--8} (\bibinfo {year}
  {2022})}\BibitemShut {NoStop}%
\bibitem [{\citenamefont {Falasco}\ and\ \citenamefont
  {Esposito}(2024)}]{falasco2024RMP}%
  \BibitemOpen
  \bibfield  {author} {\bibinfo {author} {\bibfnamefont {Gianmaria}\
  \bibnamefont {Falasco}}\ and\ \bibinfo {author} {\bibfnamefont
  {Massimiliano}\ \bibnamefont {Esposito}},\ }\bibfield  {title} {\enquote
  {\bibinfo {title} {Macroscopic stochastic thermodynamics},}\ }\href@noop {}
  {\bibfield  {journal} {\bibinfo  {journal} {to appear}\ } (\bibinfo {year}
  {2024})}\BibitemShut {NoStop}%
\bibitem [{\citenamefont {Kardar}(2007)}]{kardar2007statistical}%
  \BibitemOpen
  \bibfield  {author} {\bibinfo {author} {\bibfnamefont {Mehran}\ \bibnamefont
  {Kardar}},\ }\href@noop {} {\emph {\bibinfo {title} {Statistical physics of
  particles}}}\ (\bibinfo  {publisher} {Cambridge University Press},\ \bibinfo
  {year} {2007})\BibitemShut {NoStop}%
\bibitem [{\citenamefont {Lundstrom}(2002)}]{lundstrom2002fundamentals}%
  \BibitemOpen
  \bibfield  {author} {\bibinfo {author} {\bibfnamefont {Mark}\ \bibnamefont
  {Lundstrom}},\ }\href@noop {} {\emph {\bibinfo {title} {Fundamentals of
  carrier transport}}}\ (\bibinfo  {publisher} {IOP Publishing},\ \bibinfo
  {year} {2002})\BibitemShut {NoStop}%
\bibitem [{\citenamefont {Bergmann}\ and\ \citenamefont
  {Lebowitz}(1955)}]{bergmann1955new}%
  \BibitemOpen
  \bibfield  {author} {\bibinfo {author} {\bibfnamefont {Peter~G}\ \bibnamefont
  {Bergmann}}\ and\ \bibinfo {author} {\bibfnamefont {Joel~L}\ \bibnamefont
  {Lebowitz}},\ }\bibfield  {title} {\enquote {\bibinfo {title} {New approach
  to nonequilibrium processes},}\ }\href
  {https://journals.aps.org/pr/abstract/10.1103/PhysRev.99.578} {\bibfield
  {journal} {\bibinfo  {journal} {Physical Review}\ }\textbf {\bibinfo {volume}
  {99}},\ \bibinfo {pages} {578} (\bibinfo {year} {1955})}\BibitemShut
  {NoStop}%
\bibitem [{\citenamefont {Lebowitz}\ and\ \citenamefont
  {Bergmann}(1957)}]{lebowitz1957irreversible}%
  \BibitemOpen
  \bibfield  {author} {\bibinfo {author} {\bibfnamefont {Joel~L}\ \bibnamefont
  {Lebowitz}}\ and\ \bibinfo {author} {\bibfnamefont {Peter~G}\ \bibnamefont
  {Bergmann}},\ }\bibfield  {title} {\enquote {\bibinfo {title} {Irreversible
  gibbsian ensembles},}\ }\href
  {https://www.sciencedirect.com/science/article/abs/pii/0003491657900027}
  {\bibfield  {journal} {\bibinfo  {journal} {Annals of Physics}\ }\textbf
  {\bibinfo {volume} {1}},\ \bibinfo {pages} {1--23} (\bibinfo {year}
  {1957})}\BibitemShut {NoStop}%
\bibitem [{\citenamefont {Van~den Broeck}\ and\ \citenamefont
  {Toral}(2015)}]{van2015stochastic}%
  \BibitemOpen
  \bibfield  {author} {\bibinfo {author} {\bibfnamefont {C}~\bibnamefont
  {Van~den Broeck}}\ and\ \bibinfo {author} {\bibfnamefont {Ra{\'u}l}\
  \bibnamefont {Toral}},\ }\bibfield  {title} {\enquote {\bibinfo {title}
  {Stochastic thermodynamics for linear kinetic equations},}\ }\href
  {https://journals.aps.org/pre/abstract/10.1103/PhysRevE.92.012127} {\bibfield
   {journal} {\bibinfo  {journal} {Physical Review E}\ }\textbf {\bibinfo
  {volume} {92}},\ \bibinfo {pages} {012127} (\bibinfo {year}
  {2015})}\BibitemShut {NoStop}%
\bibitem [{\citenamefont {Horowitz}\ and\ \citenamefont
  {Esposito}(2016)}]{horowitz2016work}%
  \BibitemOpen
  \bibfield  {author} {\bibinfo {author} {\bibfnamefont {Jordan~M}\
  \bibnamefont {Horowitz}}\ and\ \bibinfo {author} {\bibfnamefont
  {Massimiliano}\ \bibnamefont {Esposito}},\ }\bibfield  {title} {\enquote
  {\bibinfo {title} {Work producing reservoirs: Stochastic thermodynamics with
  generalized gibbs ensembles},}\ }\href
  {https://journals.aps.org/pre/abstract/10.1103/PhysRevE.94.020102} {\bibfield
   {journal} {\bibinfo  {journal} {Physical Review E}\ }\textbf {\bibinfo
  {volume} {94}},\ \bibinfo {pages} {020102} (\bibinfo {year}
  {2016})}\BibitemShut {NoStop}%
\bibitem [{\citenamefont {Landau}\ and\ \citenamefont
  {Lifshitz}(1987)}]{landau1987fluid}%
  \BibitemOpen
  \bibfield  {author} {\bibinfo {author} {\bibfnamefont {Lev~Davidovich}\
  \bibnamefont {Landau}}\ and\ \bibinfo {author} {\bibfnamefont
  {Evgenii~Mikhailovich}\ \bibnamefont {Lifshitz}},\ }\href@noop {} {\emph
  {\bibinfo {title} {Fluid Mechanics: Course of Theoretical Physics}}},\
  Vol.~\bibinfo {volume} {6}\ (\bibinfo  {publisher} {Pergamon Press},\
  \bibinfo {year} {1987})\BibitemShut {NoStop}%
\bibitem [{\citenamefont {Callen}(1948)}]{callen1948application}%
  \BibitemOpen
  \bibfield  {author} {\bibinfo {author} {\bibfnamefont {Herbert~B}\
  \bibnamefont {Callen}},\ }\bibfield  {title} {\enquote {\bibinfo {title} {The
  application of onsager's reciprocal relations to thermoelectric,
  thermomagnetic, and galvanomagnetic effects},}\ }\href
  {https://journals.aps.org/pr/abstract/10.1103/PhysRev.73.1349} {\bibfield
  {journal} {\bibinfo  {journal} {Physical Review}\ }\textbf {\bibinfo {volume}
  {73}},\ \bibinfo {pages} {1349} (\bibinfo {year} {1948})}\BibitemShut
  {NoStop}%
\bibitem [{\citenamefont {Glansdorff}\ \emph {et~al.}(1974)\citenamefont
  {Glansdorff}, \citenamefont {Nicolis},\ and\ \citenamefont
  {Prigogine}}]{glansdorff1974thermodynamic}%
  \BibitemOpen
  \bibfield  {author} {\bibinfo {author} {\bibfnamefont {Paul}\ \bibnamefont
  {Glansdorff}}, \bibinfo {author} {\bibfnamefont {G}~\bibnamefont {Nicolis}},
  \ and\ \bibinfo {author} {\bibfnamefont {I}~\bibnamefont {Prigogine}},\
  }\bibfield  {title} {\enquote {\bibinfo {title} {The thermodynamic stability
  theory of non-equilibrium states},}\ }\href
  {https://www.pnas.org/content/71/1/197.short} {\bibfield  {journal} {\bibinfo
   {journal} {Proceedings of the National Academy of Sciences}\ }\textbf
  {\bibinfo {volume} {71}},\ \bibinfo {pages} {197--199} (\bibinfo {year}
  {1974})}\BibitemShut {NoStop}%
\bibitem [{\citenamefont {Glansdorff}\ and\ \citenamefont
  {Prigogine}(1971)}]{glansdorff1971thermodynamic}%
  \BibitemOpen
  \bibfield  {author} {\bibinfo {author} {\bibfnamefont {Paul}\ \bibnamefont
  {Glansdorff}}\ and\ \bibinfo {author} {\bibfnamefont {Ilya}\ \bibnamefont
  {Prigogine}},\ }\href@noop {} {\emph {\bibinfo {title} {Thermodynamic theory
  of structure, stability and fluctuations}}}\ (\bibinfo  {publisher}
  {Wiley-Interscience},\ \bibinfo {year} {1971})\BibitemShut {NoStop}%
\bibitem [{\citenamefont {Landau}\ and\ \citenamefont
  {Lifshitz}(2013)}]{landau2013statistical}%
  \BibitemOpen
  \bibfield  {author} {\bibinfo {author} {\bibfnamefont {Lev~Davidovich}\
  \bibnamefont {Landau}}\ and\ \bibinfo {author} {\bibfnamefont
  {Evgenii~Mikhailovich}\ \bibnamefont {Lifshitz}},\ }\href@noop {} {\emph
  {\bibinfo {title} {Statistical Physics: Course of Theoretical Physics}}},\
  Vol.~\bibinfo {volume} {5}\ (\bibinfo  {publisher} {Elsevier},\ \bibinfo
  {year} {2013})\BibitemShut {NoStop}%
\bibitem [{\citenamefont {Bejan}(2016)}]{bejan2016advanced}%
  \BibitemOpen
  \bibfield  {author} {\bibinfo {author} {\bibfnamefont {Adrian}\ \bibnamefont
  {Bejan}},\ }\href@noop {} {\emph {\bibinfo {title} {Advanced engineering
  thermodynamics}}}\ (\bibinfo  {publisher} {John Wiley \& Sons},\ \bibinfo
  {year} {2016})\BibitemShut {NoStop}%
\bibitem [{\citenamefont {Kotas}(2012)}]{kotas2012exergy}%
  \BibitemOpen
  \bibfield  {author} {\bibinfo {author} {\bibfnamefont {Tadeusz~J}\
  \bibnamefont {Kotas}},\ }\href@noop {} {\emph {\bibinfo {title} {The exergy
  method of thermal plant analysis}}}\ (\bibinfo  {publisher} {Paragon
  Publishing},\ \bibinfo {year} {2012})\BibitemShut {NoStop}%
\bibitem [{\citenamefont {Yang}\ \emph {et~al.}(2020)\citenamefont {Yang},
  \citenamefont {Chong}, \citenamefont {Wang}, \citenamefont {Verzicco},
  \citenamefont {Shishkina},\ and\ \citenamefont
  {Lohse}}]{yang2020periodically}%
  \BibitemOpen
  \bibfield  {author} {\bibinfo {author} {\bibfnamefont {Rui}\ \bibnamefont
  {Yang}}, \bibinfo {author} {\bibfnamefont {Kai~Leong}\ \bibnamefont {Chong}},
  \bibinfo {author} {\bibfnamefont {Qi}~\bibnamefont {Wang}}, \bibinfo {author}
  {\bibfnamefont {Roberto}\ \bibnamefont {Verzicco}}, \bibinfo {author}
  {\bibfnamefont {Olga}\ \bibnamefont {Shishkina}}, \ and\ \bibinfo {author}
  {\bibfnamefont {Detlef}\ \bibnamefont {Lohse}},\ }\bibfield  {title}
  {\enquote {\bibinfo {title} {Periodically modulated thermal convection},}\
  }\href {\doibase 10.1103/PhysRevLett.125.154502} {\bibfield  {journal}
  {\bibinfo  {journal} {Physical Review Letters}\ }\textbf {\bibinfo {volume}
  {125}},\ \bibinfo {pages} {154502} (\bibinfo {year} {2020})}\BibitemShut
  {NoStop}%
\bibitem [{\citenamefont {Urban}\ \emph {et~al.}(2022)\citenamefont {Urban},
  \citenamefont {Hanzelka}, \citenamefont {Kr\'alik}, \citenamefont
  {Musilov\'a},\ and\ \citenamefont {Skrbek}}]{urban2022thermal}%
  \BibitemOpen
  \bibfield  {author} {\bibinfo {author} {\bibfnamefont {P.}~\bibnamefont
  {Urban}}, \bibinfo {author} {\bibfnamefont {P.}~\bibnamefont {Hanzelka}},
  \bibinfo {author} {\bibfnamefont {T.}~\bibnamefont {Kr\'alik}}, \bibinfo
  {author} {\bibfnamefont {V.}~\bibnamefont {Musilov\'a}}, \ and\ \bibinfo
  {author} {\bibfnamefont {L.}~\bibnamefont {Skrbek}},\ }\bibfield  {title}
  {\enquote {\bibinfo {title} {Thermal waves and heat transfer efficiency
  enhancement in harmonically modulated turbulent thermal convection},}\ }\href
  {\doibase 10.1103/PhysRevLett.128.134502} {\bibfield  {journal} {\bibinfo
  {journal} {Physical Review Letters}\ }\textbf {\bibinfo {volume} {128}},\
  \bibinfo {pages} {134502} (\bibinfo {year} {2022})}\BibitemShut {NoStop}%
\bibitem [{\citenamefont {Maes}\ and\ \citenamefont
  {Neto{\v{c}}n{\`y}}(2015)}]{maes2015revisiting}%
  \BibitemOpen
  \bibfield  {author} {\bibinfo {author} {\bibfnamefont {Christian}\
  \bibnamefont {Maes}}\ and\ \bibinfo {author} {\bibfnamefont {Karel}\
  \bibnamefont {Neto{\v{c}}n{\`y}}},\ }\bibfield  {title} {\enquote {\bibinfo
  {title} {Revisiting the glansdorff--prigogine criterion for stability within
  irreversible thermodynamics},}\ }\href
  {https://link.springer.com/article/10.1007/s10955-015-1239-4} {\bibfield
  {journal} {\bibinfo  {journal} {Journal of Statistical Physics}\ }\textbf
  {\bibinfo {volume} {159}},\ \bibinfo {pages} {1286--1299} (\bibinfo {year}
  {2015})}\BibitemShut {NoStop}%
\bibitem [{\citenamefont {Oono}\ and\ \citenamefont
  {Paniconi}(1998)}]{oono1998steady}%
  \BibitemOpen
  \bibfield  {author} {\bibinfo {author} {\bibfnamefont {Yoshitsugu}\
  \bibnamefont {Oono}}\ and\ \bibinfo {author} {\bibfnamefont {Marco}\
  \bibnamefont {Paniconi}},\ }\bibfield  {title} {\enquote {\bibinfo {title}
  {Steady state thermodynamics},}\ }\href
  {https://academic.oup.com/ptps/article/doi/10.1143/PTPS.130.29/1842398}
  {\bibfield  {journal} {\bibinfo  {journal} {Progress of Theoretical Physics
  Supplement}\ }\textbf {\bibinfo {volume} {130}},\ \bibinfo {pages} {29--44}
  (\bibinfo {year} {1998})}\BibitemShut {NoStop}%
\bibitem [{\citenamefont {Cover}(1999)}]{cover1999elements}%
  \BibitemOpen
  \bibfield  {author} {\bibinfo {author} {\bibfnamefont {Thomas~M}\
  \bibnamefont {Cover}},\ }\href@noop {} {\emph {\bibinfo {title} {Elements of
  information theory}}}\ (\bibinfo  {publisher} {John Wiley \& Sons},\ \bibinfo
  {year} {1999})\BibitemShut {NoStop}%
\bibitem [{\citenamefont {Kirkwood}(1946)}]{kirkwood1946statistical}%
  \BibitemOpen
  \bibfield  {author} {\bibinfo {author} {\bibfnamefont {John~G}\ \bibnamefont
  {Kirkwood}},\ }\bibfield  {title} {\enquote {\bibinfo {title} {The
  statistical mechanical theory of transport processes i. general theory},}\
  }\href {https://aip.scitation.org/doi/abs/10.1063/1.1724117} {\bibfield
  {journal} {\bibinfo  {journal} {The Journal of Chemical Physics}\ }\textbf
  {\bibinfo {volume} {14}},\ \bibinfo {pages} {180--201} (\bibinfo {year}
  {1946})}\BibitemShut {NoStop}%
\bibitem [{\citenamefont {Shear}(1967)}]{shear1967analog}%
  \BibitemOpen
  \bibfield  {author} {\bibinfo {author} {\bibfnamefont {David}\ \bibnamefont
  {Shear}},\ }\bibfield  {title} {\enquote {\bibinfo {title} {An analog of the
  boltzmann h-theorem (a liapunov function) for systems of coupled chemical
  reactions},}\ }\href
  {https://www.sciencedirect.com/science/article/abs/pii/0022519367900057}
  {\bibfield  {journal} {\bibinfo  {journal} {Journal of Theoretical Biology}\
  }\textbf {\bibinfo {volume} {16}},\ \bibinfo {pages} {212--228} (\bibinfo
  {year} {1967})}\BibitemShut {NoStop}%
\bibitem [{\citenamefont {Higgins}(1968)}]{higgins1968some}%
  \BibitemOpen
  \bibfield  {author} {\bibinfo {author} {\bibfnamefont {J}~\bibnamefont
  {Higgins}},\ }\bibfield  {title} {\enquote {\bibinfo {title} {Some remarks on
  shear's liapunov function for systems of chemical reactions},}\ }\href
  {https://www.sciencedirect.com/science/article/abs/pii/0022519368901173}
  {\bibfield  {journal} {\bibinfo  {journal} {Journal of Theoretical Biology}\
  }\textbf {\bibinfo {volume} {21}},\ \bibinfo {pages} {293--304} (\bibinfo
  {year} {1968})}\BibitemShut {NoStop}%
\bibitem [{\citenamefont {Kawai}\ \emph {et~al.}(2007)\citenamefont {Kawai},
  \citenamefont {Parrondo},\ and\ \citenamefont {den
  Broeck}}]{kawai2007dissipation}%
  \BibitemOpen
  \bibfield  {author} {\bibinfo {author} {\bibfnamefont {R.}~\bibnamefont
  {Kawai}}, \bibinfo {author} {\bibfnamefont {J.~M.~R.}\ \bibnamefont
  {Parrondo}}, \ and\ \bibinfo {author} {\bibfnamefont {C.~Van}\ \bibnamefont
  {den Broeck}},\ }\bibfield  {title} {\enquote {\bibinfo {title} {Dissipation:
  The phase-space perspective},}\ }\href {\doibase
  10.1103/PhysRevLett.98.080602} {\bibfield  {journal} {\bibinfo  {journal}
  {Physical Review Letters}\ }\textbf {\bibinfo {volume} {98}},\ \bibinfo
  {pages} {080602} (\bibinfo {year} {2007})}\BibitemShut {NoStop}%
\bibitem [{\citenamefont {Chakraborti}\ \emph {et~al.}(2022)\citenamefont
  {Chakraborti}, \citenamefont {Dhar}, \citenamefont {Goldstein}, \citenamefont
  {Kundu},\ and\ \citenamefont {Lebowitz}}]{chakraborti2022entropy}%
  \BibitemOpen
  \bibfield  {author} {\bibinfo {author} {\bibfnamefont {Subhadip}\
  \bibnamefont {Chakraborti}}, \bibinfo {author} {\bibfnamefont {Abhishek}\
  \bibnamefont {Dhar}}, \bibinfo {author} {\bibfnamefont {Sheldon}\
  \bibnamefont {Goldstein}}, \bibinfo {author} {\bibfnamefont {Anupam}\
  \bibnamefont {Kundu}}, \ and\ \bibinfo {author} {\bibfnamefont {Joel~L}\
  \bibnamefont {Lebowitz}},\ }\bibfield  {title} {\enquote {\bibinfo {title}
  {Entropy growth during free expansion of an ideal gas},}\ }\href
  {https://iopscience.iop.org/article/10.1088/1751-8121/ac8a7e/meta} {\bibfield
   {journal} {\bibinfo  {journal} {Journal of Physics A: Mathematical and
  Theoretical}\ }\textbf {\bibinfo {volume} {55}},\ \bibinfo {pages} {394002}
  (\bibinfo {year} {2022})}\BibitemShut {NoStop}%
\bibitem [{\citenamefont {Struchtrup}\ and\ \citenamefont
  {Torrilhon}(2007)}]{struchtrup2007h}%
  \BibitemOpen
  \bibfield  {author} {\bibinfo {author} {\bibfnamefont {Henning}\ \bibnamefont
  {Struchtrup}}\ and\ \bibinfo {author} {\bibfnamefont {Manuel}\ \bibnamefont
  {Torrilhon}},\ }\bibfield  {title} {\enquote {\bibinfo {title} {H theorem,
  regularization, and boundary conditions for linearized 13 moment
  equations},}\ }\href
  {https://journals.aps.org/prl/abstract/10.1103/PhysRevLett.99.014502}
  {\bibfield  {journal} {\bibinfo  {journal} {Physical Review Letters}\
  }\textbf {\bibinfo {volume} {99}},\ \bibinfo {pages} {014502} (\bibinfo
  {year} {2007})}\BibitemShut {NoStop}%
\bibitem [{\citenamefont {Grad}(1963)}]{grad1963asymptotic}%
  \BibitemOpen
  \bibfield  {author} {\bibinfo {author} {\bibfnamefont {Harold}\ \bibnamefont
  {Grad}},\ }\bibfield  {title} {\enquote {\bibinfo {title} {Asymptotic theory
  of the boltzmann equation},}\ }\href
  {https://aip.scitation.org/doi/abs/10.1063/1.1706716} {\bibfield  {journal}
  {\bibinfo  {journal} {The Physics of Fluids}\ }\textbf {\bibinfo {volume}
  {6}},\ \bibinfo {pages} {147--181} (\bibinfo {year} {1963})}\BibitemShut
  {NoStop}%
\bibitem [{\citenamefont {Katz}\ \emph {et~al.}(1983)\citenamefont {Katz},
  \citenamefont {Lebowitz},\ and\ \citenamefont {Spohn}}]{katz1983phase}%
  \BibitemOpen
  \bibfield  {author} {\bibinfo {author} {\bibfnamefont {Sheldon}\ \bibnamefont
  {Katz}}, \bibinfo {author} {\bibfnamefont {Joel~L}\ \bibnamefont {Lebowitz}},
  \ and\ \bibinfo {author} {\bibfnamefont {H}~\bibnamefont {Spohn}},\
  }\bibfield  {title} {\enquote {\bibinfo {title} {Phase transitions in
  stationary nonequilibrium states of model lattice systems},}\ }\href
  {https://journals.aps.org/prb/abstract/10.1103/PhysRevB.28.1655} {\bibfield
  {journal} {\bibinfo  {journal} {Physical Review B}\ }\textbf {\bibinfo
  {volume} {28}},\ \bibinfo {pages} {1655} (\bibinfo {year}
  {1983})}\BibitemShut {NoStop}%
\bibitem [{\citenamefont {Jou}\ \emph {et~al.}(1996)\citenamefont {Jou},
  \citenamefont {Casas-V{\'a}zquez}, \citenamefont {Lebon}, \citenamefont
  {Jou}, \citenamefont {Casas-V{\'a}zquez},\ and\ \citenamefont
  {Lebon}}]{jou1996extended}%
  \BibitemOpen
  \bibfield  {author} {\bibinfo {author} {\bibfnamefont {David}\ \bibnamefont
  {Jou}}, \bibinfo {author} {\bibfnamefont {Jos{\'e}}\ \bibnamefont
  {Casas-V{\'a}zquez}}, \bibinfo {author} {\bibfnamefont {Georgy}\ \bibnamefont
  {Lebon}}, \bibinfo {author} {\bibfnamefont {David}\ \bibnamefont {Jou}},
  \bibinfo {author} {\bibfnamefont {Jos{\'e}}\ \bibnamefont
  {Casas-V{\'a}zquez}}, \ and\ \bibinfo {author} {\bibfnamefont {Georgy}\
  \bibnamefont {Lebon}},\ }\href@noop {} {\emph {\bibinfo {title} {Extended
  irreversible thermodynamics}}}\ (\bibinfo  {publisher} {Springer},\ \bibinfo
  {year} {1996})\BibitemShut {NoStop}%
\bibitem [{\citenamefont {Sasa}(2014)}]{sasa2014derivation}%
  \BibitemOpen
  \bibfield  {author} {\bibinfo {author} {\bibfnamefont {Shin-ichi}\
  \bibnamefont {Sasa}},\ }\bibfield  {title} {\enquote {\bibinfo {title}
  {Derivation of hydrodynamics from the hamiltonian description of particle
  systems},}\ }\href
  {https://journals.aps.org/prl/abstract/10.1103/PhysRevLett.112.100602}
  {\bibfield  {journal} {\bibinfo  {journal} {Physical Review Letters}\
  }\textbf {\bibinfo {volume} {112}},\ \bibinfo {pages} {100602} (\bibinfo
  {year} {2014})}\BibitemShut {NoStop}%
\bibitem [{\citenamefont {Mabillard}\ and\ \citenamefont
  {Gaspard}(2020)}]{mabillard2020microscopic}%
  \BibitemOpen
  \bibfield  {author} {\bibinfo {author} {\bibfnamefont {Jo{\"e}l}\
  \bibnamefont {Mabillard}}\ and\ \bibinfo {author} {\bibfnamefont {Pierre}\
  \bibnamefont {Gaspard}},\ }\bibfield  {title} {\enquote {\bibinfo {title}
  {Microscopic approach to the macrodynamics of matter with broken
  symmetries},}\ }\href
  {https://iopscience.iop.org/article/10.1088/1742-5468/abb0e0/meta} {\bibfield
   {journal} {\bibinfo  {journal} {Journal of Statistical Mechanics: Theory and
  Experiment}\ }\textbf {\bibinfo {volume} {2020}},\ \bibinfo {pages} {103203}
  (\bibinfo {year} {2020})}\BibitemShut {NoStop}%
\bibitem [{\citenamefont {Saito}\ \emph {et~al.}(2021)\citenamefont {Saito},
  \citenamefont {Hongo}, \citenamefont {Dhar},\ and\ \citenamefont
  {Sasa}}]{saito2021microscopic}%
  \BibitemOpen
  \bibfield  {author} {\bibinfo {author} {\bibfnamefont {Keiji}\ \bibnamefont
  {Saito}}, \bibinfo {author} {\bibfnamefont {Masaru}\ \bibnamefont {Hongo}},
  \bibinfo {author} {\bibfnamefont {Abhishek}\ \bibnamefont {Dhar}}, \ and\
  \bibinfo {author} {\bibfnamefont {Shin-ichi}\ \bibnamefont {Sasa}},\
  }\bibfield  {title} {\enquote {\bibinfo {title} {Microscopic theory of
  fluctuating hydrodynamics in nonlinear lattices},}\ }\href
  {https://journals.aps.org/prl/abstract/10.1103/PhysRevLett.127.010601}
  {\bibfield  {journal} {\bibinfo  {journal} {Physical Review Letters}\
  }\textbf {\bibinfo {volume} {127}},\ \bibinfo {pages} {010601} (\bibinfo
  {year} {2021})}\BibitemShut {NoStop}%
\bibitem [{\citenamefont {Bouchet}(2020)}]{bouchet2020boltzmann}%
  \BibitemOpen
  \bibfield  {author} {\bibinfo {author} {\bibfnamefont {Freddy}\ \bibnamefont
  {Bouchet}},\ }\bibfield  {title} {\enquote {\bibinfo {title} {Is the
  boltzmann equation reversible? a large deviation perspective on the
  irreversibility paradox},}\ }\href
  {https://link.springer.com/article/10.1007/s10955-020-02588-y} {\bibfield
  {journal} {\bibinfo  {journal} {Journal of Statistical Physics}\ }\textbf
  {\bibinfo {volume} {181}},\ \bibinfo {pages} {515--550} (\bibinfo {year}
  {2020})}\BibitemShut {NoStop}%
\bibitem [{\citenamefont {Prigogine}(1947)}]{prigogine1947etude}%
  \BibitemOpen
  \bibfield  {author} {\bibinfo {author} {\bibfnamefont {I.}~\bibnamefont
  {Prigogine}},\ }\href@noop {} {\emph {\bibinfo {title} {Etude thermodynamique
  des phénomènes irréversibles}}}\ (\bibinfo  {publisher} {Dunod, Paris},\
  \bibinfo {year} {1947})\BibitemShut {NoStop}%
\bibitem [{\citenamefont {Hill}(1977)}]{hill2005free}%
  \BibitemOpen
  \bibfield  {author} {\bibinfo {author} {\bibfnamefont {Terrell~L}\
  \bibnamefont {Hill}},\ }\href@noop {} {\emph {\bibinfo {title} {Free energy
  transduction in Biology}}}\ (\bibinfo  {publisher} {Academic Press},\
  \bibinfo {year} {1977})\BibitemShut {NoStop}%
\bibitem [{\citenamefont {Rao}\ and\ \citenamefont
  {Esposito}(2016)}]{rao2016nonequilibrium}%
  \BibitemOpen
  \bibfield  {author} {\bibinfo {author} {\bibfnamefont {Riccardo}\
  \bibnamefont {Rao}}\ and\ \bibinfo {author} {\bibfnamefont {Massimiliano}\
  \bibnamefont {Esposito}},\ }\bibfield  {title} {\enquote {\bibinfo {title}
  {Nonequilibrium thermodynamics of chemical reaction networks: wisdom from
  stochastic thermodynamics},}\ }\href
  {https://journals.aps.org/prx/abstract/10.1103/PhysRevX.6.041064} {\bibfield
  {journal} {\bibinfo  {journal} {Physical Review X}\ }\textbf {\bibinfo
  {volume} {6}},\ \bibinfo {pages} {041064} (\bibinfo {year}
  {2016})}\BibitemShut {NoStop}%
\bibitem [{\citenamefont {Avanzini}\ \emph {et~al.}(2021)\citenamefont
  {Avanzini}, \citenamefont {Penocchio}, \citenamefont {Falasco},\ and\
  \citenamefont {Esposito}}]{avanzini2021nonequilibrium}%
  \BibitemOpen
  \bibfield  {author} {\bibinfo {author} {\bibfnamefont {Francesco}\
  \bibnamefont {Avanzini}}, \bibinfo {author} {\bibfnamefont {Emanuele}\
  \bibnamefont {Penocchio}}, \bibinfo {author} {\bibfnamefont {Gianmaria}\
  \bibnamefont {Falasco}}, \ and\ \bibinfo {author} {\bibfnamefont
  {Massimiliano}\ \bibnamefont {Esposito}},\ }\bibfield  {title} {\enquote
  {\bibinfo {title} {Nonequilibrium thermodynamics of non-ideal chemical
  reaction networks},}\ }\href
  {https://aip.scitation.org/doi/full/10.1063/5.0041225} {\bibfield  {journal}
  {\bibinfo  {journal} {The Journal of Chemical Physics}\ }\textbf {\bibinfo
  {volume} {154}},\ \bibinfo {pages} {094114} (\bibinfo {year}
  {2021})}\BibitemShut {NoStop}%
\bibitem [{\citenamefont {Avanzini}\ and\ \citenamefont
  {Esposito}(2022)}]{avanzini2022}%
  \BibitemOpen
  \bibfield  {author} {\bibinfo {author} {\bibfnamefont {Francesco}\
  \bibnamefont {Avanzini}}\ and\ \bibinfo {author} {\bibfnamefont
  {Massimiliano}\ \bibnamefont {Esposito}},\ }\bibfield  {title} {\enquote
  {\bibinfo {title} {{Thermodynamics of concentration vs flux control in
  chemical reaction networks}},}\ }\href {https://doi.org/10.1063/5.0076134}
  {\bibfield  {journal} {\bibinfo  {journal} {J. Chem. Phys.}\ }\textbf
  {\bibinfo {volume} {156}} (\bibinfo {year} {2022})}\BibitemShut {NoStop}%
\bibitem [{\citenamefont {Batchelor}(1967)}]{batchelor1967introduction}%
  \BibitemOpen
  \bibfield  {author} {\bibinfo {author} {\bibfnamefont {George~Keith}\
  \bibnamefont {Batchelor}},\ }\href@noop {} {\emph {\bibinfo {title} {An
  introduction to fluid dynamics}}}\ (\bibinfo  {publisher} {Cambridge
  university press},\ \bibinfo {year} {1967})\BibitemShut {NoStop}%
\end{thebibliography}%
\end{document}